\documentclass[12pt,preprint]{aastex}
\usepackage{empheq}
\usepackage{multirow}
\usepackage{arydshln}

\shorttitle{GRB X-Ray Flare Properties among Different GRB Subclasses}

\shortauthors{Liu \& Mao}

\begin{document}

\title{\textbf{GRB X-Ray Flare Properties among Different GRB Subclasses}}
\author{Chuanxi Liu\altaffilmark{1,2,3} and Jirong Mao\altaffilmark{1,2,4}\\[5mm]
jirongmao@mail.ynao.ac.cn}
\altaffiltext{1}{Yunnan Observatories, Chinese Academy of Sciences, 650011 Kunming, Yunnan Province, China}
\altaffiltext{2}{Center for Astronomical Mega-Science, Chinese Academy of Sciences, 20A Datun Road, Chaoyang District, 100012 Beijing, China}
\altaffiltext{3}{University of Chinese Academy of Sciences, 100049 Beijing, China}
\altaffiltext{4}{Key Laboratory for the Structure and Evolution of Celestial Objects, Chinese Academy of Sciences, 650011 Kunming, China}

\begin{abstract}
Gamma-ray bursts (GRBs) can be divided into three subclasses: X-ray flash (XRF), X-ray rich (XRR), and classical GRB (C-GRB). An X-ray flare is the rebrightening emission shown in the early X-ray afterglow of some GRBs. In this paper, we comprehensively examine the X-ray flare properties among XRF, XRR, and C-GRB subclasses. We utilize the XRF, XRR, and C-GRB subclass samples obtained from the {\it Swift}-BAT3 catalog, and the X-ray flare observational properties are collected from Falcone et al., Chincarini et al., and Yi et al. We find that XRFs and XRRs have more bright X-ray flares than C-GRBs. The ratio of the X-ray flare fluence to the prompt emission fluence
has different distributions between XRF and C-GRB subclasses. The linear correlation between the duration and the peak time of the X-ray flares is also different between XRF and C-GRB subclasses.
We are inclined to identify the GRBs with the bright X-ray flares as XRFs or XRRs. We discuss some issues that are related to the XRF/XRR/C-GRB classification. We also caution the selection effects and the instrument bias in our investigation. Large samples are required in the future to further confirm our results.
\end{abstract}

\keywords{gamma rays: general - radiation mechanisms: non-thermal}

\section{Introduction}
Gamma-ray bursts (GRBs) are violent explosions in the gamma-ray band. They are specified by two types based on the burst duration time, notated as $T_{90}$ \citep{kou93}. A burst is a identified as long GRB when $T_{90}$ is more than 2 s, while a burst is identified as a short GRB when $T_{90}$ is less than 2 s. Besides $T_{90}$, the hardness ratio is another important parameter for the long/short GRB classification. de Ugarte Postigo et al. (2011) found that some GRBs in the {\it Swift}-detected sample are neither long nor short, and these GRBs can be classified as intermediate GRBs. The GRBs detected by BATSE and {\it Fermi}-GBM were further examined in the duration-hardness plane (Tarnopolski 2019). On the other hand, the GRB prompt spectrum can be described by the Band function. The Band function has three parameters: low-energy index $\alpha$, peak energy $E_p$, and high-energy index $\beta$ \citep{Band1993ApJ}. \cite{Preece2000ApJ} found that the peak energy spreads around 250 keV for the GRBs detected by BATSE. We recently classified GRBs into X-Ray Flashes (XRFs), X-Ray Riches (XRRs), and Classical GRBs (C-GRBs) using the latest {\it Swift}-BAT3 catalog \citep{Bi2018ApJ}. In particular, the peak energy of XRFs is located around 24 keV, which is consistent with the results of \citet{Sakamoto2005ApJ}. It is also noted that some XRFs are associated with supernovae \citep{fynbo04,levan05,Bersier2006ApJ,soderberg06}. The relation between GRB 120422A and SN 2012bz indicates that low-luminosity GRBs are relevant to supernovae at low redshift \citep{Schulze2014AA}. \cite{Bi2018ApJ} took 6 XRFs, 14 XRRs, and 3 C-GRBs that are related to the supernova explosion into account in the {\it Swift}-BAT3 catalog.

Some models have been proposed to reveal the physical origins for XRFs. Different baryon loading can have an effect on the GRB bulk Lorentz factor \citep{dermer99}. \citet{barraud05} suggested the different bulk Lorentz factors of the shock waves for XRFs and C-GRBs. In general, jet off-axis/beaming effects have been accepted to explain the XRF origin \citep{yamazaki02,granot05,Lamb2005ApJ,salafia16}. \cite{Matzner2013ApJ} proposed that the nonradial motion of the GRB outflow near the progenitor stellar surface can suppress the photon flash so that the low energetic GRBs can be discovered. \cite{Ciolfi2016ApJ} suggested that XRF 020903 can originate from the neutron star spindown if this XRF was formed by the binary neutron star or the core-collapse supernova, because the modeling spectrum mainly falls in the X-ray band and the modeling luminosity is similar to the XRF luminosity. This suggests that XRFs can be related to Poynting-dominated jets. Moreover, photospheric radiation can also be applied for the explanation of the XRF properties (e.g., Pe'er et al. 2006).

X-ray flares are the rebrightening emission shown in some GRB X-ray afterglows \citep{falcone06}.
Some statistical analysis results from the {\it Swift} samples have been achieved. Chincarini et al. (2007) performed one temporal investigation on 69 X-ray flares, and they concluded that the X-ray flare and the prompt emission have the same origin. Falcone et al. (2007) performed one spectral investigation on 77 X-ray flares, and the Band function can be a good fitting to the X-ray flare spectra. \cite{Chincarini2010MN} systematically studied the temporal properties of the X-ray flares with a large sample. They found that the X-ray flare duration decreases with the energy as $w\propto E^{-0.5}$ and linearly increases with the X-ray flare peak time as $w\sim0.2 t_{p}$, where $w$ is the duration, $E$ is the isotropic energy in the 0.3-10 keV band, and $t_p$ is the peak time. \cite{Shuangxi2016ApJ} enlarged the X-ray flare sample from the updated {\it Swift}-XRT dataset.
From a statistical point of view, about 30\% of GRBs have X-ray flares \citep{Chincarini2010MN,Margutti2011MN-2,Shuangxi2016ApJ}.

There are some models to explain X-ray flare origin. X-ray flares that are contemporaneous with the early X-ray afterglow are likely due to a late time manifestation of the same emission mechanism as the prompt emission \citep{Zhang2006ApJ}. \cite{Ioka2005ApJ} and \cite{Lazzati2007MN} proposed several possibilities to explain the GRB X-ray flare. \cite{Chincarini2010MN} put the data on the
log($\Delta F/F$)-log($w/t_{pk}$) plane to examine the different models suggested by \cite{Ioka2005ApJ}, where $\Delta F$ is the fluence of the X-ray flare, and F is the underlying fluence. It looks like that X-ray flares have an internal origin from a long-active central engine. Mu et al. (2018) indicated that the X-ray flares of short GRBs may also have an internal origin.
\cite{Jin2010ApJ} estimated the Lorentz factor of the X-ray flare in their model, and the Lorentz factor of the X-ray flare is smaller than that of the general GRB outflow. \cite{Mu2016ApJ} took use of the curvature effect to further constrain the Lorentz factor of the X-ray flare. It indicated long-time activity from the GRB central engine to explain the observed GRB X-ray flares. \cite{Jia2016ApJ} examined the temporal decay index and the spectral index of the X-ray flares, and the results suggested a Poynting-dominated outflow for the X-ray flare production \citep{uhm16}. \citet{geng17} further calculated an anisotropic effect in the X-ray flare radiation.
\cite{Lazzati2011MN} simulated GRB jet propagation. It was inferred that the X-ray flare originates from some jet propagation instabilities in different cases of the jet opening angle.
\cite{Giannios2012MN} built a dissipative photospheric model to obtain the bulk Lorentz factor that could be used to explain both the spectrum and the light curve for the X-ray flares. \cite{Beniamini2016MN} examined the small emission radius of the X-ray flares and further considered the photosphere model for the X-ray flare origin. It is also possible for a remnant disk accretion to produce X-ray flares if the magnetic coupling is involved between the inner disk and the central black hole \citep{Luo2013ApJ}.

We attempt to comprehensively examine the X-ray flare properties among XRF, XRR, and C-GRB subclasses using the {\it Swift} samples in this paper. We utilize our statistical results presented by \cite{Bi2018ApJ}, in which XRFs, XRRs, and C-GRBs were classified from the latest \textit{Swift}-BAT3 catalog. We collect the X-ray flare data from both \cite{Chincarini2010MN} and \cite{Shuangxi2016ApJ} for the temporal analysis and the X-ray flare data from Falcone et al. (2007) for the spectral analysis. Because some similar explanations in the former descriptions have been proposed for both XRFs and X-ray flares, it indicates that X-ray flares may have some special associations to XRFs. In fact, some GRBs, such as GRB 040916, GRB 050502B, and GRB 050406, with prominent X-ray flares, were identified as XRFs or XRRs \citep{Burrows2005Sci, Arimoto2007PASJ}. We expect some statistical differences of the X-ray flare properties between XRF and C-GRB subclasses in this paper. However, the long/short GRB identification and the XRF/XRR/C-GRB classification are important to our results in this paper. For example, GRB 050502B is classified as an XRR, but its prompt emission has a power-law spectral index of 1.6 over the primary time of 6 s. Thus, it is very similar to a C-GRB. We provide a  comprehensive discussion on the GRB classifications in this paper.

We present the data selection in Section 2. In section 3, we show some general properties of XRFs, XRRs, and C-GRBs. Then, we illustrate the properties of the X-ray flare fluence for XRFs, XRRs, and C-GRBs. We investigate the temporal properties of the X-ray flares for XRFs, XRRs, and C-GRBs. Some properties of the redshift-corrected parameters are presented. The linear correlation between the duration and the peak time of the X-ray flares for XFRs, XRRs, and C-GRBs are given. We plot the log($\Delta F/F$)-log($w/t_p$) distribution of the X-ray flares for XRFs, XRRs, and C-GRBs. Finally, the spectral properties of the X-ray flares among XRF, XRR, and C-GRB subclasses are illustrated.
In section 4, we comprehensively discuss some issues, such as sample classification, selection effect, and instrument bias, which affect our results. We list the conclusions in Section 5.

We use the standard cosmological parameters: $H_{0}=72~\rm{km~s^{-1}~Mpc^{-1}}$, $\Omega_{\Lambda}=0.7$, and $\Omega_{M}=0.3$.

\section{Data Description}
We utilize the GRB X-ray flare temporal data from both \cite{Chincarini2010MN} and \cite{Shuangxi2016ApJ}. The X-ray flares presented by \cite{Chincarini2010MN} were observed from 2005 April to 2008 March, while those of \cite{Shuangxi2016ApJ} were observed from 2005 April to 2015 March. The two samples cover the same observational period, from 2005 April to 2008 March.
In this period, some GRBs having X-ray flares identified in \cite{Shuangxi2016ApJ} are not shown in \cite{Chincarini2010MN}. These GRBs are GRB 050712, GRB 050724, GRB 050803, GRB 050820A, GRB 050904, GRB 050915A, GRB 050916, GRB 060124, GRB 060223A, GRB 060510B, GRB 060522, GRB 060926, GRB 061121, GRB 070103, GRB 070129, GRB 070616, GRB 070714A, GRB 071112C, GRB 071122, GRB 080229A, GRB 080319A, and GRB 080325. In the same period, some GRBs having X-ray flares identified in \cite{Chincarini2010MN} are not shown in \cite{Shuangxi2016ApJ}. These GRBs are GRB 051210, GRB 051227, GRB 060729, GRB 060814, GRB 060908, GRB 070220, GRB 070306, and GRB 070621.

We also consider the spectral properties of the X-ray flares among XRF, XRR, and C-GRB subclasses. We utilize the data from Falcone et al. (2007). Thirty-three GRBs were observed from 2005 February to 2006 January, and 77 X-ray flares with the spectral analysis are included in the sample.

We have classified GRBs into XRFs, XRRs, and C-GRBs from the {\it Swift}-BAT3 catalog \citep{Bi2018ApJ}. The criteria for the classification of XRFs, CRRs, and C-GRBs were introduced by Sakamoto et al. (2008). We select the XRFs, XRRs, and C-GRBs that have the X-ray flares identified from Falcone et al. (2007), \cite{Chincarini2010MN}, and \cite{Shuangxi2016ApJ} to be the data sample in this paper. We list each GRB having the X-ray flare temporal properties in Table \ref{Tab:infor}. In general, we have 7 XRFs, 34 XRRs, and 15 C-GRBs having X-ray flares from the sample of Chincarini et al. (2010), where 1 XRR and 1 C-GRB are short bursts. We have 16 XRFs, 118 XRRs, and 64 C-GRBs having X-ray flares from the sample of Yi et al. (2016), where 1 XRR and 2 C-GRBs are short bursts.

We note that some GRBs do not have fluence values in the {\it Swift}-BAT sample. They are GRB 060602, GRB 071112C, GRB 081028, GRB 090516, GRB 090807, and GRB 090809. In addition, the GRB 050714B X-ray flare has no measurements on the peak time and the duration in \cite{Chincarini2010MN}.

\section{Results}
\subsection{General Properties of XRFs, XRRs, and C-GRBs}
\subsubsection{$T_{90}$ Distributions}
\begin{figure}[h!]
    \centering
    \includegraphics[height=.6\linewidth - 0.25mm]{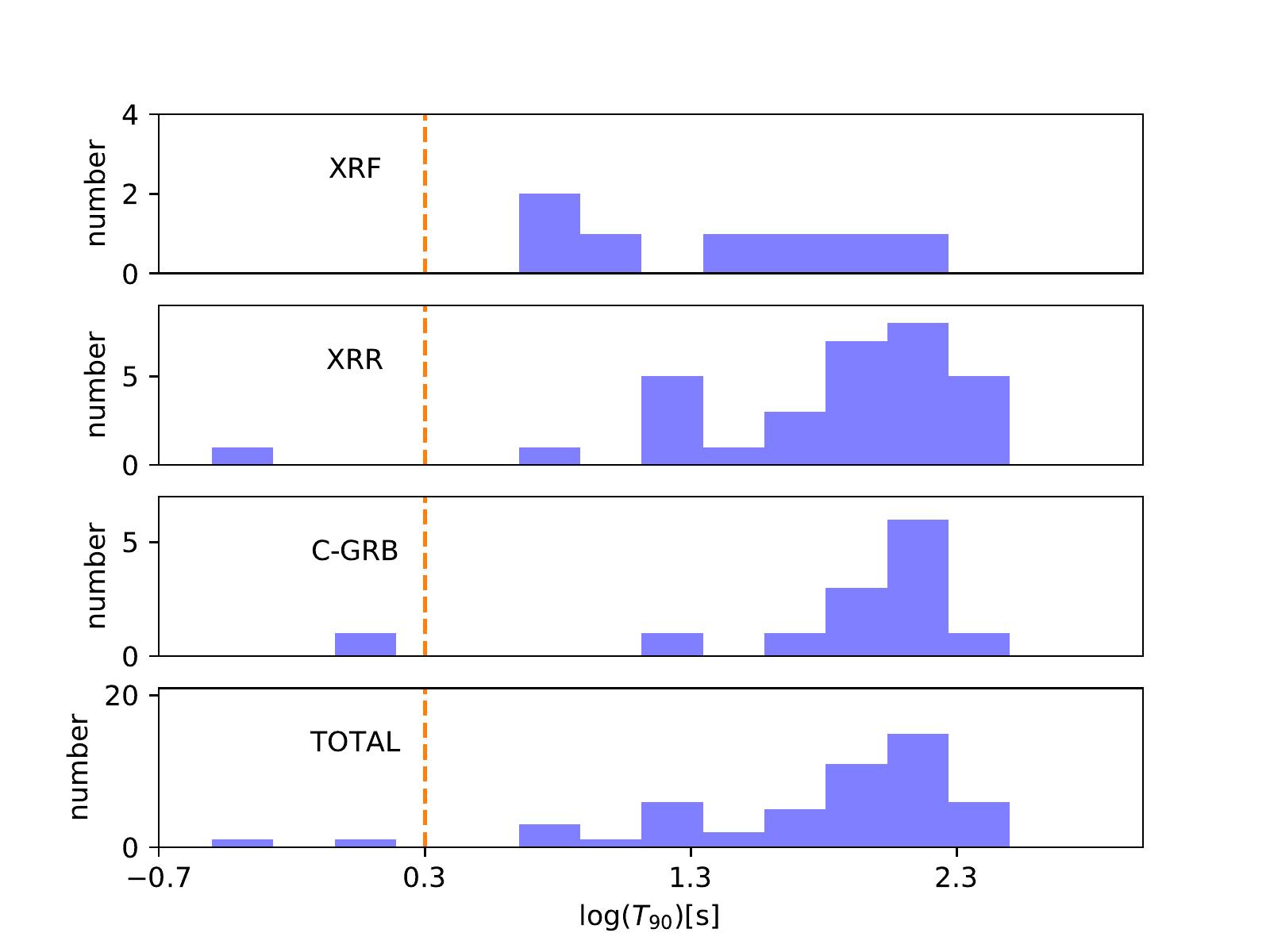}
    \includegraphics[height=.6\linewidth - 0.25mm]{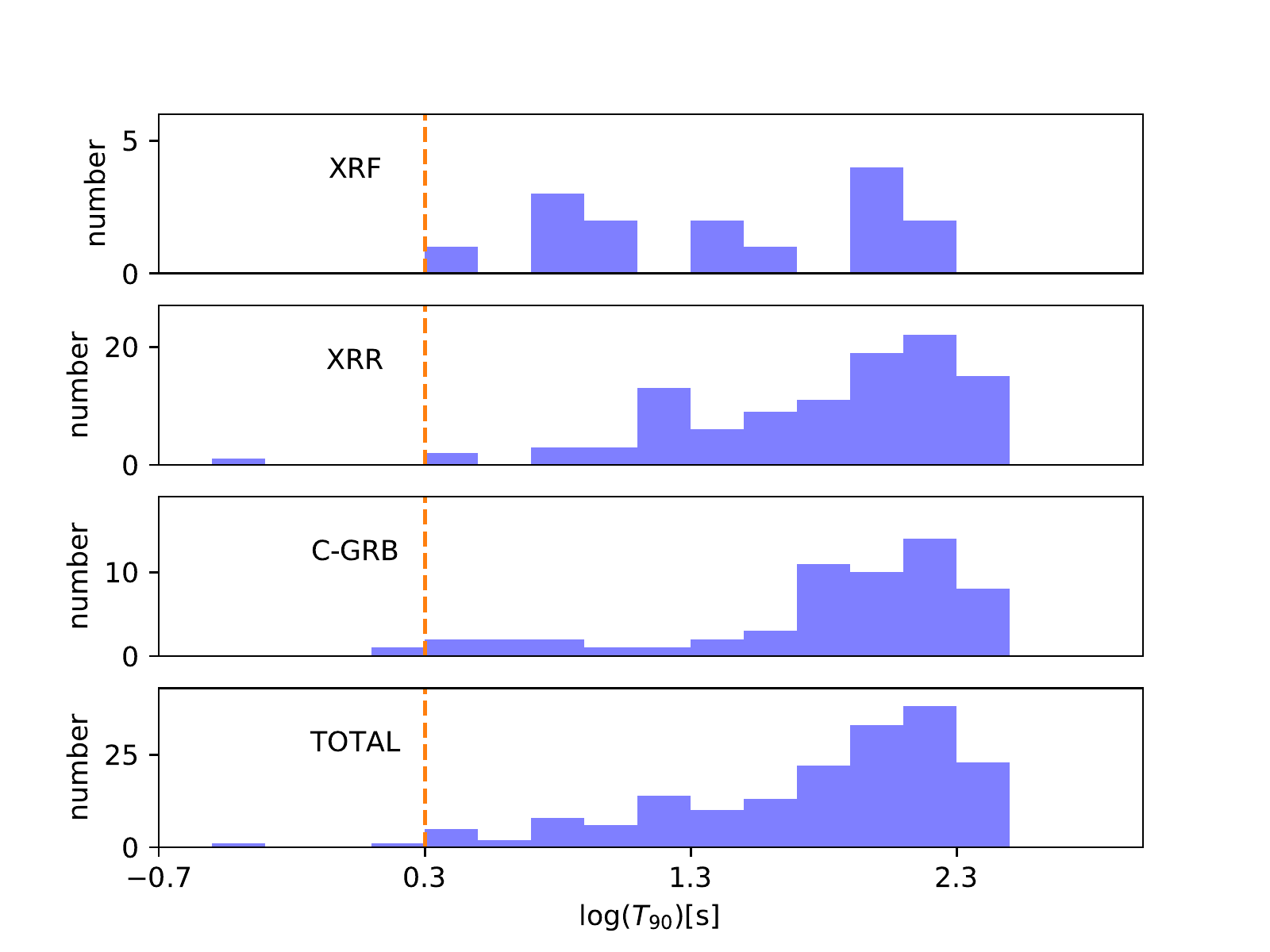}
    \caption{The $T_{90}$ distributions of XRFs, XRRs, and C-GRBs. The dashed line separates the sources to short and long GRB catalogs. We find that all of the XRFs with X-ray flares in this sample are long GRBs. Top panel: the X-ray flares are collected from \cite{Chincarini2010MN}. Bottom panel: the X-ray flares are collected from \cite{Shuangxi2016ApJ}.}
    \label{Fig:T90}
\end{figure}

We examine the $T_{90}$ distributions for the XRFs, XRRs, and C-GRBs with the temporal properties of the X-ray flares. We show the results in Figure \ref{Fig:T90}.
Although it seems that all the XRFs with X-ray flares are long GRBs, we may consider some XRFs with X-ray flares that are short bursts found in the future observations. We also note that some XRFs are identified as intermediate GRBs that are neither short nor long \citep{ho10}. We list the intermediate GRBs in Table 1. The sources were identified by de Ugarte Postigo et al. (2011)\footnote{The intermediate GRBs in the sample of de Ugarte Postigo et al. (2011) occurred from 2004 December to 2008 December. The sample is not large enough, so we cannot use it to perform further analysis. We also note some issues with the intermediate GRBs in Section 4.}.

In order to examine the $T_{90}$ distribution differences among XRF, XRR, and C-GRB subclasses, we further perform the Kolmogorov-Smirnov (K-S) test on the $T_{90}$ distributions. We first take the GRBs that have the X-ray flares from Chincarini et al. (2010). We obtain the $p$-value 0.11 that indicates the $T_{90}$ distribution difference between XRF and C-GRB subclasses. The $p$-value 0.84 is obtained to indicate the $T_{90}$ distribution difference between XRR and C-GRB subclasses, and the $p$-value 0.26 is obtained to indicate the $T_{90}$ distribution difference between XRR and XRF subclasses. If we exclude the short bursts in XRFs, XRRs, and C-GRBs, the numbers of the $p$-value are 0.06, 0.67, and 0.21 to indicate $T_{90}$ distribution differences of XRF and C-GRB subclasses, XRR and C-GRB subclasses, and XRF and XRR subclasses, respectively. Then, we take the GRBs that have the X-ray flares from Yi et al. (2016). We obtain the numbers of the $p$-value as 0.10, 0.82, and 0.08 to indicate the $T_{90}$ distribution differences of XRF and C-GRB subclasses, XRR and C-GRB subclasses, and XRF and XRR subclasses, respectively. If we exclude short GRBs in XRFs, XRRs, and C-GRBs, we obtain the numbers of the $p$-value as 0.06, 0.63, and 0.07 to indicate the $T_{90}$ distribution differences of XRF and C-GRB subclasses, XRR and C-GRB subclasses, and XRF and XRR subclasses, respectively. We write the $p$-value numbers for the difference cases in Table 2.

It seems that $T_{90}$ distributions among XRF, XRR, and C-GRB subclasses have high similarity. Thus, there are no significant differences of $T_{90}$ distribution among XRF, XRR, and C-GRB subclasses. This result is consistent with that of \cite{Sakamoto2005ApJ}.

\subsubsection{Hardness-duration and Peak Energy Distributions of XRFs, XRRs, and C-GRBs}
\begin{figure}[h!]
    \centering
    \includegraphics[width=.98\textwidth]{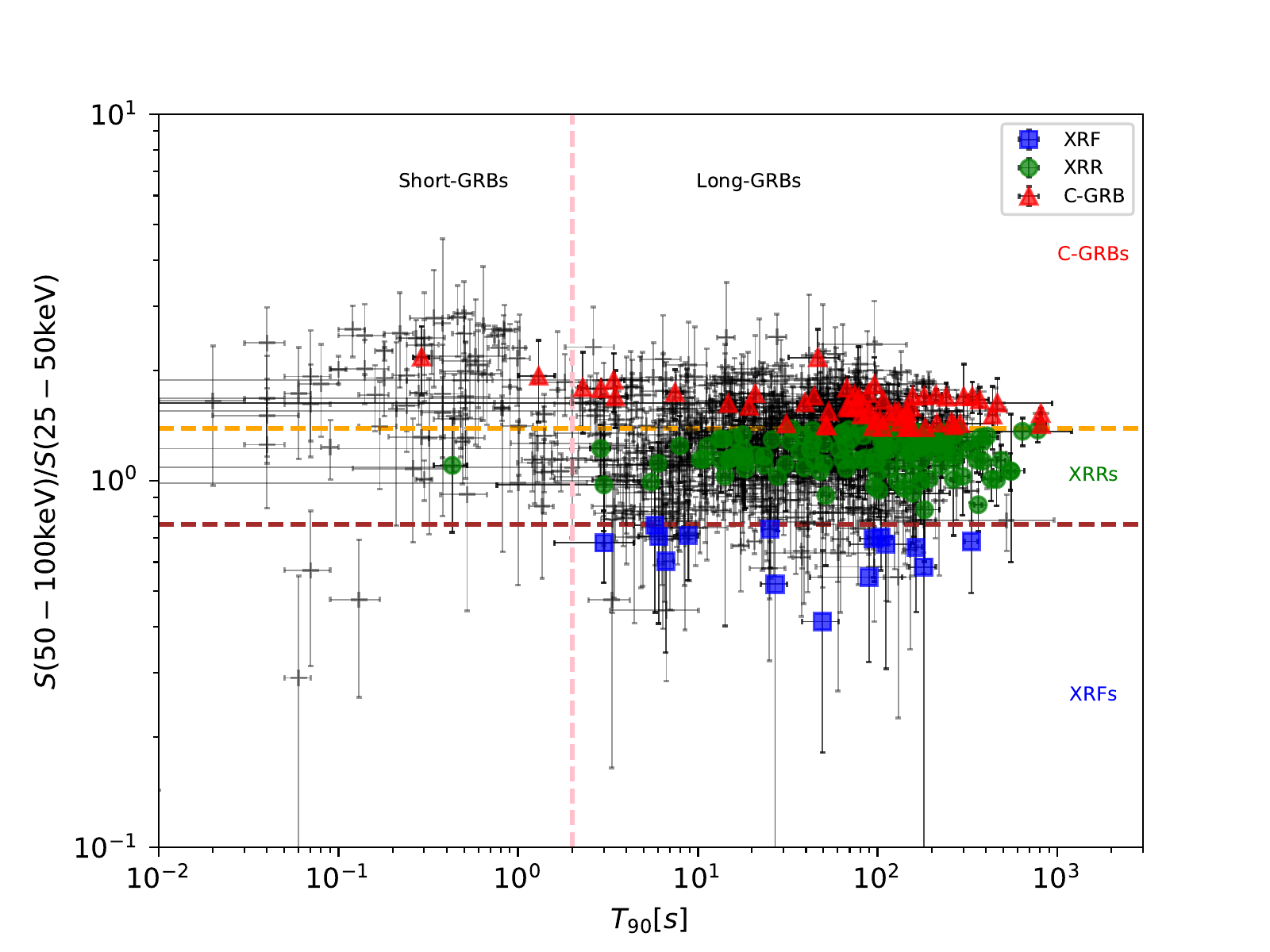}
    \caption{The hardness$-T_{90}$ distributions of XRFs, XRRs, and C-GRBs having X-ray flares in our sample. There are only three sources that fall into the short burst region. They are XRR 070724A, C-GRB 051210, and C-GRB 100117. We also plot all the GRBs without X-ray flares marked as black crosses in the Figure.}
    \label{Fig:hardness_T90}
\end{figure}

In order to further examine the classification of XRFs, XRRs, and C-GRBs in our sample, we first investigate the hardness-duration distribution for XRFs, XRRs, and C-GRBs in Figure \ref{Fig:hardness_T90}. Although the classification can successfully separate XRFs from C-GRBs, we see the overlaps between XRR and C-GRB subclasses, and between XRF and XRR subclasses.

\begin{figure}[h!]
    \centering
    \includegraphics[width=.98\textwidth]{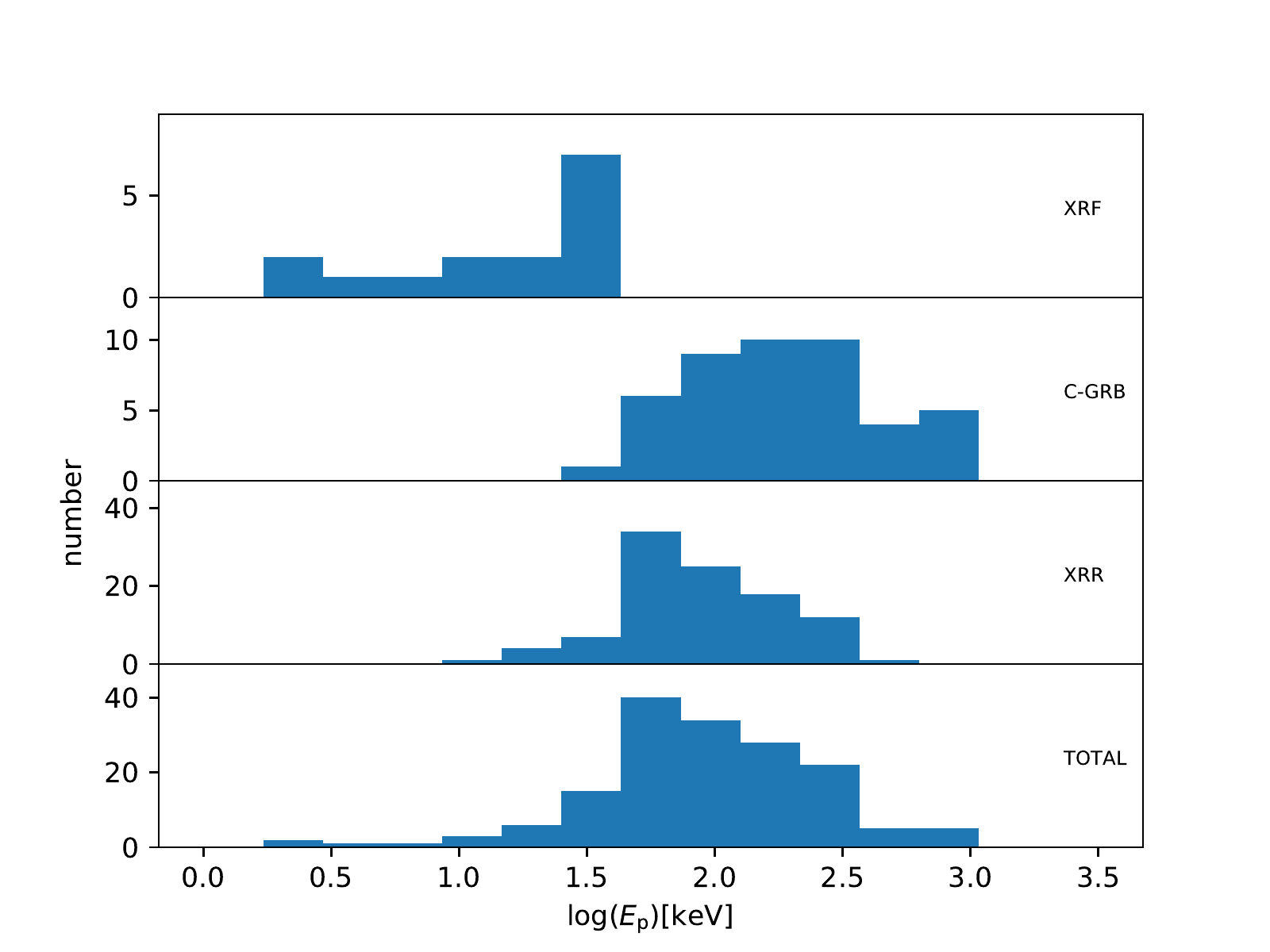}
    \caption{The peak energy $E_{\rm{peak}}$ distributions of XRFs, XRRs, and C-GRBs with X-ray flares in our sample.}
    \label{Fig:GRBEp}
\end{figure}

We then investigate the peak energy of XRFs, XRRs, and C-GRBs in our sample. The peak energy distributions are shown in Figure \ref{Fig:GRBEp}. We perform the K-S test for the distributions of XRF and C-GRB subclasses, XRF and XRR subclasses, and XRR and C-GRB subclasses, respectively. The results are given in Table 2. We clearly see that the $E_p$ distributions among XRF, XRR, and C-GRB subclasses are different. However, we note that the distribution overlaps are also shown in Figure \ref{Fig:GRBEp}.

\subsection{X-Ray Flare Fluence Properties among XRF, XRR, and C-GRB Subclasses}
\subsubsection{Fluence Ratio between X-Ray Flare and Prompt Emission}
We perform a detailed analysis on the X-ray flare fluence properties for XRFs, XRRs, and C-GRBs in this subsection. We obtain the fluence of each individual X-ray flare $S_{i,\rm{flare}}$ from \cite{Chincarini2010MN} and \cite{Shuangxi2016ApJ}. The fluence of the GRB prompt emission $S_{\rm{prompt}}$ can be obtained from the {\it Swift}-BAT3 Catalog. In order to properly understand the energy released from the X-ray flare that can be compared to the energy released from the prompt emission, we define a parameter of $r_i$, and it can be identified as the fluence ratio between one single flare and the prompt emission for each GRB. Some GRBs have multiple X-ray flares, then we accumulate all the $S_{i,\rm{flare}}$ numbers to get the total fluence of the X-ray flares $S_{t,\rm{flare}}$ in a single GRB. A parameter of $r_t$ can be identified as the fluence ratio between the total flares and the prompt emission for each of these GRBs. Thus, we write the following formula for a certain GRB as:
\begin{align}
	&r_i = S_{i,flare}/S_{prompt}, \\
	&S_{t, flare} = \sum_i S_{i, flare}, \\
	&r_t = S_{t,flare}/S_{prompt}.
\end{align}

\begin{figure}[h!]
    \centering
    \includegraphics[height=.6\linewidth - 0.25mm]{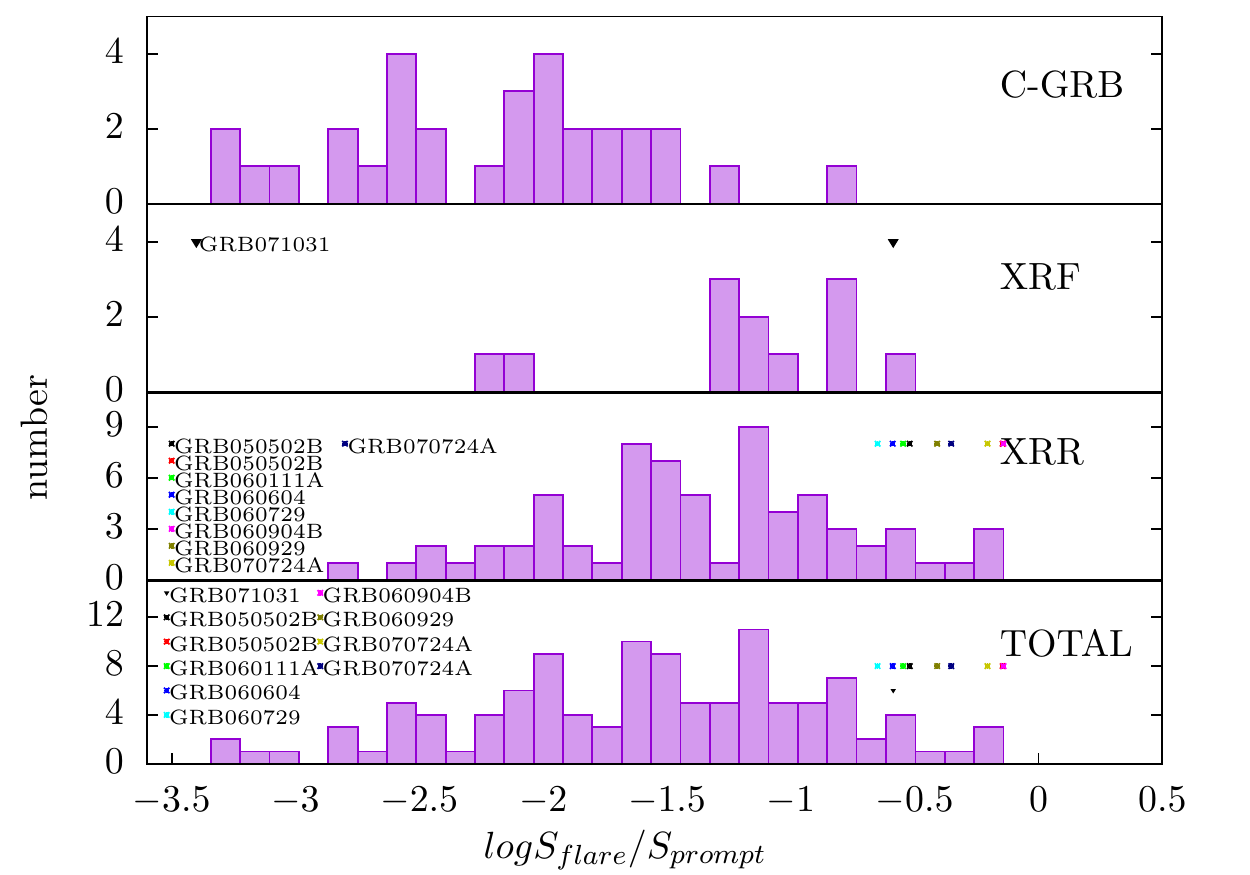}
    \includegraphics[height=.6\linewidth - 0.25mm]{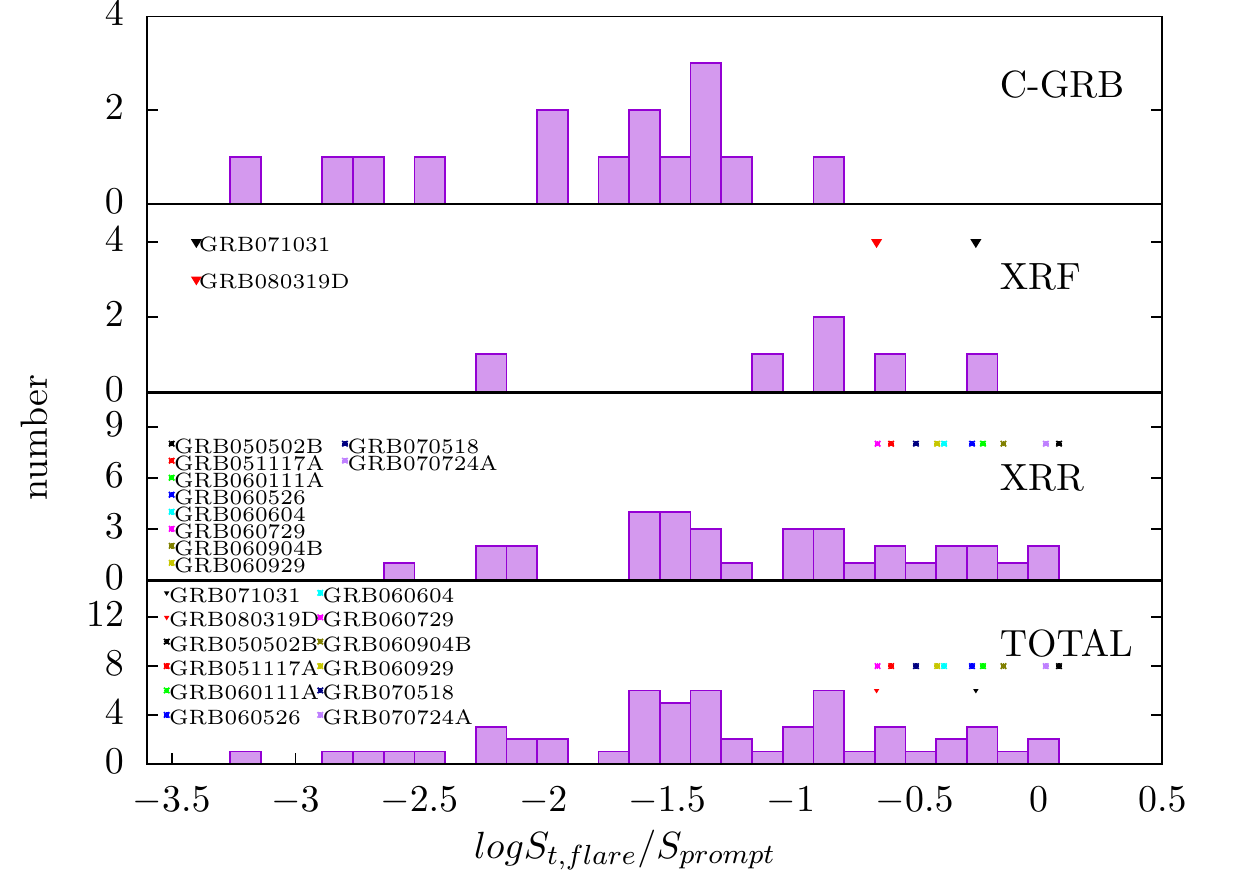}
    \vglue -0.5cm
    \caption{Distributions of the X-ray flare fluence ratio for XRFs, XRRs, and C-GRBs. The X-ray flare data are from \cite{Chincarini2010MN}. We mark the GRBs with bright X-ray flares. Top panel: $r_i$ distribution. Bottom panel: $r_t$ distribution.
}
    \label{Fig:distribution:ratio_t:Chin}
\end{figure}

\begin{figure}[h!]
    \centering
    \includegraphics[height=.6\linewidth - 0.25mm]{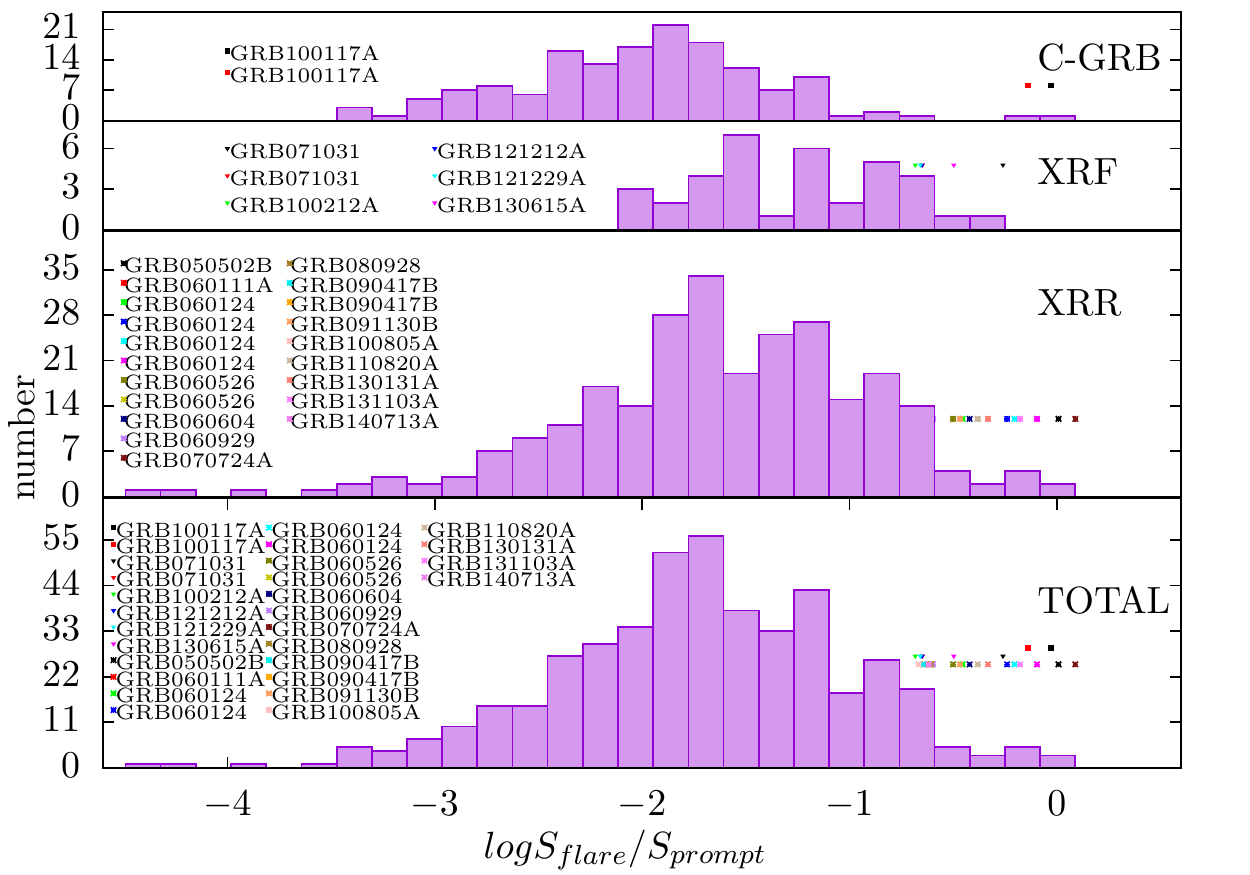}
    \includegraphics[height=.6\linewidth - 0.25mm]{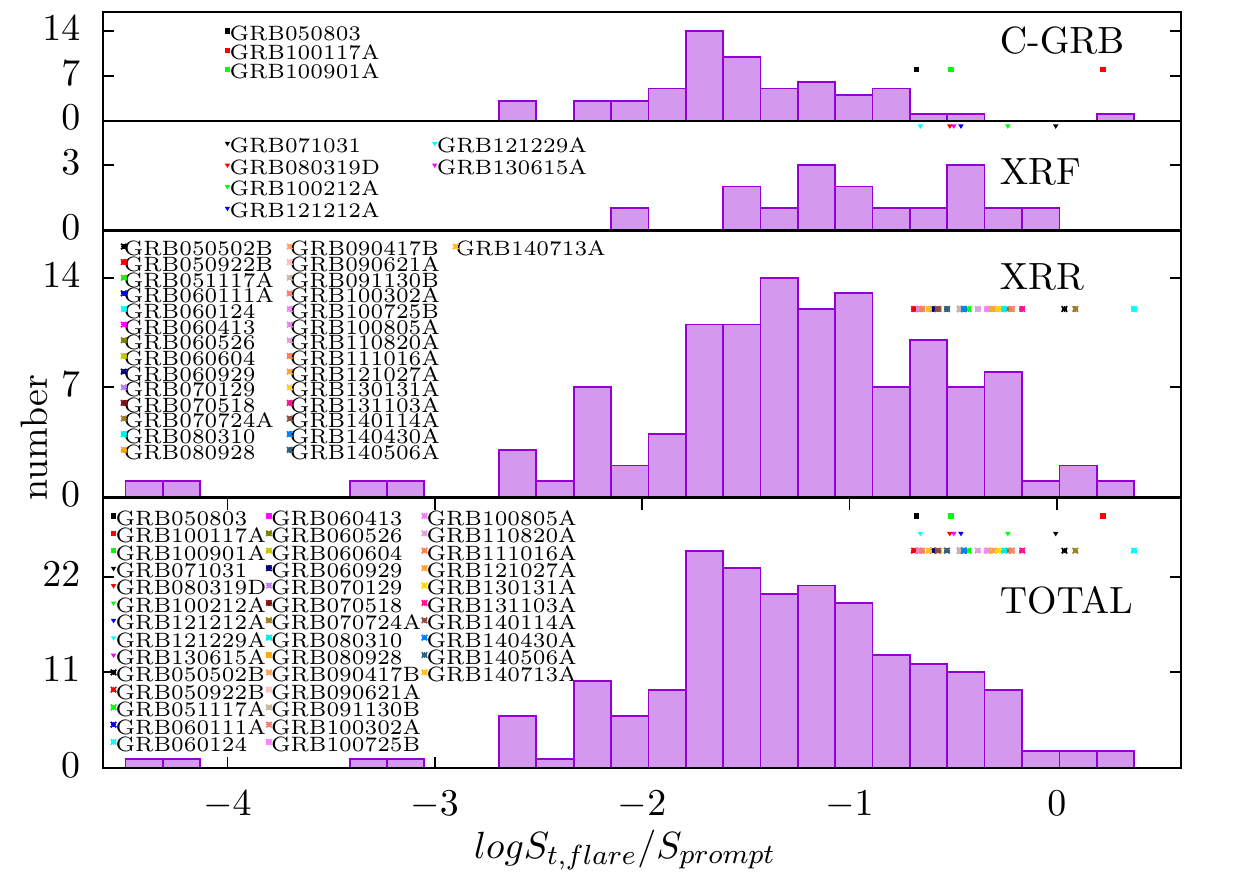}
    \vglue -0.5cm
    \caption{Distributions of the X-ray flare fluence ratio for XRFs, XRRs, and C-GRBs. The X-ray flare data are from Yi et al. (2016). We mark the GRBs with bright X-ray flares. Top panel: $r_i$ distribution. Bottom panel: $r_t$ distribution.
    }
    \label{Fig:distribution:ratio_t:Shuang}
\end{figure}

We select the X-ray flare data of \cite{Chincarini2010MN} and investigate the $r_i$ and $r_t$ distributions for XRFs, XRRs, and C-GRBs. The distributions are shown in Figure \ref{Fig:distribution:ratio_t:Chin}. We further perform the K-S test, and the results are written in Table \ref{Tab:K-S}. We also take the X-ray flare data of \cite{Shuangxi2016ApJ} to investigate the $r_i$ and $r_t$ distributions for XRFs, XRRs, and C-GRBs. The distributions are shown in Figure \ref{Fig:distribution:ratio_t:Shuang}, and the K-S test results are also written in Table \ref{Tab:K-S}. Although different $p$-value numbers are obtained among different distributions, we clearly identify that XRF and C-GRB subclasses are different on either $r_i$ or $r_t$ distributions.

\begin{figure}[h!]
    \centering
    \includegraphics[height=.6\linewidth - 0.25mm]{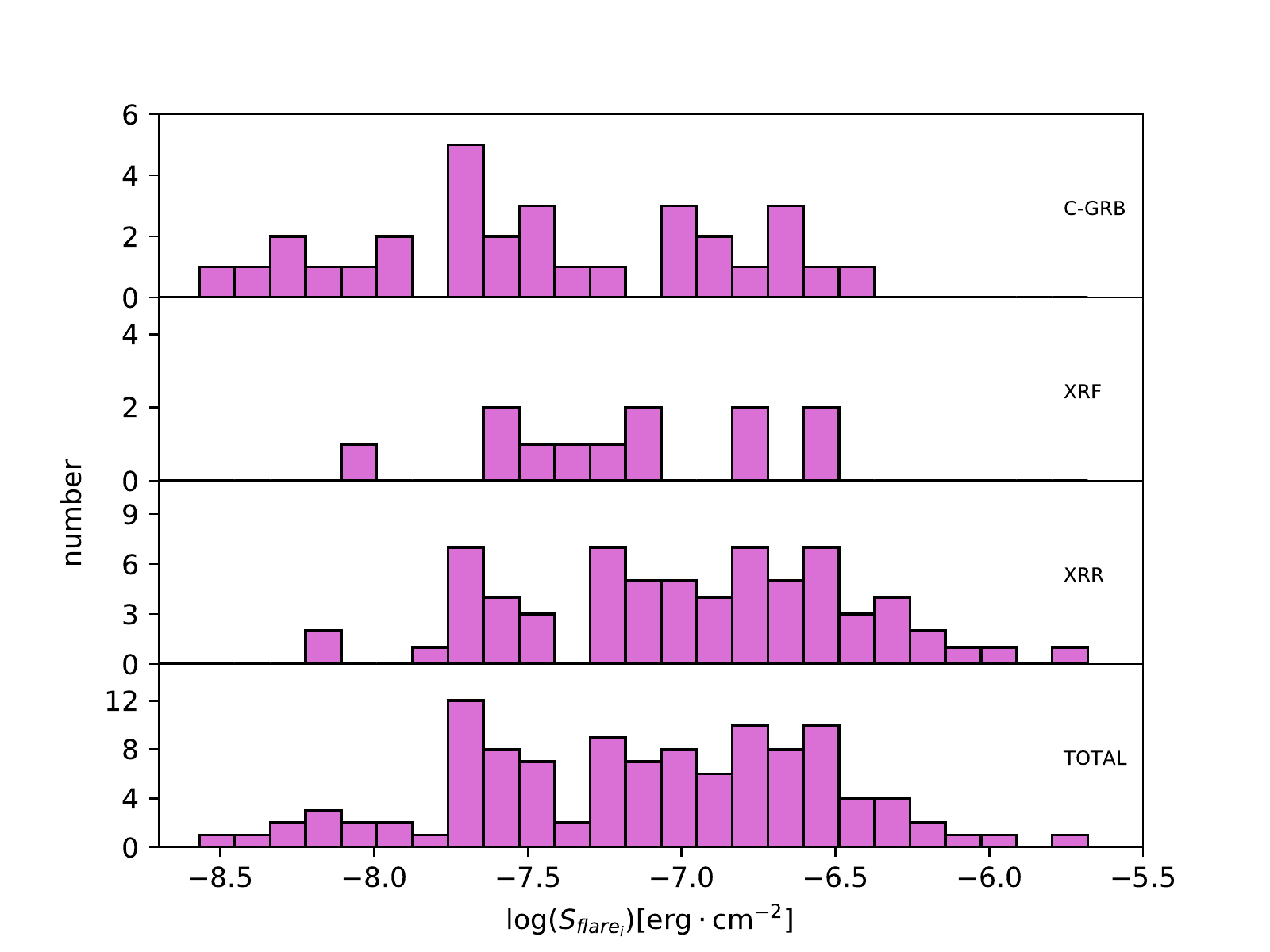}
    \includegraphics[height=.6\linewidth - 0.25mm]{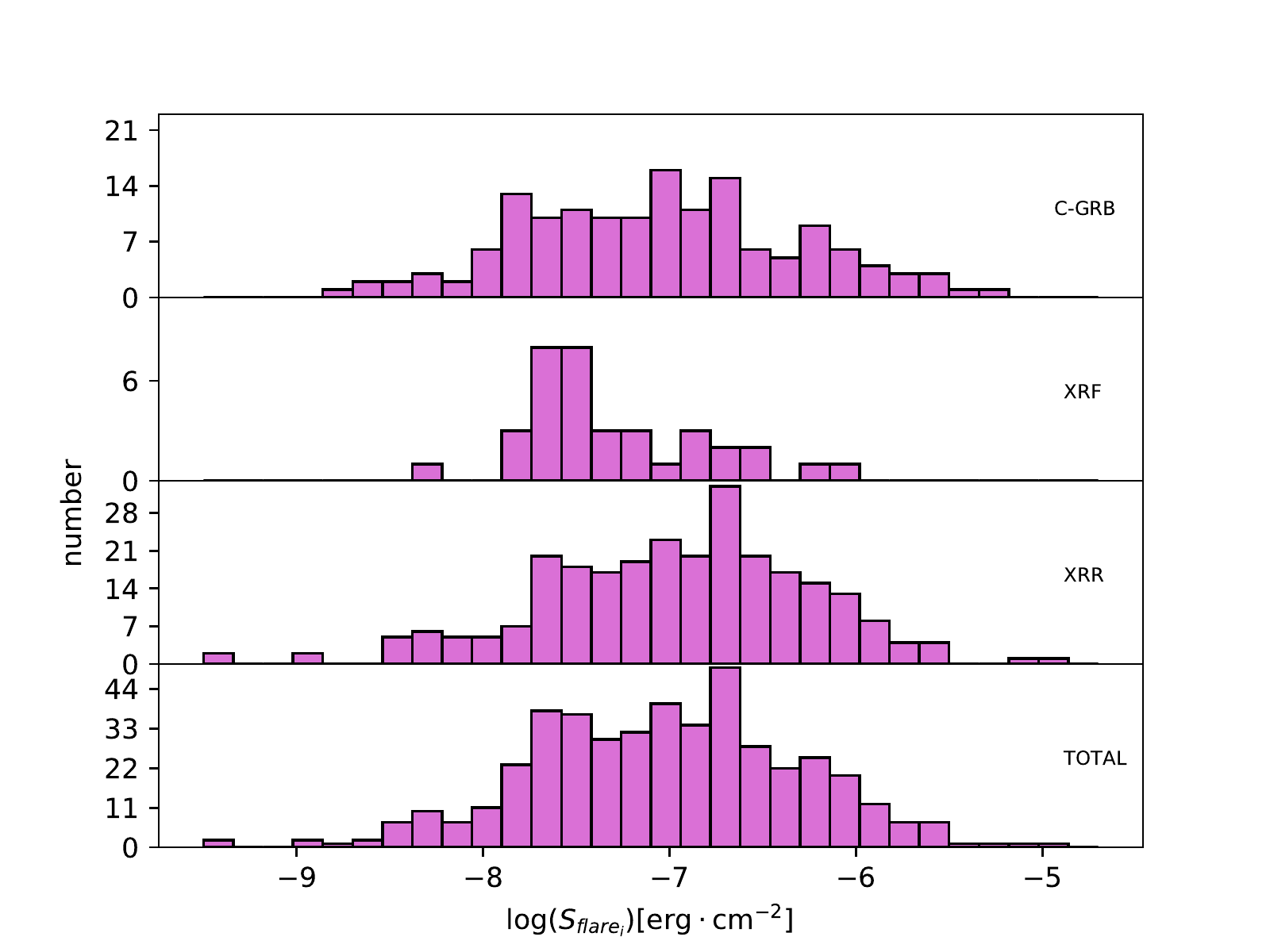}
    \vglue -0.5cm
    \caption{Distributions of the X-ray flare fluence $S_{i,\rm{flare}}$ for XRFs, XRRs, and C-GRBs. Top panel: the X-ray flare data are from Chincarini et al. (2010). Bottom panel: the X-ray flare data are from Yi et al. (2016).
}
    \label{Fig:distribution:S_flare}
\end{figure}

\begin{figure}[h!]
    \centering
    \includegraphics[height=.6\linewidth - 0.25mm]{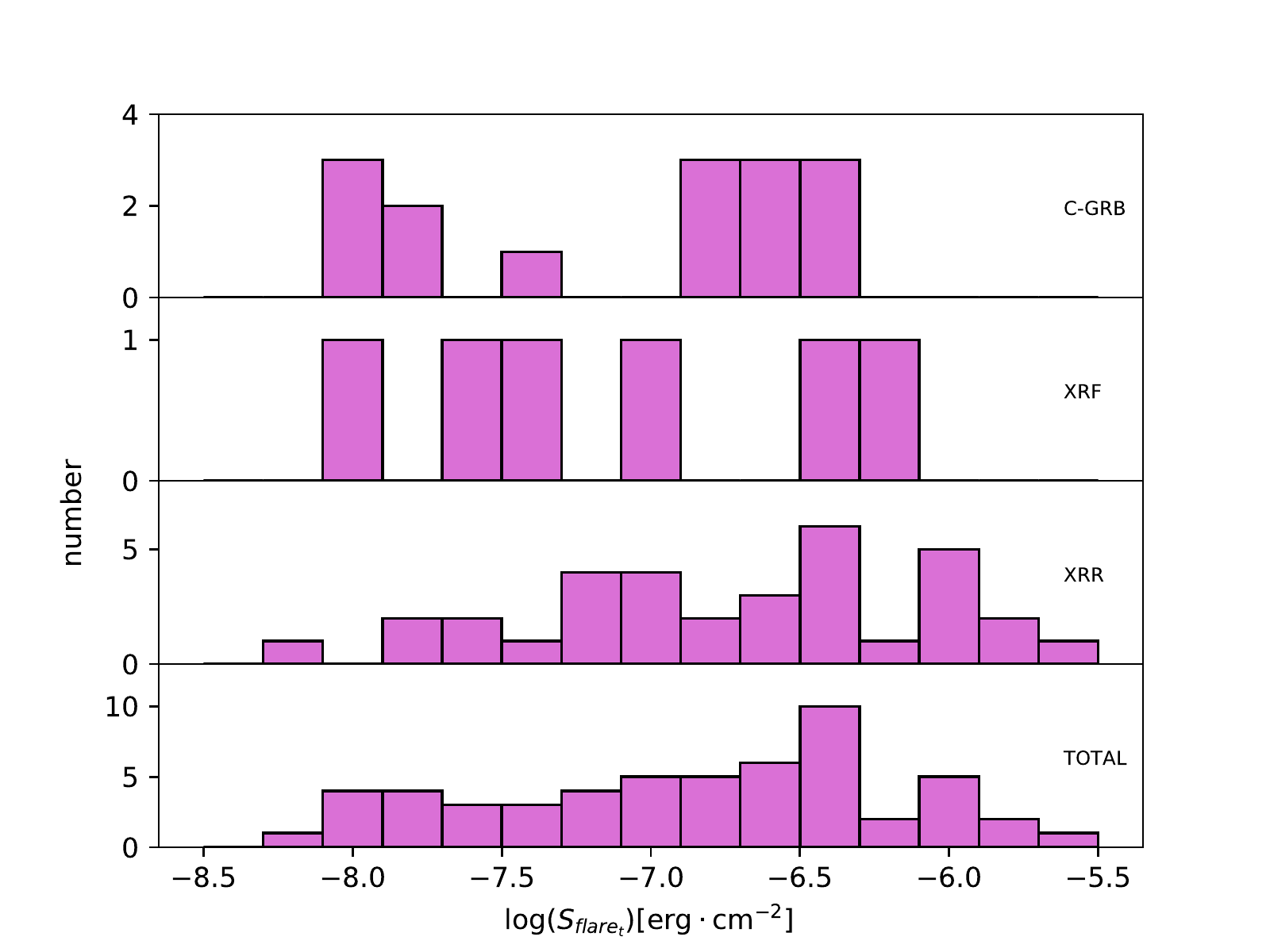}
    \includegraphics[height=.6\linewidth - 0.25mm]{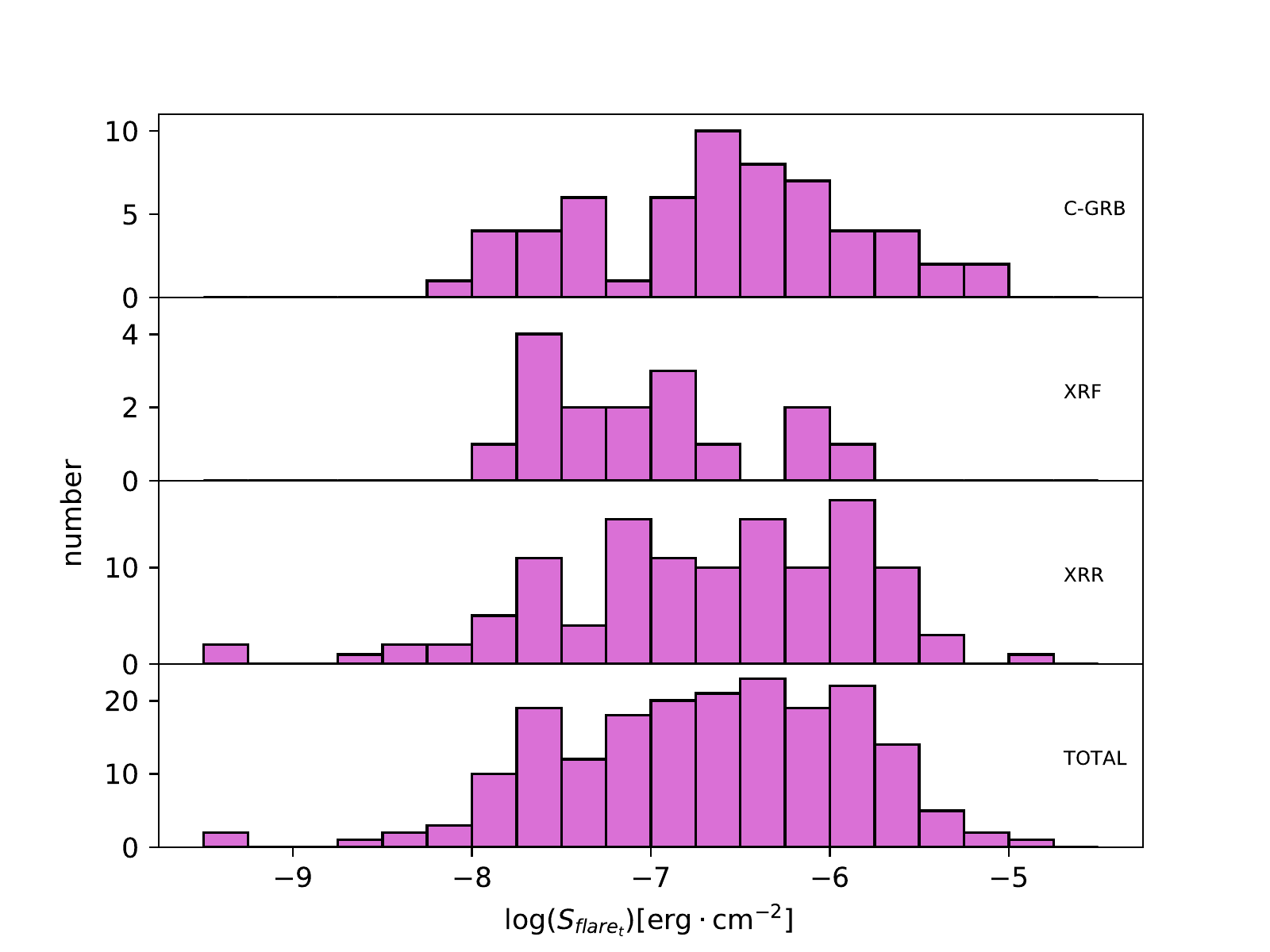}
    \vglue -0.5cm
    \caption{Distributions of the X-ray flare fluence $S_{t,\rm{flare}}$ for XRFs, XRRs, and C-GRBs. Top panel: the X-ray flare data are from Chincarini et al. (2010). Bottom panel: the X-ray flare data are from Yi et al. (2016).
}
    \label{Fig:distribution:S_flare_t}
\end{figure}

We attempt to directly compare the X-ray flare fluence among C-GRB, XRR, and XRF subclasses using the X-ray flare data from \cite{Chincarini2010MN} and \cite{Shuangxi2016ApJ}. We show the fluence distributions of the X-ray flares for C-GRBs, XRRs and XRFs in Figures \ref{Fig:distribution:S_flare} and \ref{Fig:distribution:S_flare_t}. We also perform the K-S test, and the results are written in Table \ref{Tab:K-S}. However, because some GRBs have no redshift measurements, we caution the direct fluence comparison of the X-ray flares among XRF, XRR, and C-GRB subclasses.

\subsubsection{Bright X-Ray Flares}
Some XRFs and XRRs have prominent X-ray flares \citep{Burrows2005Sci, Arimoto2007PASJ}.
We have compared the X-ray flare fluence to the prompt emission fluence in each GRB and performed the statistics on the fluence ratio in Section 3.2.1. Here, we define a threshold ratio between the X-ray flare fluence and the prompt emission fluence. The bright X-ray flare is satisfied with the condition of $r_i\ge 0.2$. When a GRB has multiple X-ray flares, the bright X-ray flares with the accumulated fluence is satisfied with the condition of $r_t\ge 0.2$. We take notes on the GRBs having the bright X-ray flares in Figures 4, 5, 11, and 12.

We take the XRFs, XRRs, and C-GRBs that have the bright X-ray flares into account, and we define three ratios as
\begin{align}
	\varepsilon_{C-GRB} &= \frac{N_{C-GRB,\,r_t\ge 0.2}}{N_{C-GRB}},\\
	\varepsilon_{XRF} &= \frac{N_{XRF,\,r_t\ge 0.2}}{N_{XRF}},\\
	\varepsilon_{XRR} &= \frac{N_{XRR,\,r_t\ge 0.2}}{N_{XRR}}.
\end{align}
We first examine the bright X-ray flares from the data of \cite{Chincarini2010MN}. We do not find any bright X-ray flares in C-GRBs. The fraction of the XRRs that have the bright X-ray flares to the total XRRs is $29.4\%$, and the fraction of the XRFs that have the bright X-ray flares to the total XRFs is $28.6\%$. When we consider only the bright X-ray flares, the mean values of $r_t$ are $0.39$ and $0.53$ for XRFs and XRRs, respectively. We then examine the bright X-ray flares from the data of \cite{Shuangxi2016ApJ}. The fraction of the C-GRBs that have bright X-ray flares out of the total C-GRBs is $4.5\%$, the fraction of the XRRs that have bright X-ray flares out of the total XRRs is $24.6\%$, and the fraction of the XRFs that have bright X-ray flares out of the total XRFs is $37.5\%$. When we consider only the bright X-ray flares, the mean values of $r_t$ are $0.46$, $0.51$, and $0.73$ for XRFs, XRRs, and C-GRBs, respectively. Therefore, for the GRBs that have bright X-ray flares, the fluence of the X-ray flare is comparable to the fluence of the prompt emission. In our sample, it seems that XRFs and XRRs have more bright X-ray flares than C-GRBs.

\subsubsection{Relation between X-Ray Flare Fluence and Prompt Emission Fluence}
\begin{figure}[h!]
    \centering
    \includegraphics[height=.5\linewidth - 0.25mm, width=.5\linewidth - 0.25mm]{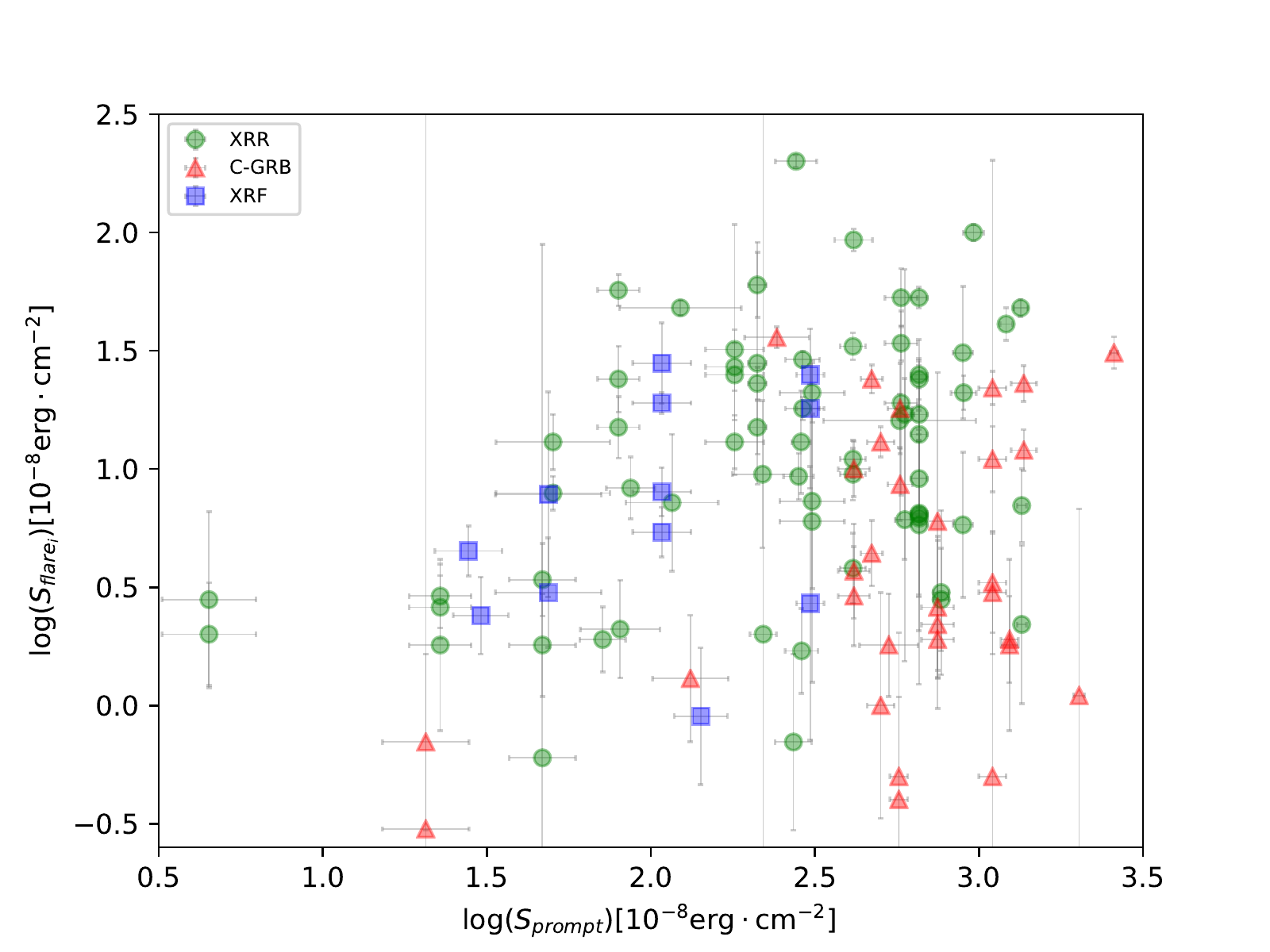}\hfill
    \includegraphics[height=.5\linewidth - 0.25mm, width=.5\linewidth - 0.25mm]{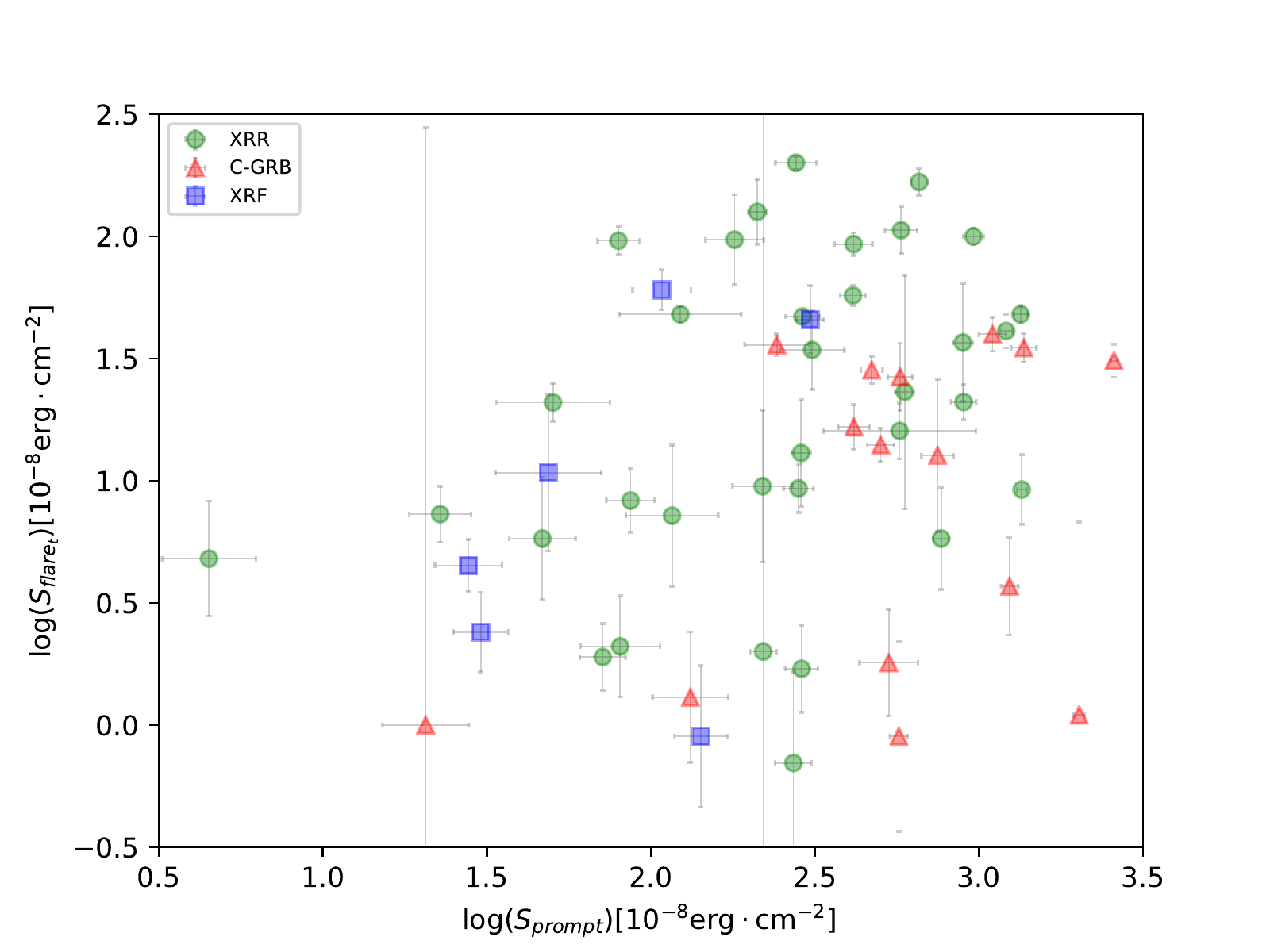}
    \includegraphics[height=.5\linewidth - 0.25mm, width=.5\linewidth - 0.25mm]{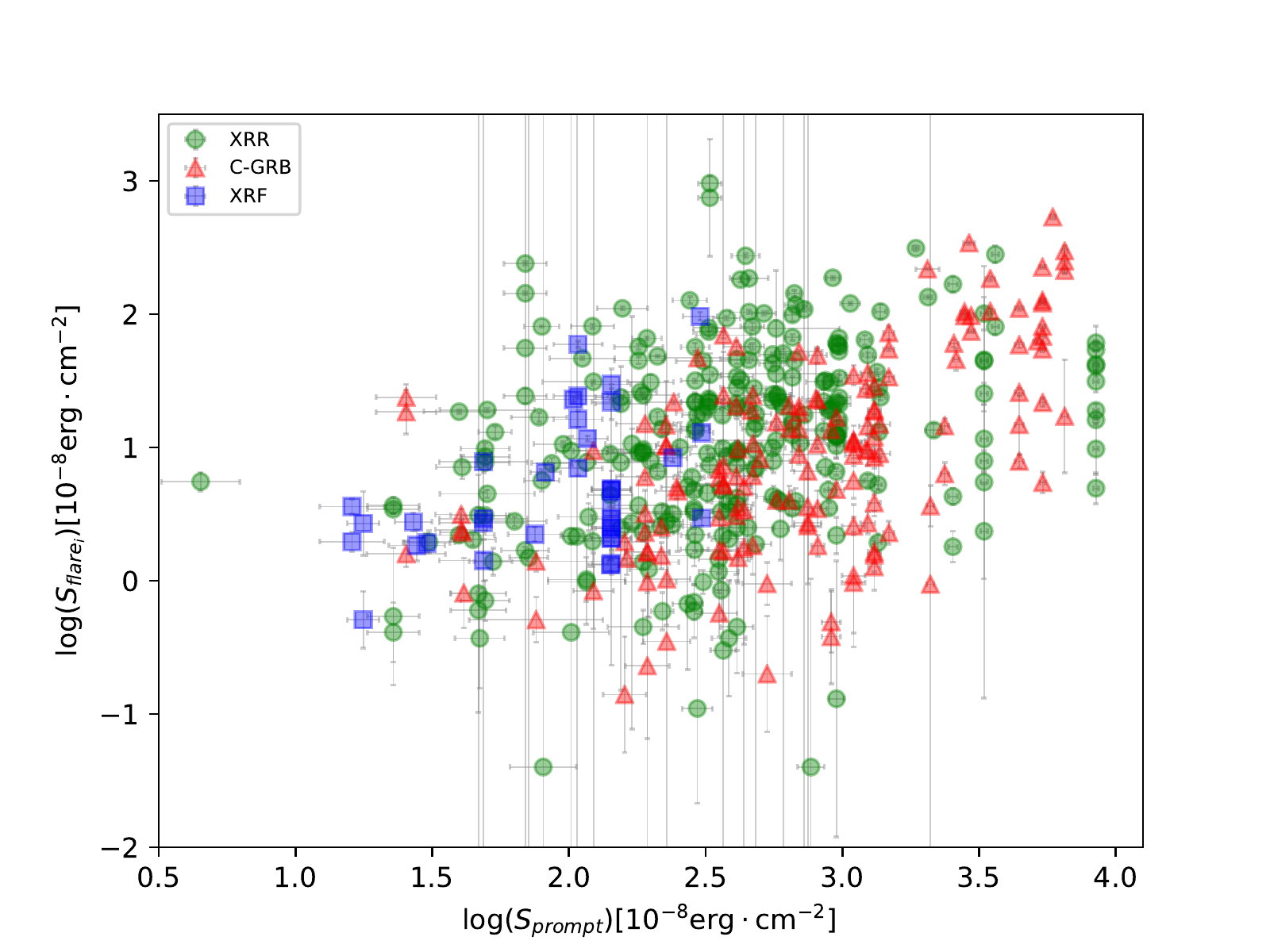}\hfill
    \includegraphics[height=.5\linewidth - 0.25mm, width=.5\linewidth - 0.25mm]{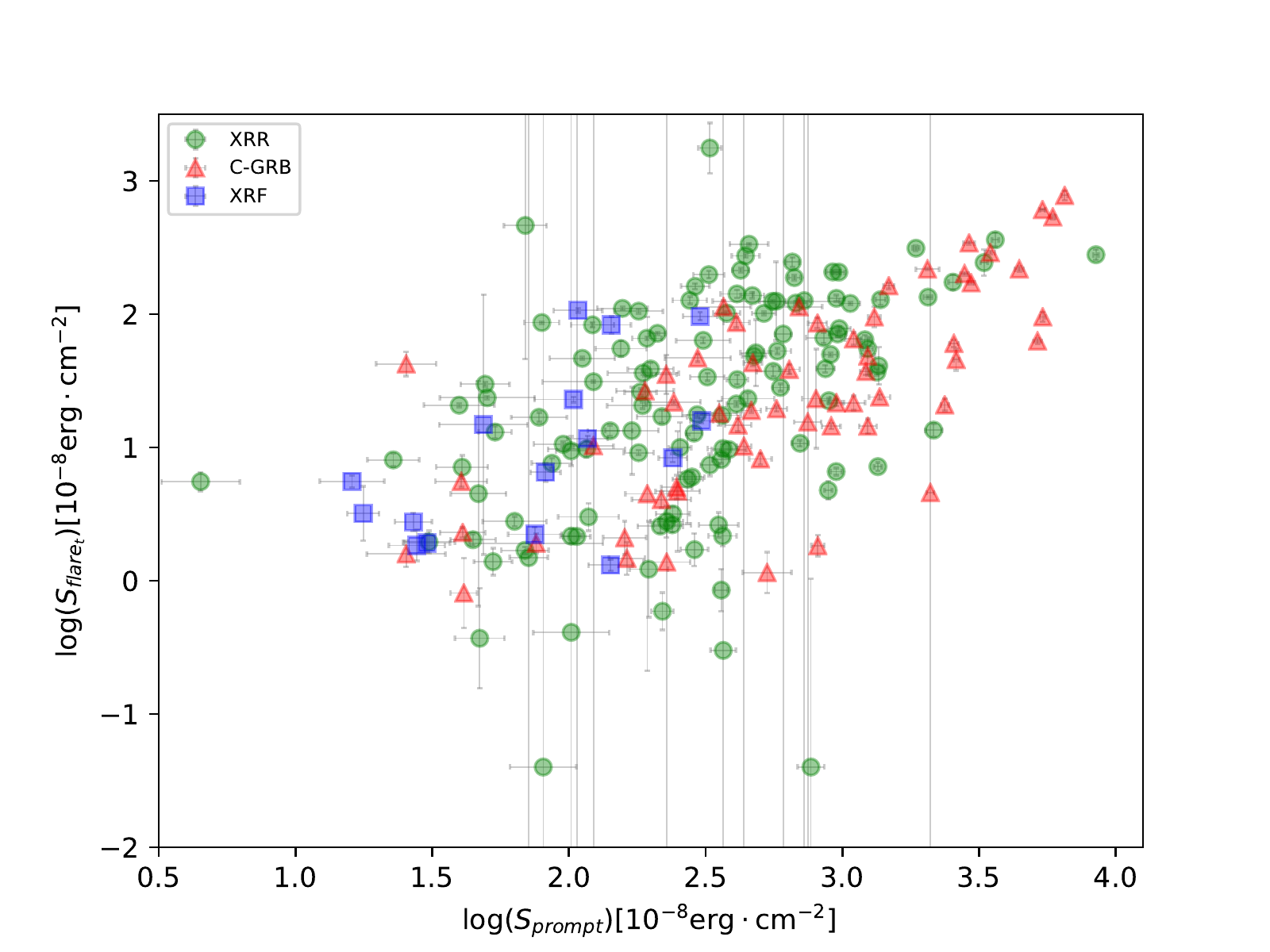}
    \caption{X-ray flare fluence vs. prompt emission fluence for XRRs, XRFs, and C-GRBs. Top left panel: single X-ray flare fluence of one GRB is used in Chincarini et al. (2010). Top right panel: total X-ray flare fluence of one GRB is used in Chincarini et al. (2010). Bottom left panel: single X-ray flare fluence of one GRB is used in Yi et al. (2016). Bottom right panel: total X-ray flare fluence of one GRB is used in Yi et al. (2016).
  }
    \label{Fig:distribution:S_flare:Chin}
\end{figure}

In order to examine energy release of the GRB X-ray flare that is related to GRB prompt emission, we further investigate the X-ray flare fluence and the prompt emission fluence among XRF, XRR, and C-GRB subclasses. The X-ray flare data samples are taken from Chincarini et al. (2010) and Yi et al. (2016). We obtain four panels of X-ray flare fluence vs. prompt emission fluence in Figure \ref{Fig:distribution:S_flare:Chin}. We cannot find a reliable correlation between the X-ray flare fluence and the prompt emission fluence due to the large data scattering.

If X-ray flares in the GRB early X-ray afterglow are originally from a late time manifestation of the same emission mechanism as the prompt emission, we may consider a correlation between the X-ray flare fluence and the prompt emission fluence. However, we cannot find a significant correlation. It seems that the correlation is not straightforward if it really exists.

\subsubsection{Isotropic Energy Release of X-Ray Flares}
\begin{figure}[h!]
    \centering
    \includegraphics[height=.6\linewidth - 0.25mm]{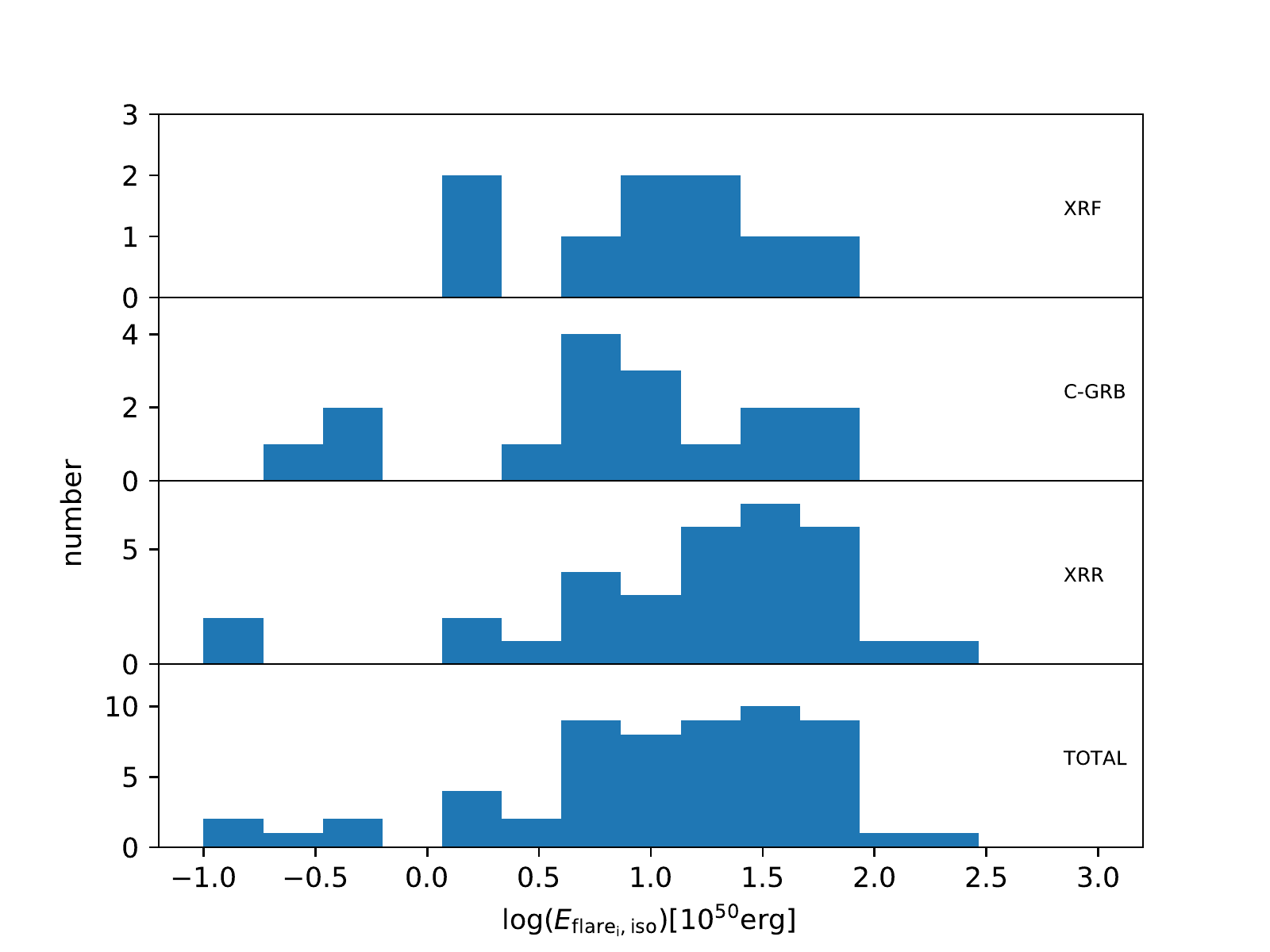}
    \includegraphics[height=.6\linewidth - 0.25mm]{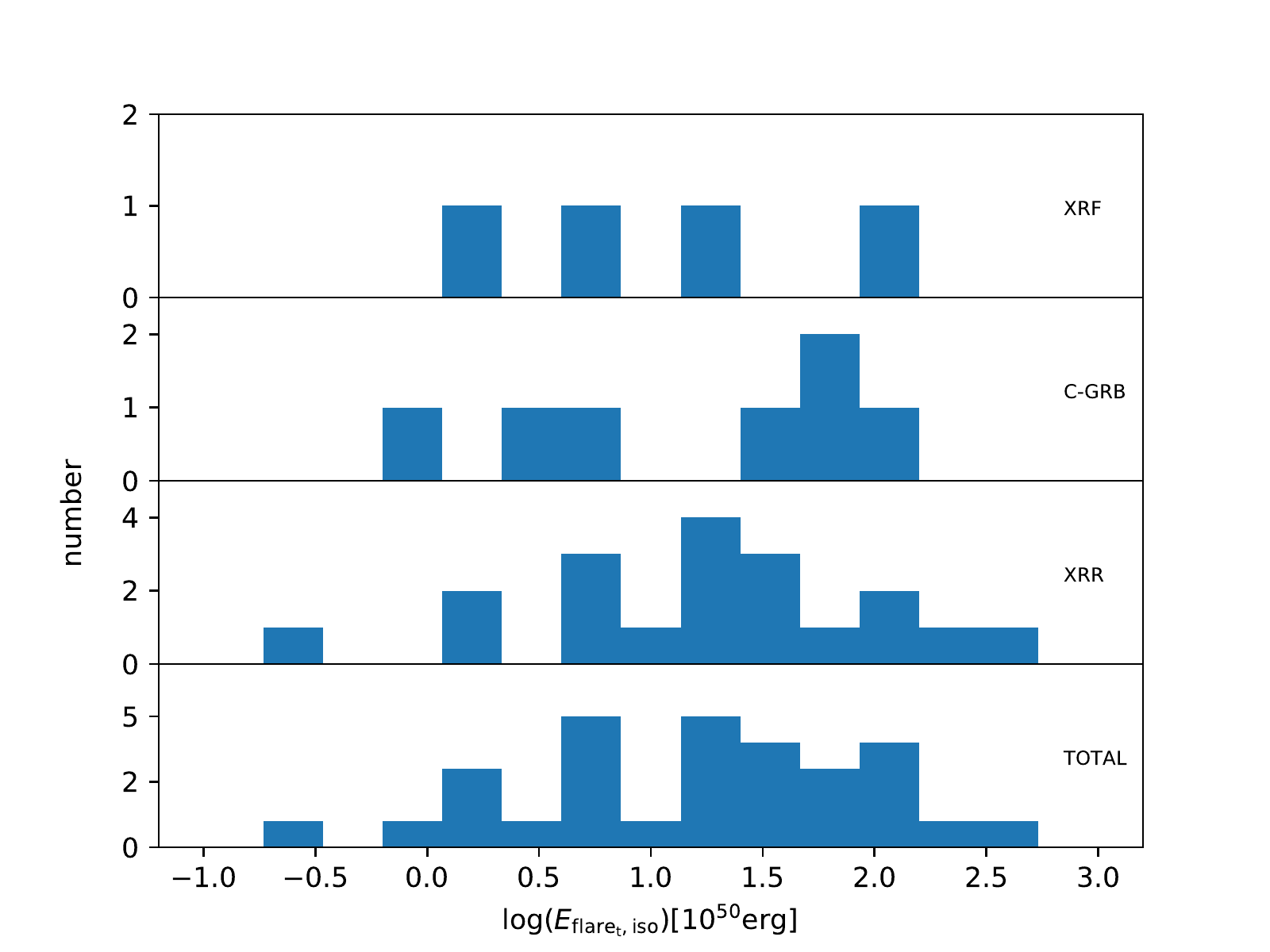}
    \caption{Distributions of the X-ray flare isotropic energy of XRFs, XRRs, and C-GRBs. The X-ray flare sample of Chincarini et al. (2010) is used. Top panel: single X-ray flare isotropic energy in one GRB is considered. Bottom panel: total X-ray flare isotropic energy in one GRB is considered.
  }
    \label{Fig:distribution:Epeak:Chin}
\end{figure}

\begin{figure}[h!]
    \centering
    \includegraphics[height=.6\linewidth - 0.25mm]{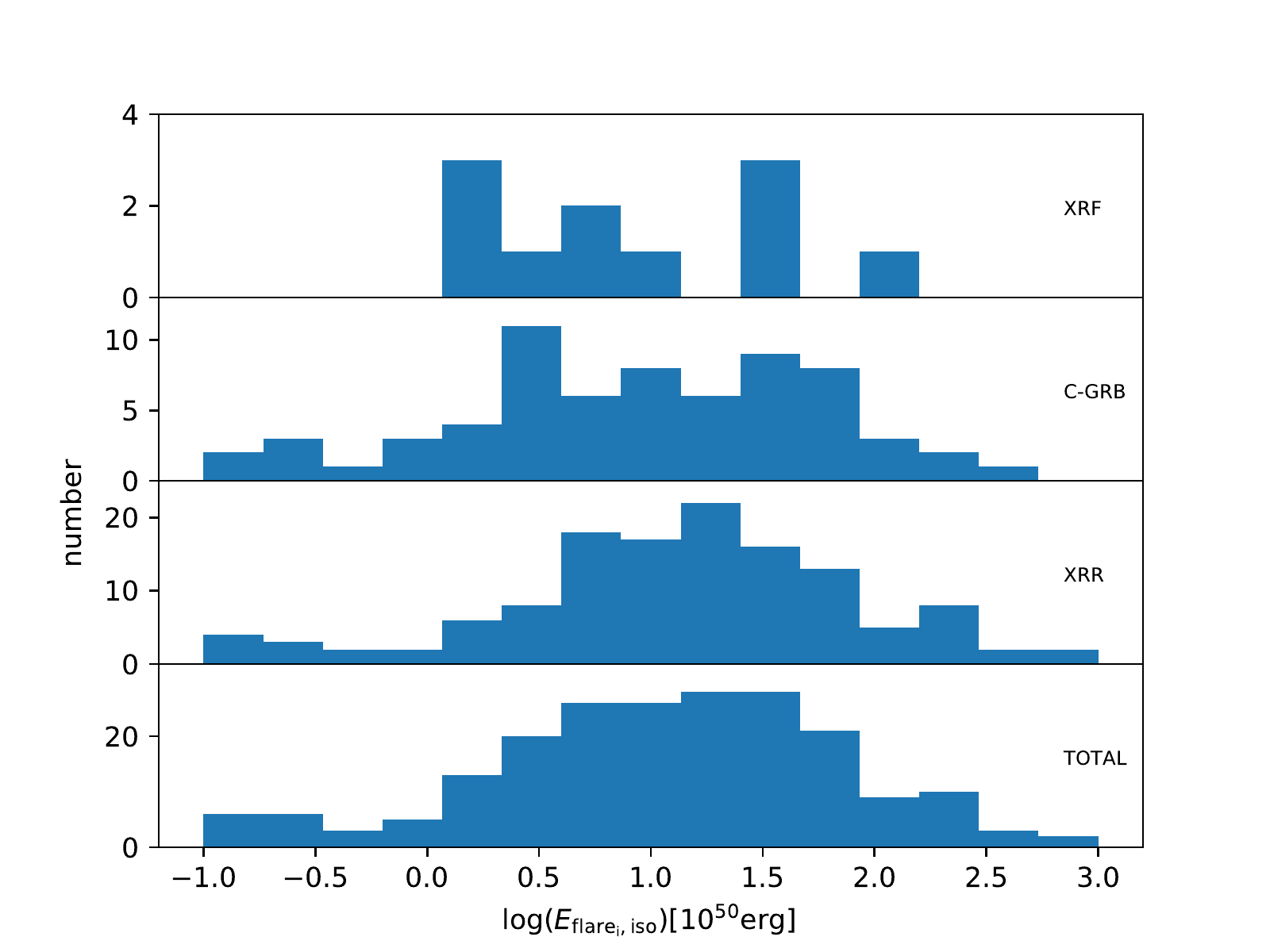}
    \includegraphics[height=.6\linewidth - 0.25mm]{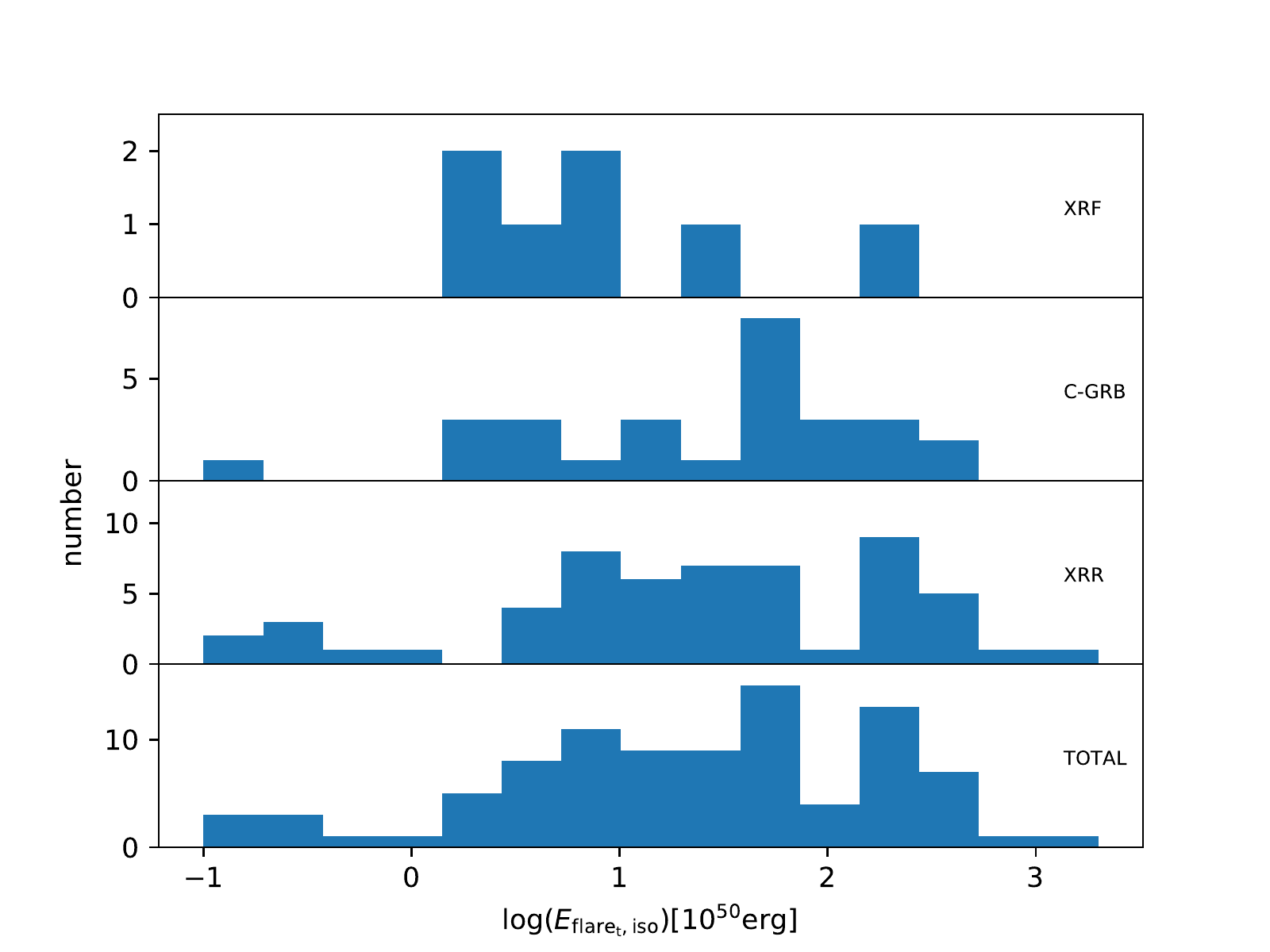}
    \caption{Distributions of the X-ray flare isotropic energy of XRFs, XRRs, and C-GRBs. The X-ray flare sample of Yi et al. (2016) is used. Top panel: single X-ray flare isotropic energy in one GRB is considered. Bottom panel: total X-ray flare isotropic energy in one GRB is considered.}
    \label{Fig:distribution:Epeak:Yi}
\end{figure}

We calculate the isotropic energy of the X-ray flare in each GRB if we have GRB redshift measurements. The distributions of the isotropic energy for XRFs, XRRs, and C-GRBs are shown in Figures \ref{Fig:distribution:Epeak:Chin} and \ref{Fig:distribution:Epeak:Yi}. The X-ray flare data samples are taken from Chincarini et al. (2010) and Yi et al. (2016). We perform the K-S test for the distributions. It seems that the distributions of the X-ray flare energy release have no significant differences among XRF, XRR, and C-GRB subclasses.

We note that a direct comparison of X-ray flare energy release among XRF, XRR, and C-GRB subclasses cannot reveal the relation between X-ray flare and prompt emission for one GRB. The results from the fluence ratios of XRFs, XRRs, and C-GRBs properly present the energy release of X-ray flare compared to the energy release of prompt emission (see Figures \ref{Fig:distribution:ratio_t:Chin} and \ref{Fig:distribution:ratio_t:Shuang}).

\subsection{Temporal Properties of X-Ray Flares among XRF, XRR, and C-GRB Subclasses}
\subsubsection{Distributions of Duration and Peak Time of X-Ray Flares among XRF, XRR, and C-GRB Subclasses}
A detailed analysis of the X-ray flare temporal properties among XRF, XRR, and C-GRB subclasses is also an important issue in this paper. Here, we ignore the Lorentz factor of the X-ray flare, and we do not have the redshift correction to the X-ray flare temporal properties.

\begin{figure}[h!]
    \centering
    \includegraphics[height=.6\linewidth - 0.25mm]{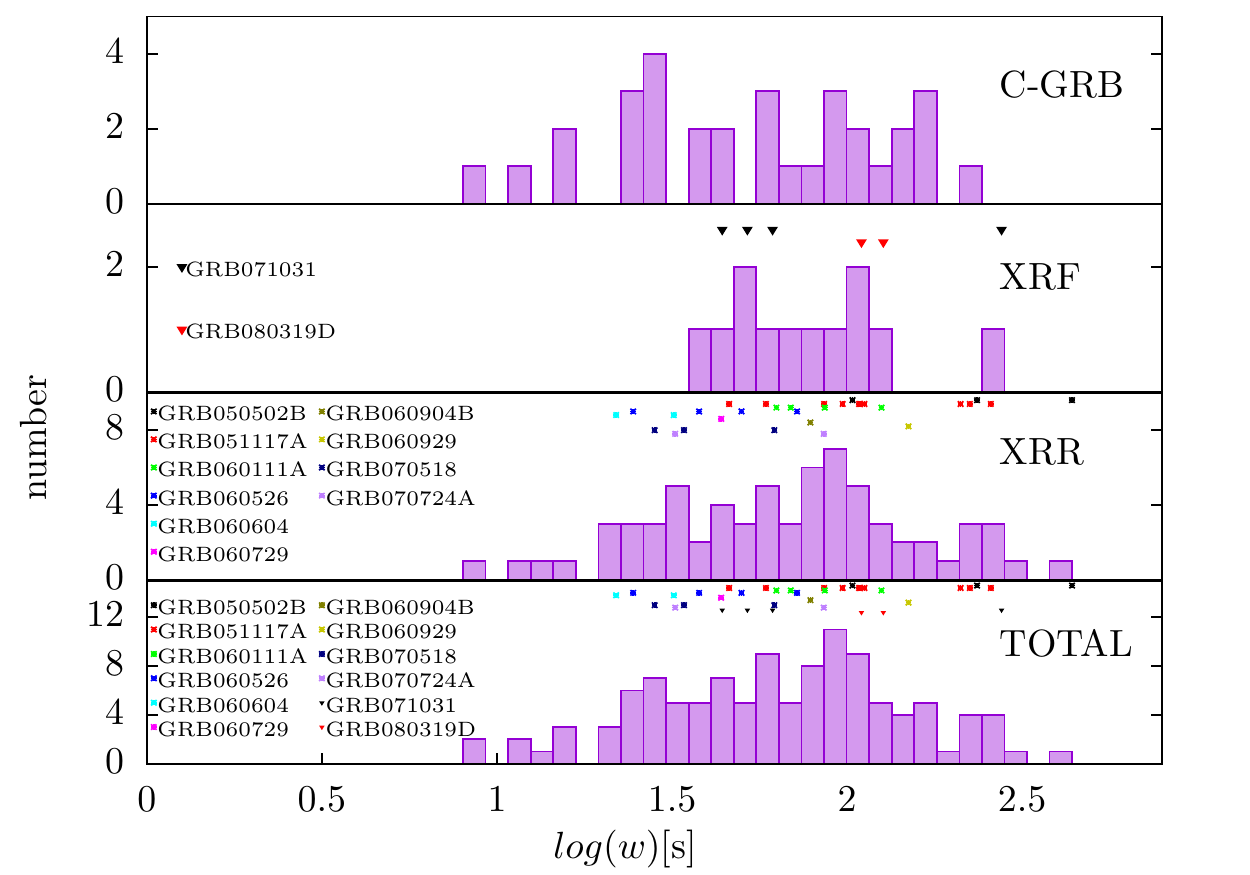}
    \includegraphics[height=.6\linewidth - 0.25mm]{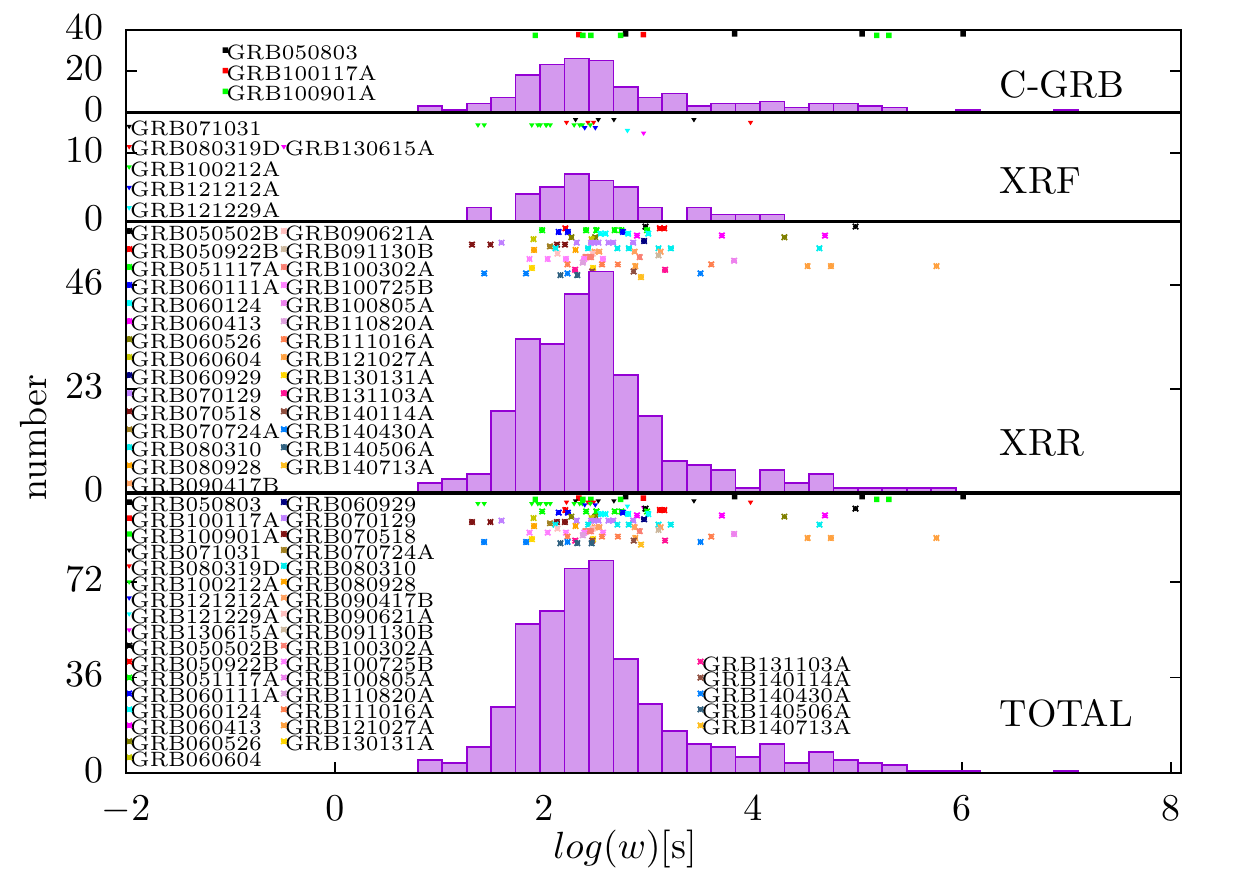}
    \vglue -0.5cm
    \caption{Distributions of the X-ray flare duration for XRFs, XRRs, and C-GRBs. We mark the GRBs having the bright X-ray flares that have $r_t\ge 0.2$. Top panel: the X-ray flare data are from Chincarini et al. (2010). Bottom panel: the X-ray flare data are from Yi et al. (2016).
}
    \label{Fig:distribution_w}
\end{figure}

We identify the width of the X-ray flare $w$ to be the X-ray flare duration. We present the distributions of the X-ray flare duration for XRFs, XRRs, and C-GRBs. The results from the X-ray flare data of \cite{Chincarini2010MN} and \cite{Shuangxi2016ApJ} are shown in Figure \ref{Fig:distribution_w}. The K-S test results are listed in Table \ref{Tab:K-S}.

\begin{figure}[h!]
    \centering
    \includegraphics[height=.6\linewidth - 0.25mm]{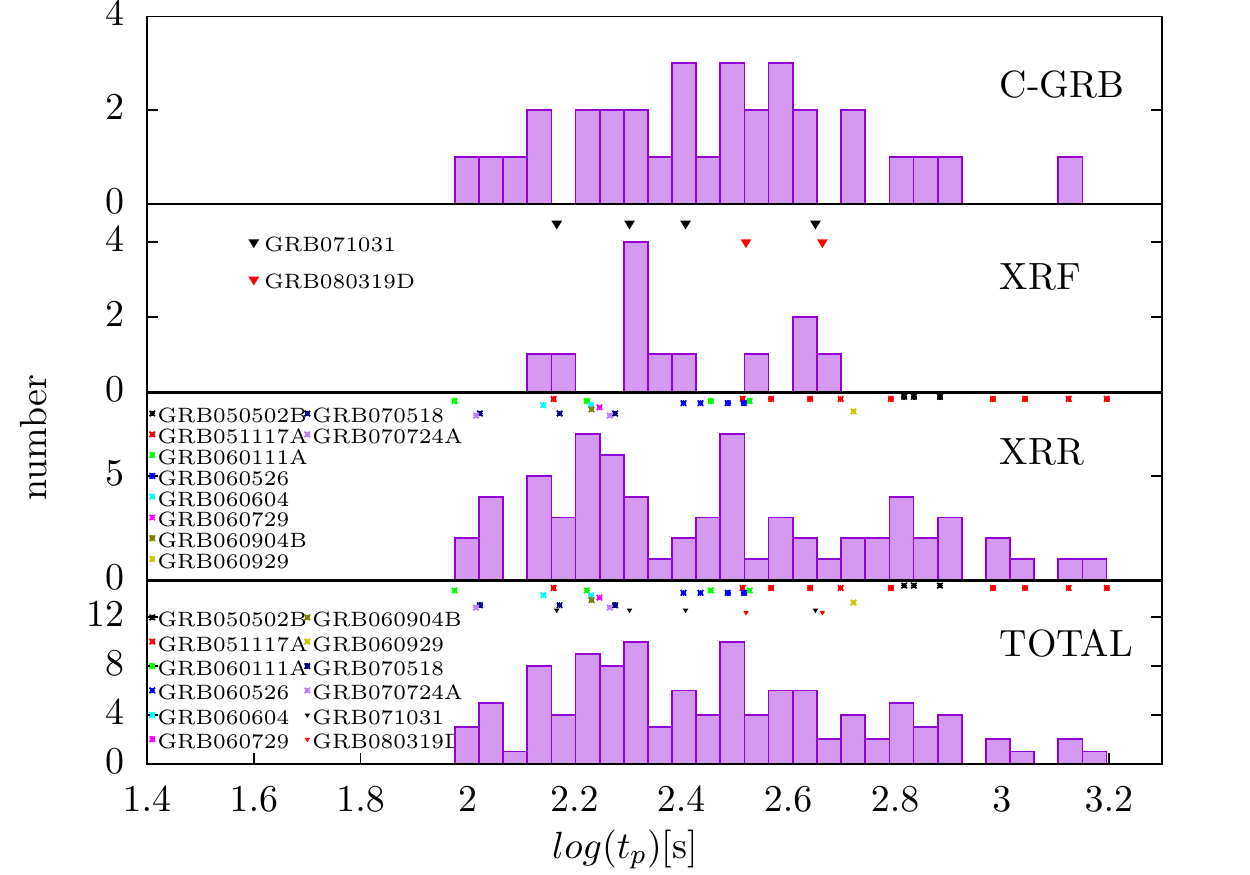}
    \includegraphics[height=.6\linewidth - 0.25mm]{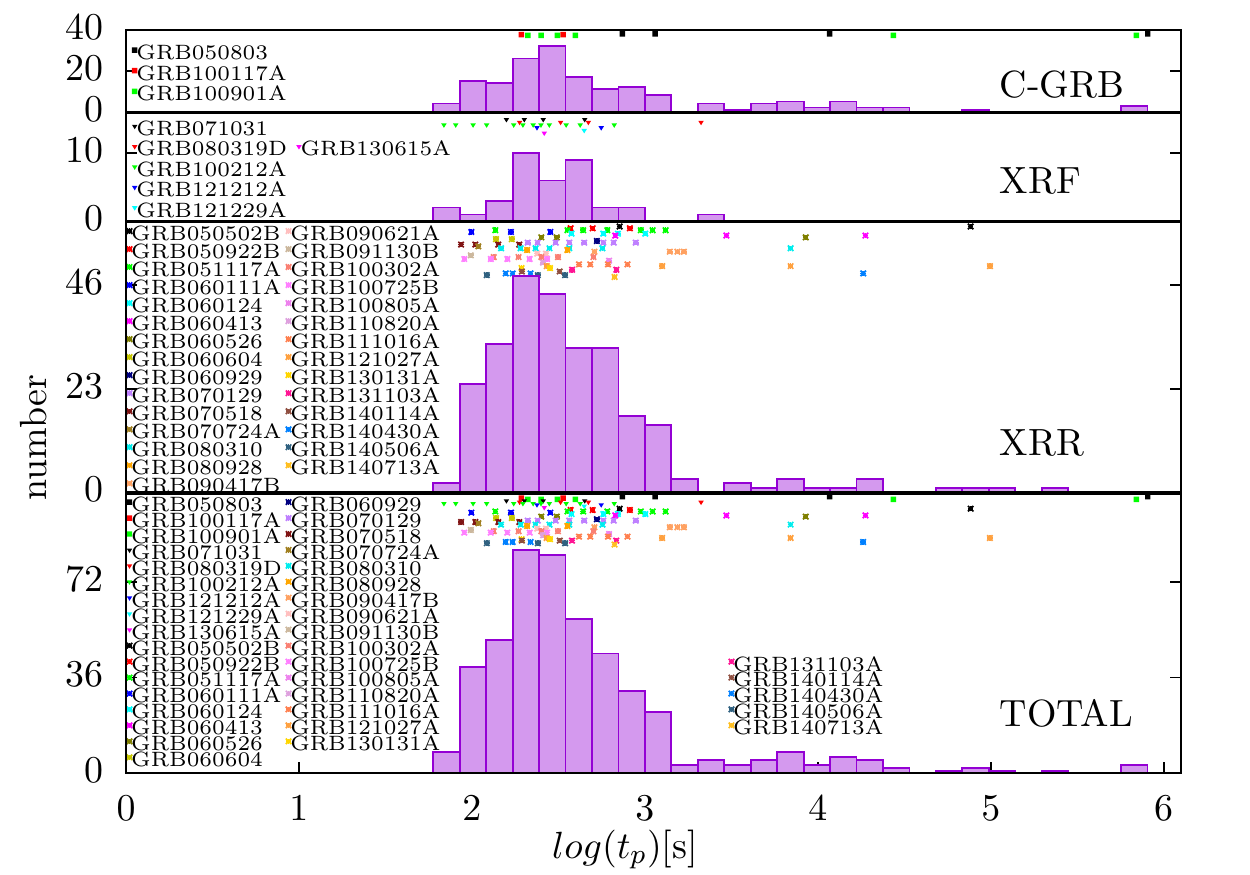}
    \vglue -0.5cm
    \caption{Distributions of the X-ray flare peak time for XRFs, XRRs, and C-GRBs. We mark the GRBs with bright X-ray flares that have $r_t\ge 0.2$. Top panel: the X-ray flare data are from Chincarini et al. (2010). Bottom panel: the X-ray flare data are from Yi et al. (2016).
}
    \label{Fig:distribution_tp}
\end{figure}

We also use the X-ray flare data from \cite{Chincarini2010MN} and \cite{Shuangxi2016ApJ} to analyze the distributions of the X-ray flare peak time $t_p$ for XRFs, XRRs, and C-GRBs. The results are in Figure \ref{Fig:distribution_tp}. The K-S test results are listed in Table \ref{Tab:K-S}.

We do not find any differences in the duration distribution and the peak time distribution of the X-ray flares among XRF, XRR, and C-GRB subclasses. However, we ignore the effect of the bulk Lorentz factor on these temporal properties, and we do not perform the redshift correction to these temporal properties. We caution the direct comparison of the X-ray flare temporal properties among XRF, XRR, and C-GRB subclasses.

\subsubsection{Distributions of Redshift-corrected Duration and Peak Time of X-Ray Flares among XRF, XRR, and C-GRB Subclasses}
\begin{figure}[h!]
    \centering
    \includegraphics[height=.6\linewidth - 0.25mm]{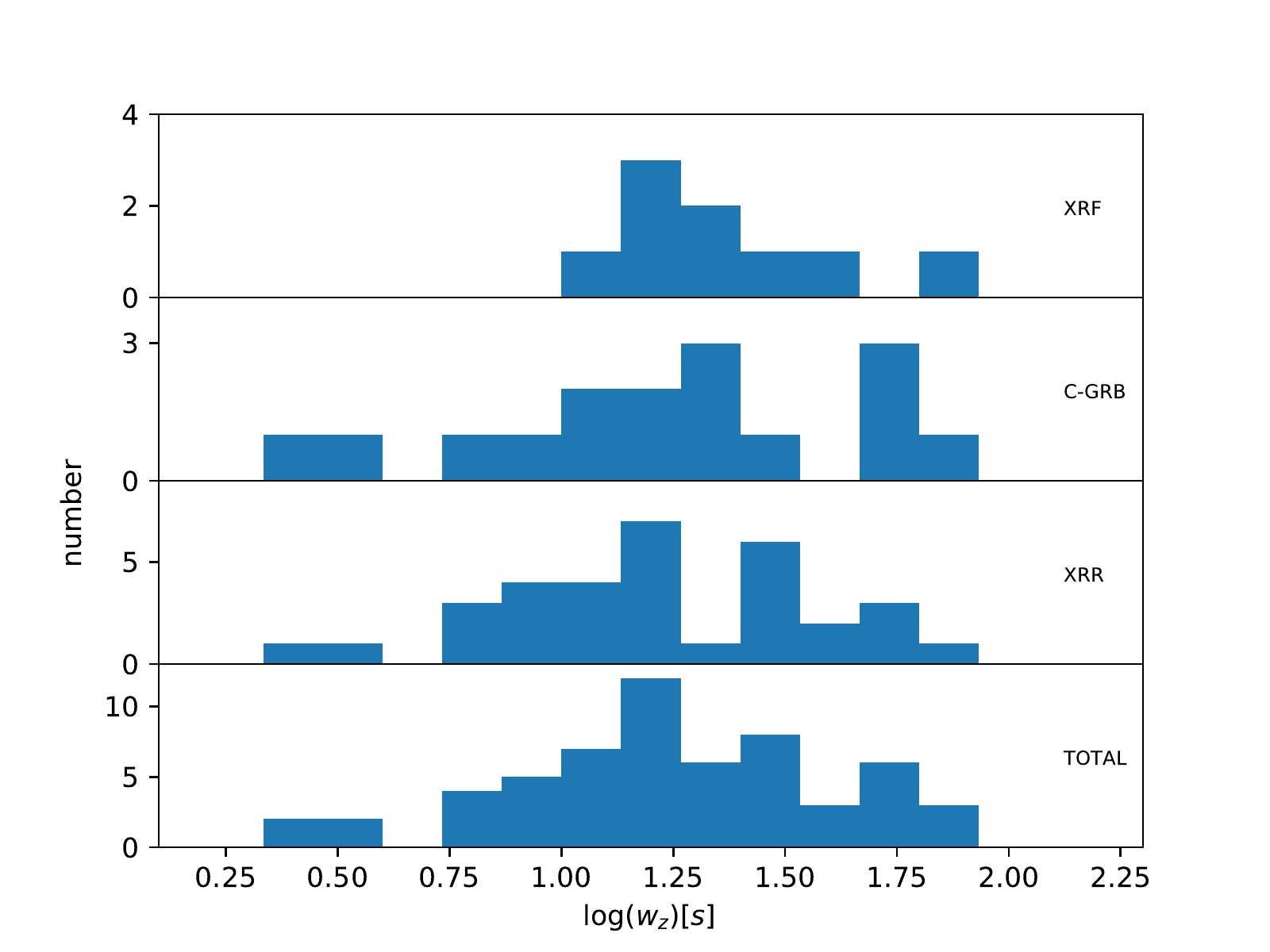}
    \includegraphics[height=.6\linewidth - 0.25mm]{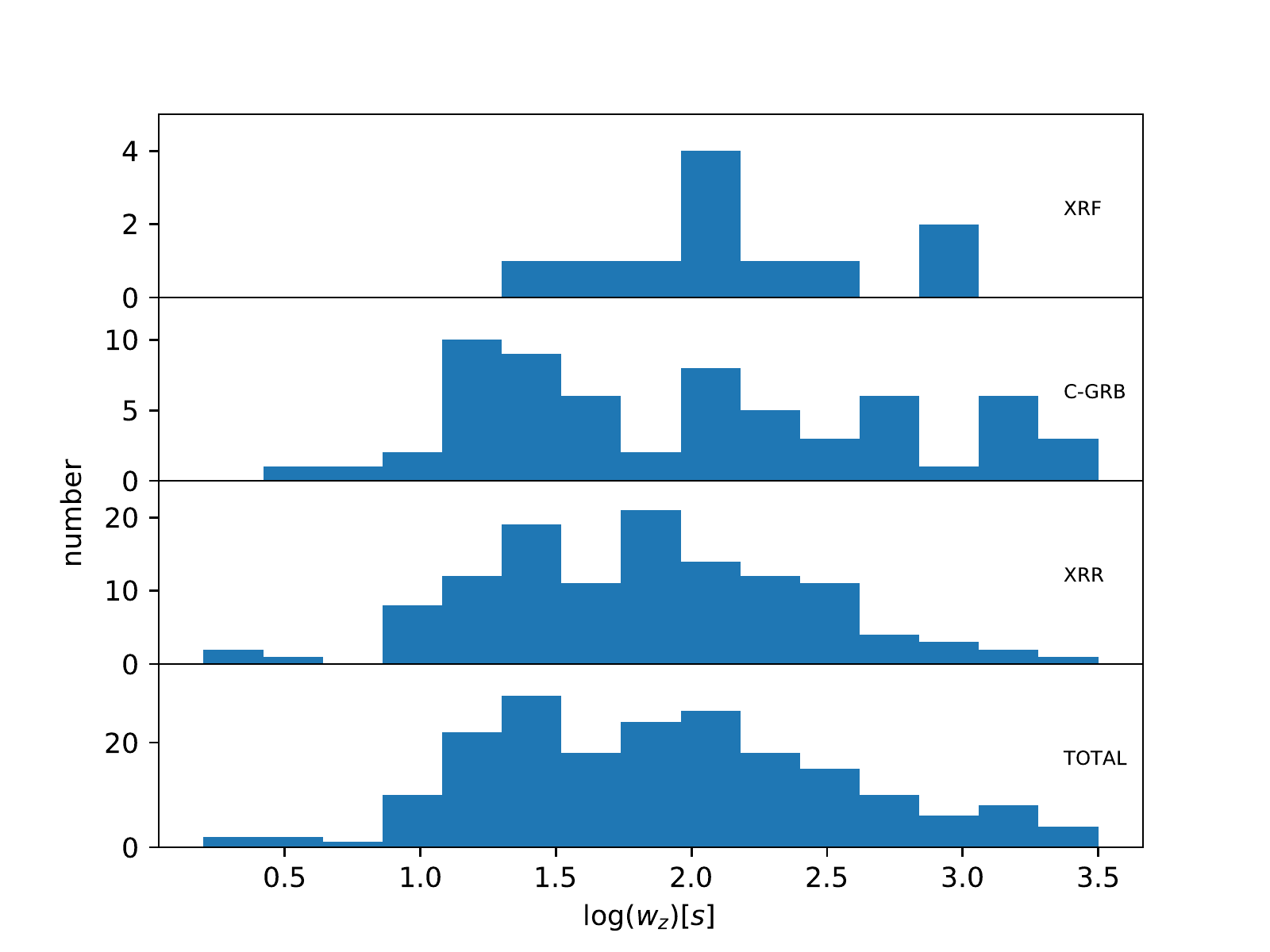}
    \caption{Distributions of the X-ray flare duration for XRFs, XRRs, and C-GRBs. The duration is redshift-corrected. Top panel: the X-ray flare sample from Chincarini et al. (2010). Bottom panel: the X-ray flare sample from Yi et al. (2016).
  }
    \label{Fig:distribution:wz}
\end{figure}

\begin{figure}[h!]
    \centering
    \includegraphics[height=.6\linewidth - 0.25mm]{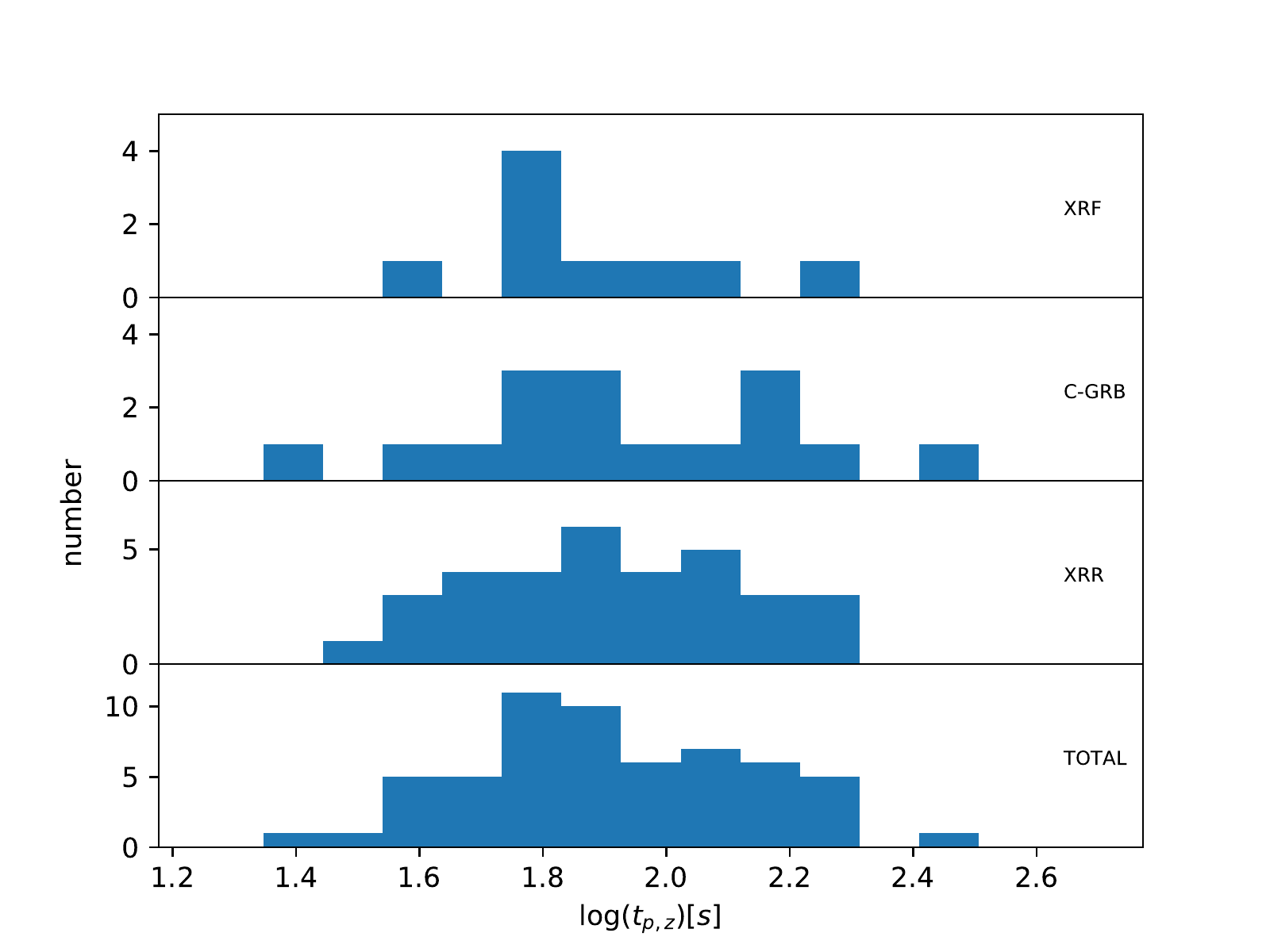}
    \includegraphics[height=.6\linewidth - 0.25mm]{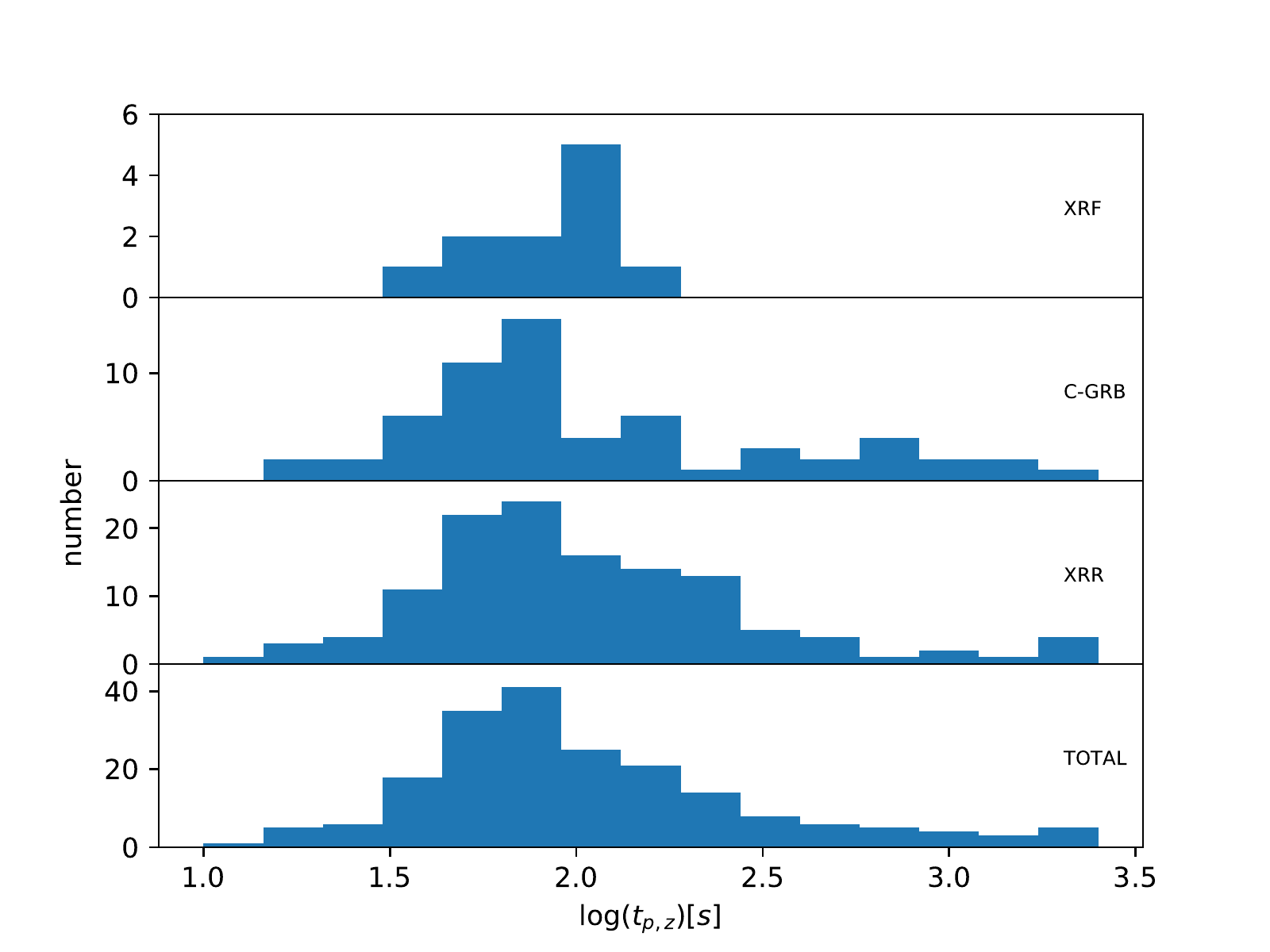}
    \caption{Distributions of the X-ray flare peak time for XRFs, XRRs, and C-GRBs. The peak time is redshift-corrected. Top panel: the X-ray flare sample from Chincarini et al. (2010). Bottom panel: the X-ray flare sample from Yi et al. (2016).
  }
    \label{Fig:distribution:tpz}
\end{figure}

In order to directly compare the temporal properties of the X-ray flares among XRF, XRR, and C-GRB classes, we perform the redshift-corrected duration and peak time of the X-ray flares in our sample. The distributions of the duration and the peak time of the X-ray flares are shown in Figures \ref{Fig:distribution:wz} and \ref{Fig:distribution:tpz}.

The distributions of the duration and the peak time do not have significant differences among XRF, XRR, and C-GRB subclasses in general. Yi et al. (2016) have larger X-ray flare duration values than Chincarini et al. (2010) because some late X-ray flares were included in Yi et al. (2016). However, we note that the issues mentioned above might have strong bias because the data samples in our statistical analysis are small. We hope that more X-ray flare data with the redshift measurements are available in the future to further investigate the distributions of the temporal properties.

\subsubsection{Correlation between Duration and Peak Time of X-Ray Flare among XRF, XRR, and C-GRB Subclasses}
We can examine the correlation between the duration and the peak time of the X-ray flares for XRFs, XRRs, and C-GRBs. When we usually perform the least-square method for a linear fitting of observational data, the data errors are not included. Thus, the extrinsic scatter of the linear fitting may exist due to the large data error, and the fitting results are not reliable. In order to consider the effect from the large data error and obtain reliable fitting results, we perform the maximum likelihood method suggested by \cite{Amati2008MN} to obtain the linear correlation fitting between the duration and the peak time of the X-ray flares.

\begin{figure}[h!]
    \centering
    \includegraphics[height=.5\linewidth - 0.25mm, width=.5\linewidth - 0.25mm]{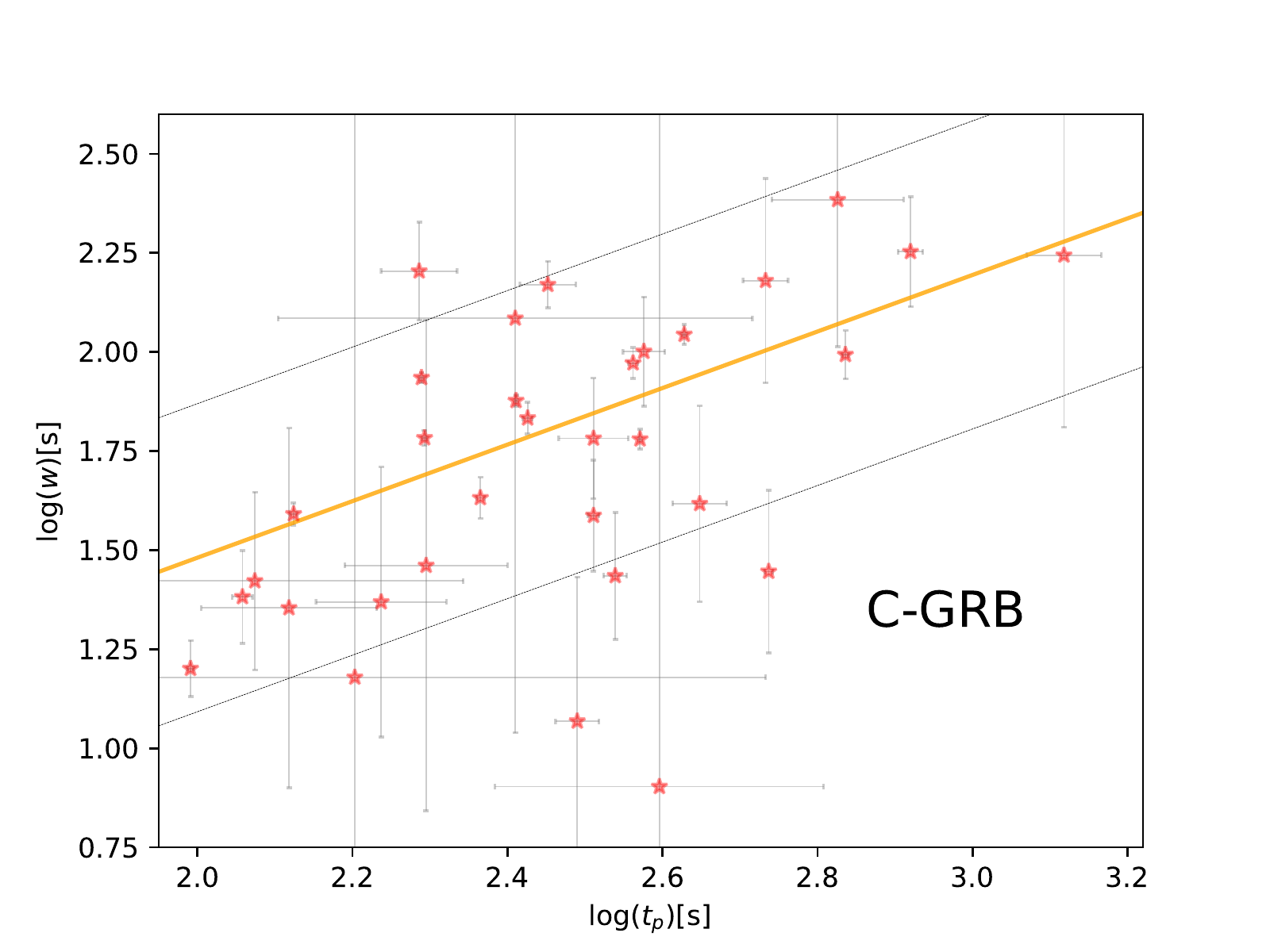}\hfill
    \includegraphics[height=.5\linewidth - 0.25mm, width=.5\linewidth - 0.25mm]{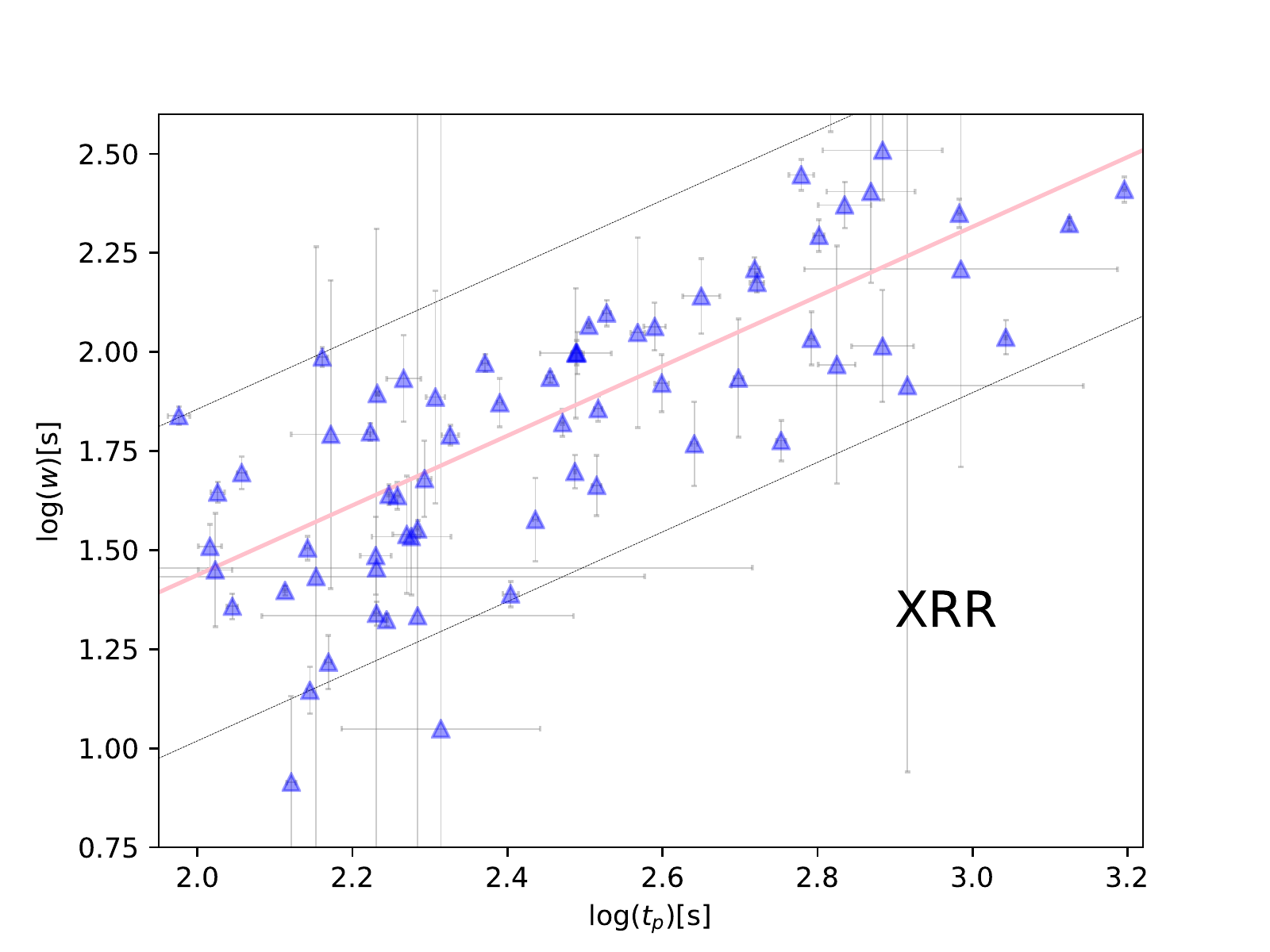}\\[0.5mm]
    \includegraphics[height=.5\linewidth - 0.25mm, width=.5\linewidth - 0.25mm]{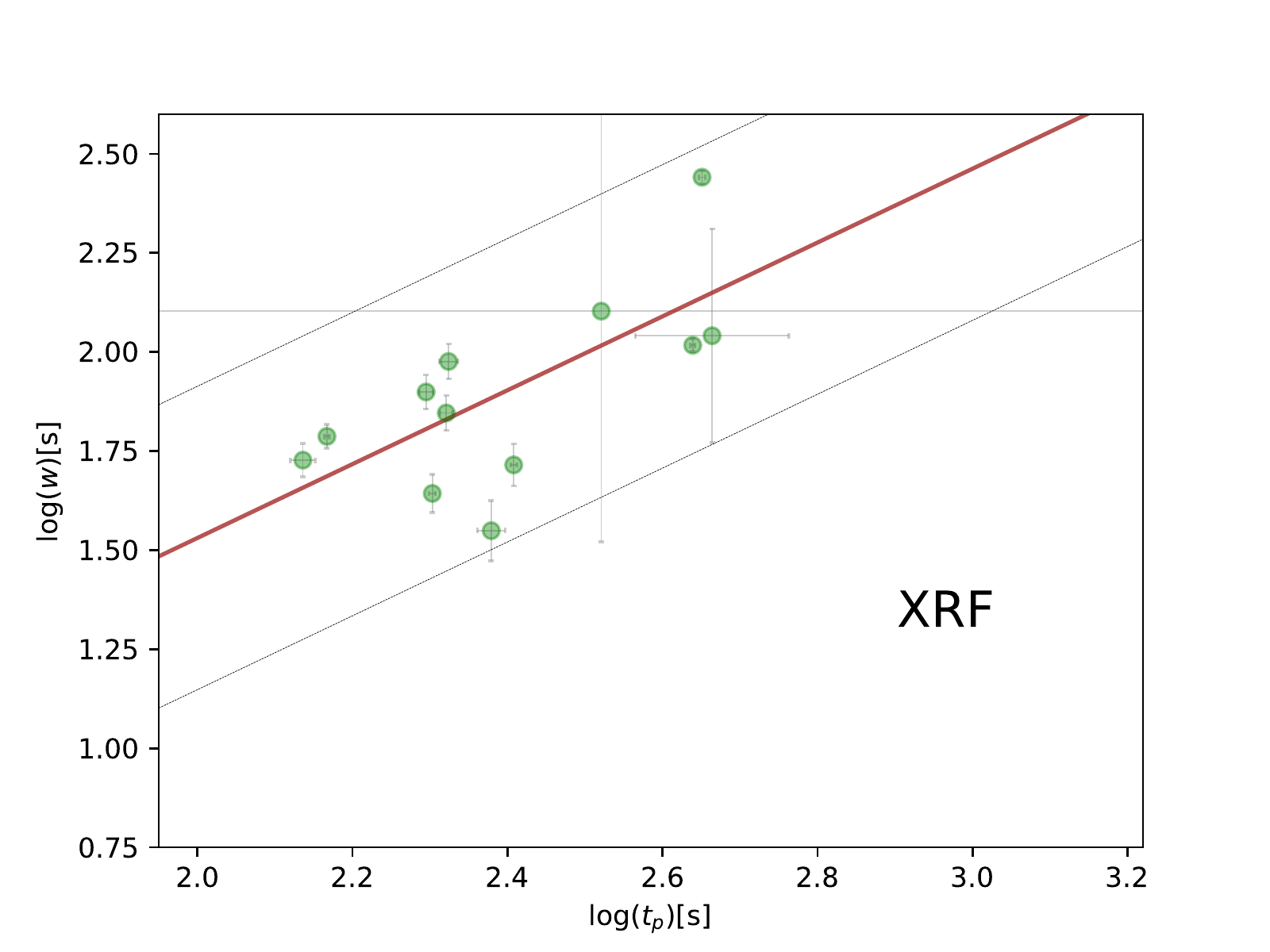}\hfill
    \includegraphics[height=.5\linewidth - 0.25mm, width=.5\linewidth - 0.25mm]{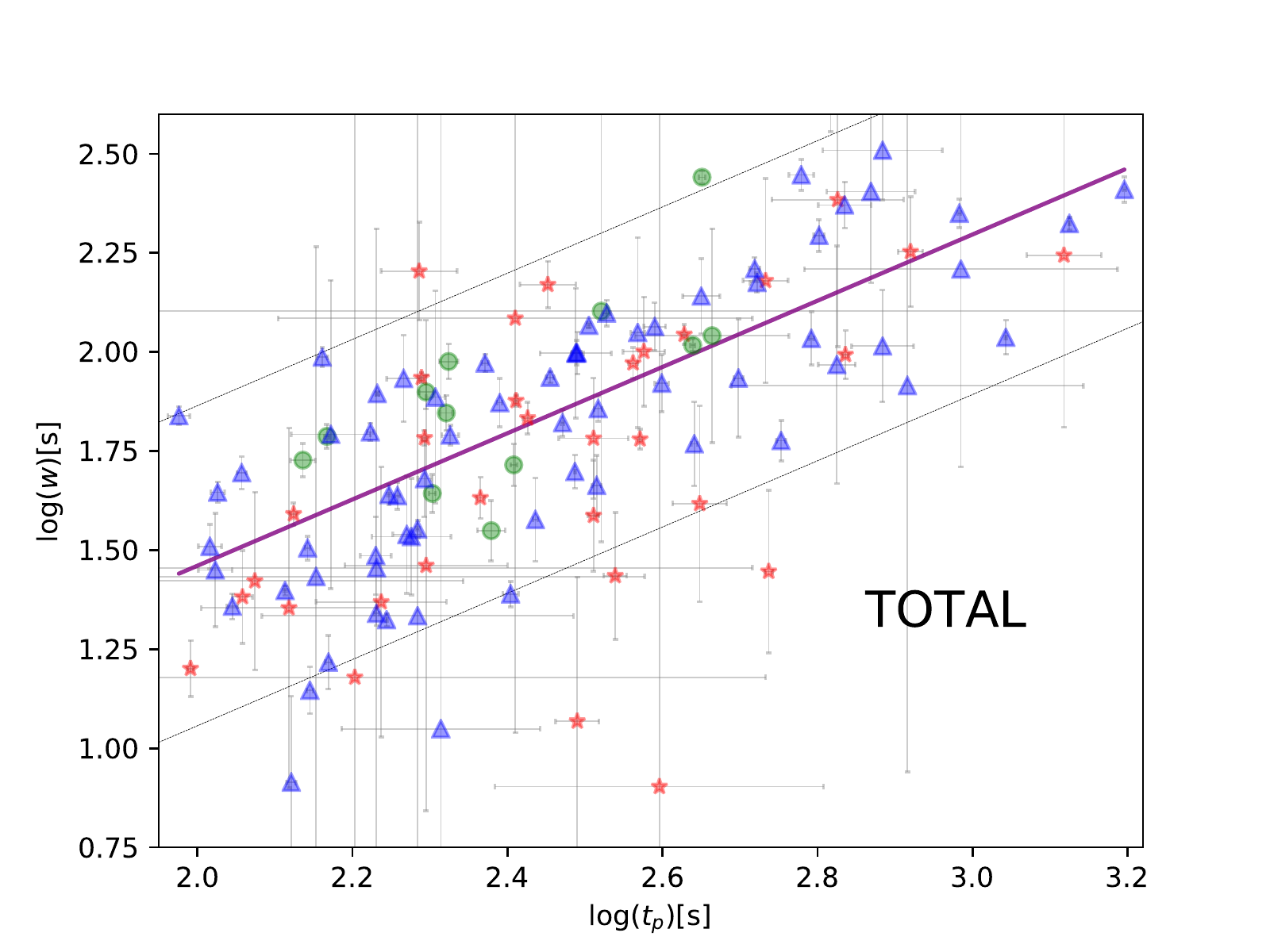}
    \vglue -0.5cm
    \caption{The $w-t_p$ correlations of the X-ray flares for XRFs, XRRs, and C-GRBs. The X-ray flare data are from \cite{Chincarini2010MN}. The green circles represent the X-flares of XRFs, the blue triangles represent the X-ray flares of XRRs, and the red stars represent the X-ray flares of C-GRB. The $w-t_p$ correlations are plotted by the solid lines. The dashed lines enclose the data within $1\sigma$. The best-fitting for XRFs is $\log \mathrm{w}=(-0.42^{+0.87}_{-0.87})+(0.97^{+0.36}_{-0.37})\log \mathrm{t}_\mathrm{p}$ with the scatter of $\sigma = 0.19^{+0.06}_{-0.04}$. The best-fitting for XRRs is $\log \mathrm{w}=(-0.34^{+0.24}_{-0.23})+(0.89^{+0.09}_{-0.09})\log \mathrm{t}_\mathrm{p}$ with the scatter of $\sigma = 0.21^{+0.02}_{-0.02}$. The best-fitting for C-GRBs is $\log\mathrm{w}=(0.05^{+0.46}_{-0.46})+(0.71^{+0.19}_{-0.19})\log \mathrm{t}_\mathrm{p}$ with the scatter of $\sigma = 0.19^{+0.05}_{-0.04}$. The best-fitting for total GRBs is $\log \mathrm{w}=(-0.22^{+0.19}_{-0.21})+(0.84^{+0.08}_{-0.08})\log \mathrm{t}_\mathrm{p}$ with the scatter of $\sigma = 0.20^{+0.02}_{-0.02}$.}
    \label{Fig:t_p_w:Chin}
\end{figure}

We use the X-ray flare data of \cite{Chincarini2010MN} to perform the $w-t_p$ correlations for XRFs, XRRs, and C-GRBs. We find the correlation of $\log w=(-0.42^{+0.87}_{-0.87})+(0.97^{+0.36}_{-0.37})\log t_p$ with $\sigma = 0.19^{+0.06}_{-0.04}$ for XRFs, the correlation of $\log w=(-0.34^{+0.24}_{-0.23})+(0.89^{+0.09}_{-0.09})\log t_p$ with $\sigma = 0.21^{+0.02}_{-0.02}$ for XRRs, and the correlation of $\log w=(0.05^{+0.46}_{-0.46})+(0.71^{+0.19}_{-0.19})\log t_p$ with $\sigma = 0.19^{+0.05}_{-0.04}$ for C-GRBs, where $\sigma$ is the extrinsic scatter. The results are shown in Figure \ref{Fig:t_p_w:Chin}.

\begin{figure}[h!]
    \centering
    \includegraphics[height=.5\linewidth - 0.25mm, width=.5\linewidth - 0.25mm]{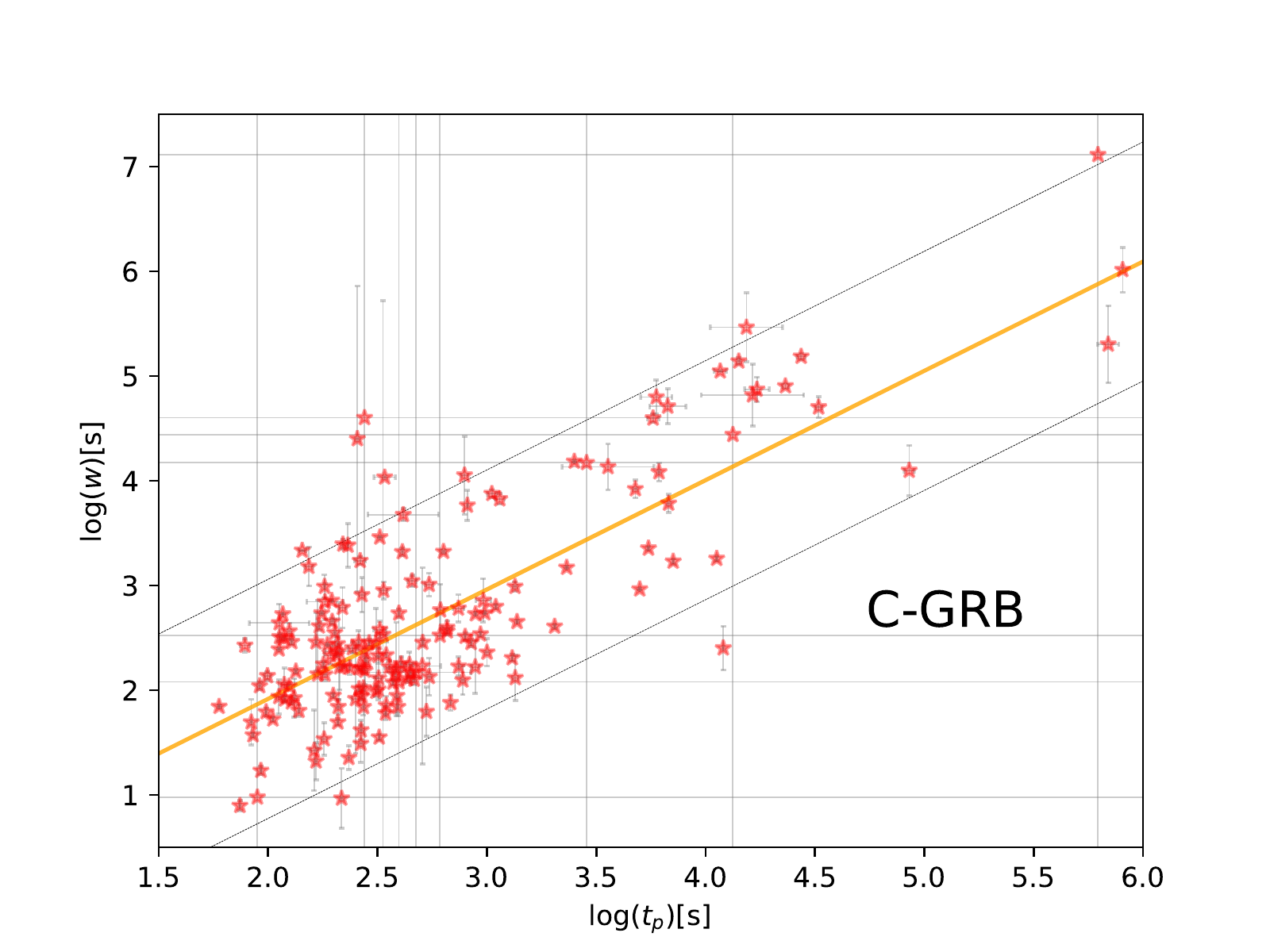}\hfill
    \includegraphics[height=.5\linewidth - 0.25mm, width=.5\linewidth - 0.25mm]{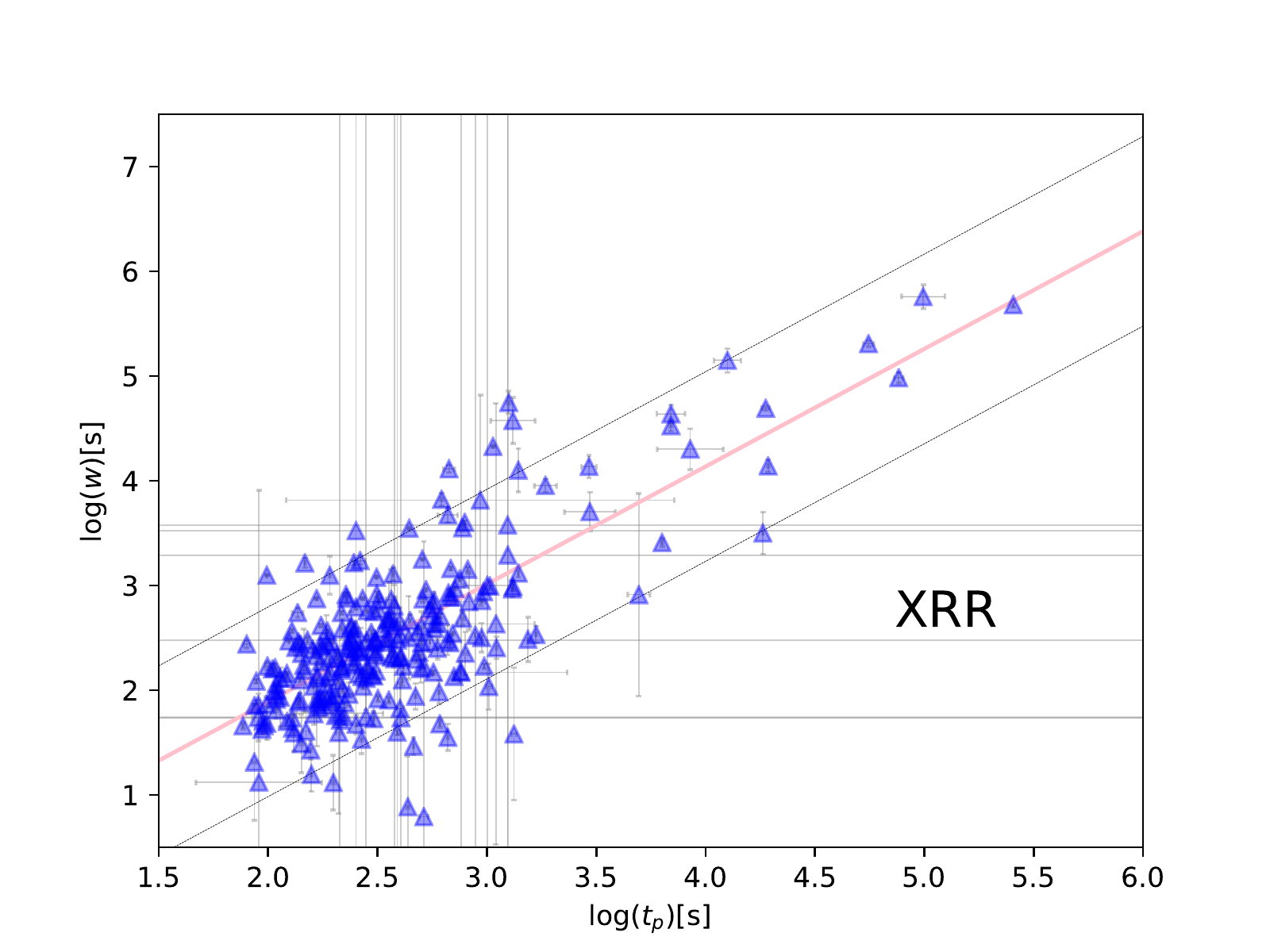}\\[0.5mm]
    \includegraphics[height=.5\linewidth - 0.25mm, width=.5\linewidth - 0.25mm]{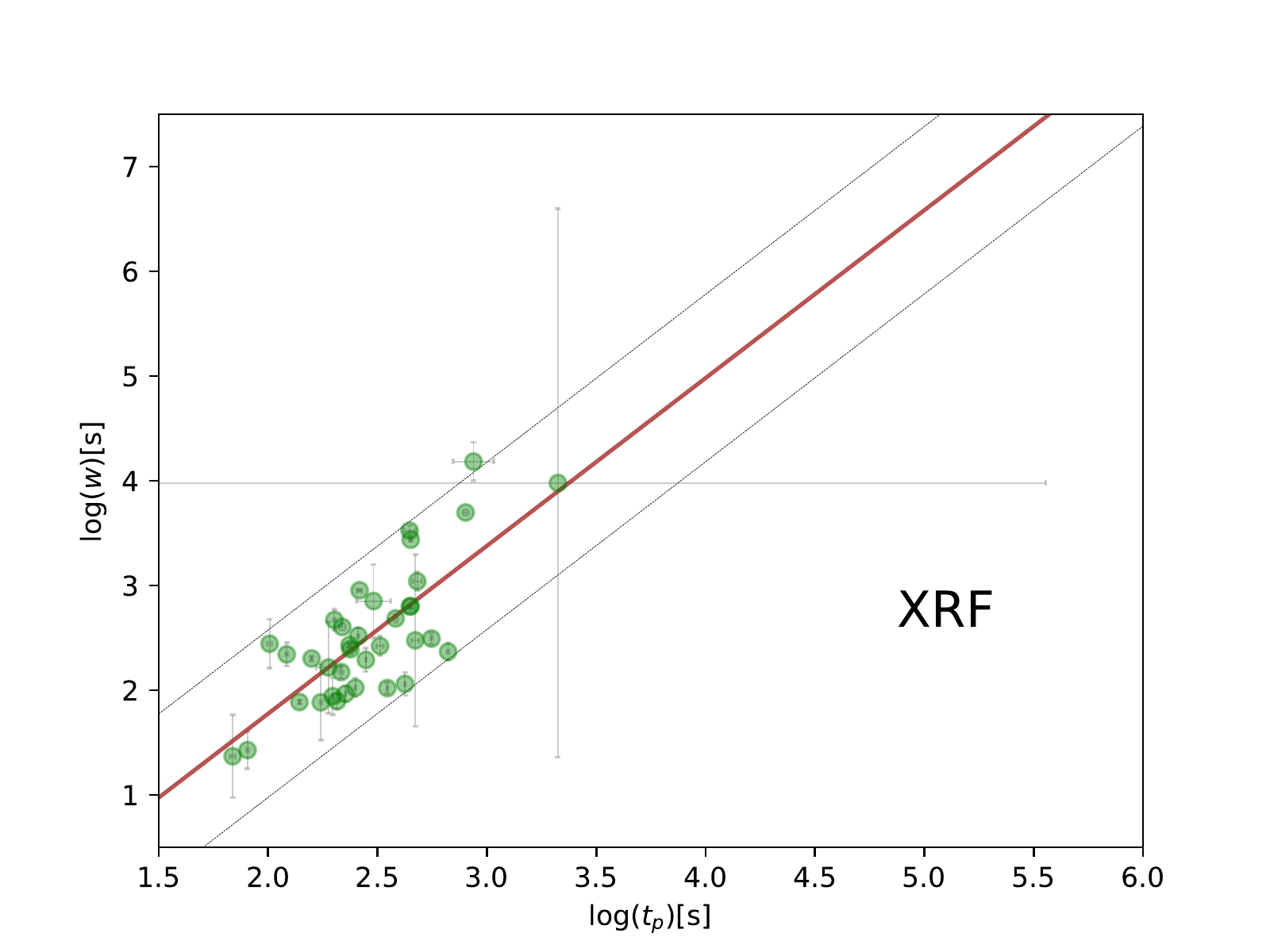}\hfill
    \includegraphics[height=.5\linewidth - 0.25mm, width=.5\linewidth - 0.25mm]{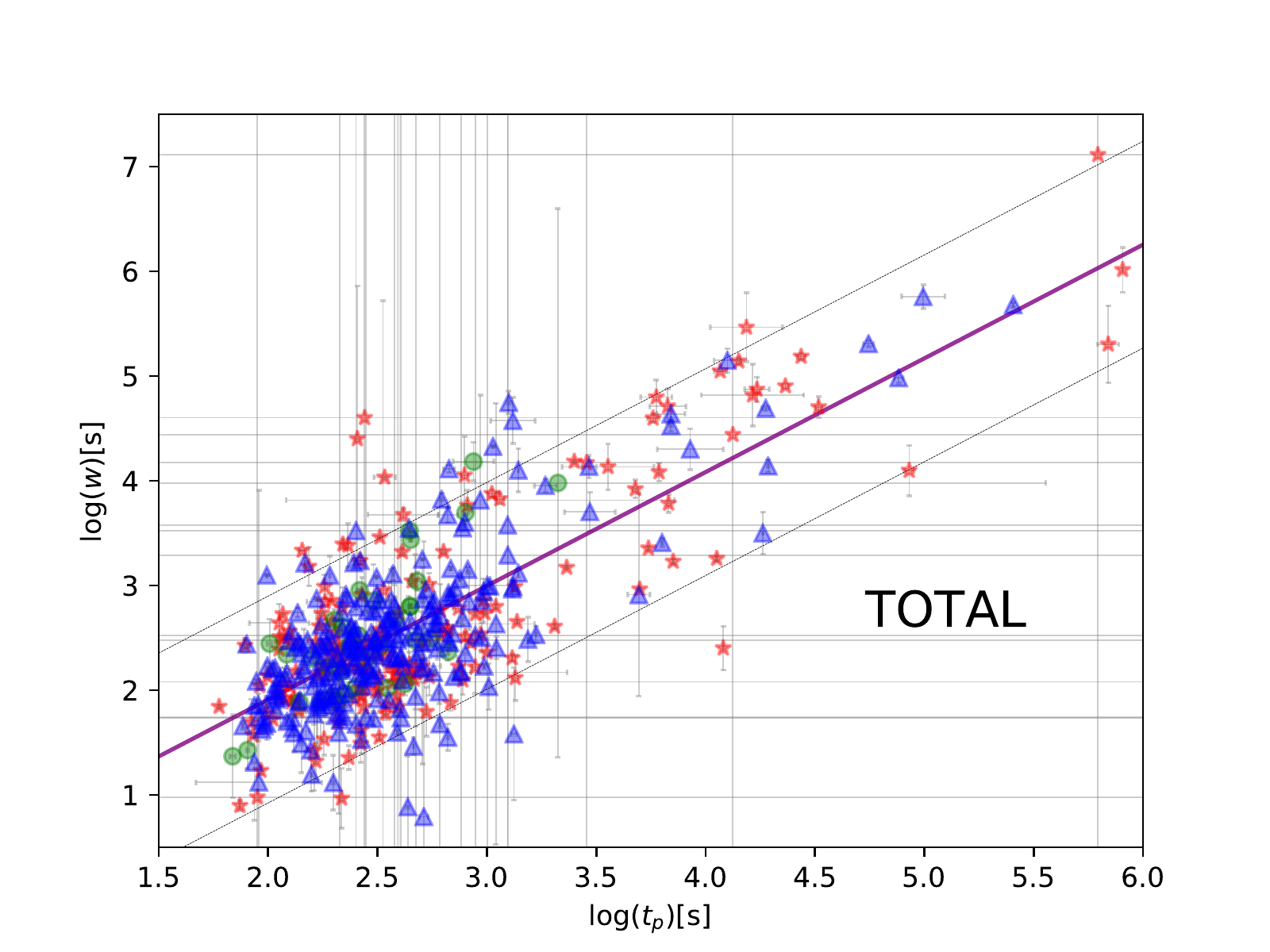}
    \vglue -0.5cm
    \caption{The $w-t_p$ correlations of the X-ray flares for XRFs, XRRs, and C-GRBs. The X-ray flare data are from \cite{Shuangxi2016ApJ}. The green circles represent the X-ray flares of XRFs, the blue triangles represent the X-ray flares of XRRs, and the red stars represent the X-ray flares of C-GRBs, The $w-t_p$ correlations are plotted by the solid lines. The dashed lines enclose the data within $1\sigma$. The best-fitting for XRFs is $\log \mathrm{w}=(-1.41^{+0.73}_{-0.73})+(1.59^{+0.30}_{-0.29})\log \mathrm{t}_\mathrm{p}$ with the scatter of $\sigma = 0.40^{+0.06}_{-0.05}$. The best-fitting for XRRs is $\log \mathrm{w}=(-0.36^{+0.14}_{-0.15})+(1.12^{+0.06}_{-0.05})\log \mathrm{t}_\mathrm{p}$ with the scatter of $\sigma = 0.45^{+0.02}_{-0.02}$. The best-fitting for C-GRBs is $\log\mathrm{w}=(-0.16^{+0.19}_{-0.18})+(1.04^{+0.06}_{-0.07})\log \mathrm{t}_\mathrm{p}$ with the scatter of $\sigma = 0.57^{+0.04}_{-0.03}$. The best-fitting for total GRBs is $\log \mathrm{w}=(-0.25^{+0.11}_{-0.11})+(1.08^{+0.04}_{-0.04})\log \mathrm{t}_\mathrm{p}$ with the scatter of $\sigma = 0.49^{+0.02}_{-0.02}$.}
    \label{Fig:t_p_w:Shuang}
\end{figure}

We also use the X-ray flare data of \cite{Shuangxi2016ApJ} to perform the $w-t_p$ correlations for XRFs, XRRs, and C-GRBs. We find the correlation of
$\log w=(-1.41^{+0.73}_{-0.73})+(1.59^{+0.30}_{-0.29})\log t_p$ with $\sigma = 0.40^{+0.06}_{-0.05}$ for XRFs, the correlation of $\log w=(-0.36^{+0.14}_{-0.15})+(1.12^{+0.06}_{-0.05})\log t_p$ with $\sigma = 0.45^{+0.02}_{-0.02}$ for XRRs, and the correlation of $\log w=(-0.16^{+0.19}_{-0.18})+(1.04^{+0.06}_{-0.07})\log t_p$ with $\sigma = 0.57^{+0.04}_{-0.03}$ for C-GRBs. The results are shown in Figure \ref{Fig:t_p_w:Shuang}.

It seems that the $w-t_p$ correlation of XRFs and that of C-GRBs are different, although the fitting errors and the extrinsic scatters are relatively large. We note that the fitting results from the data of \cite{Chincarini2010MN} and \cite{Shuangxi2016ApJ} are significantly different.

\subsection{XRFs, XRRs, and C-GRBs in the $\log(\Delta F/F)-\log (w/t_p)$ Plane}
The GRB X-ray flare has been explained by some physical models. \cite{Ioka2005ApJ} and \cite{Lazzati2007MN} suggested both internal and external shocks to explain GRB X-ray flares.
The excess flux $\Delta F$ to the underlying flux $F$ can be identified as the X-ray flare flux. It seems that most GRB X-ray flares provided by \cite{Chincarini2010MN} favor the internal origin of the central engine model.

\begin{figure}[h!]
    \centering
    \includegraphics[width=.98\textwidth]{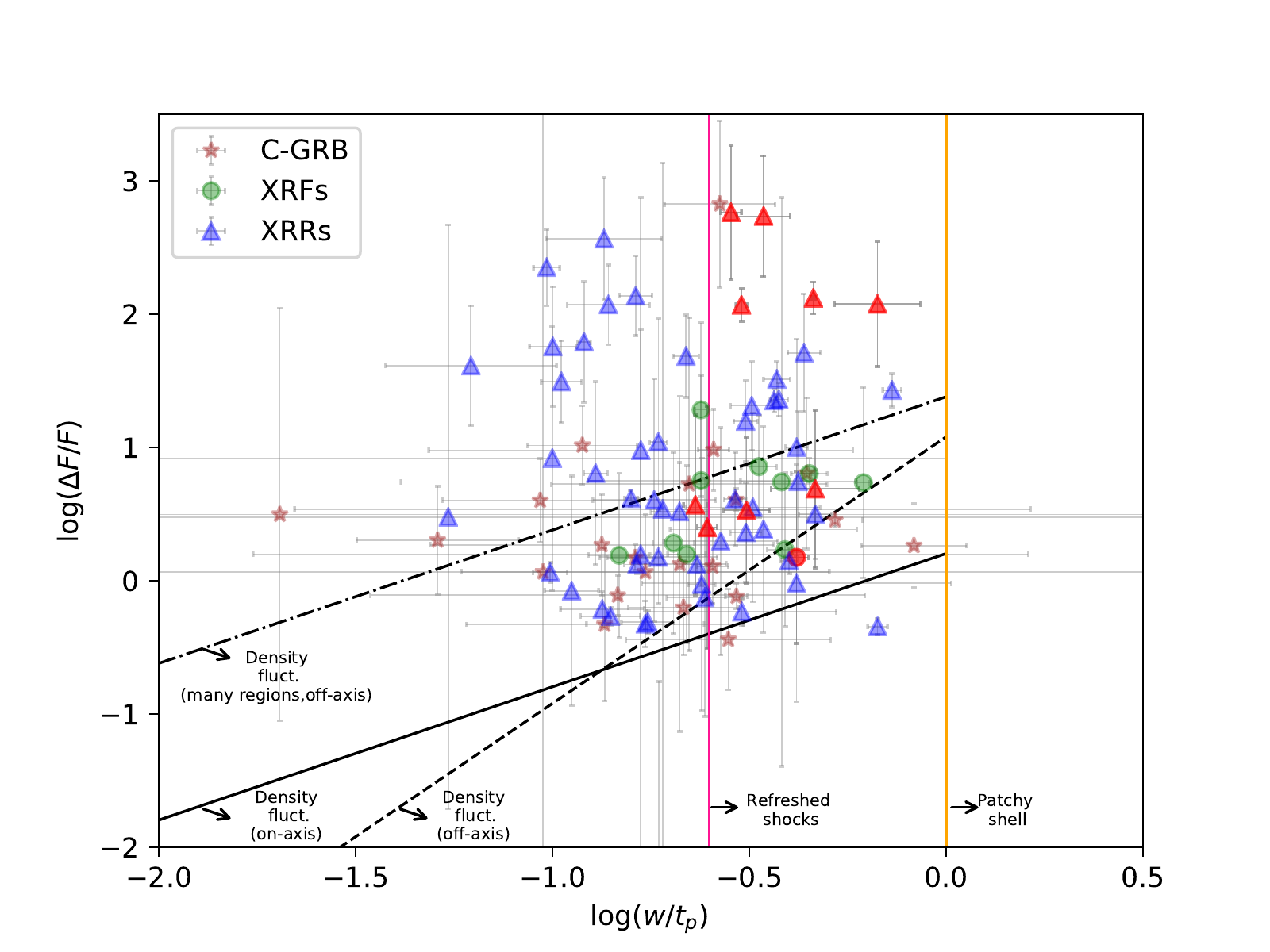}
    \caption{GRB X-ray flares in the log($\Delta F/F$)-log($w/t_p$) plane.
Some modeling constraints are presented by different lines (see Ioka et al. 2005 in detail). The green circles represent the X-ray flares of XRFs, the blue triangles represent the X-ray flares of XRRs, and the brown stars represent the X-ray flares of C-GRBs. The red symbols indicate the bright X-ray flares that have $r_i\ge 0.2$.}
    \label{Fig:origin}
\end{figure}

We can use the X-ray flare data of \cite{Chincarini2010MN} to examine the X-ray flare distribution in the $\log\Delta F/F-\log w/t_p$ plane for XRF, XRR, and C-GRB subclasses. The results are shown in Figure \ref{Fig:origin}. The constraints of some models are also shown\footnote{We do not find the $F$ values in the X-ray flare data of \cite{Shuangxi2016ApJ}, and we cannot examine the distribution of those X-ray flares in the $\log\Delta F/F-\log w/t_p$ plane.}.
It seems that there is no difference among XRF, XRR, and C-GRB subclasses on the X-ray flare distribution in the $\log\Delta F/F-\log w/t_p$ plane.

\subsection{Spectral Properties of X-Ray Flares among XRF, XRR, and C-GRB Subclasses}
\begin{figure}[h!]
    \centering
    \includegraphics[height=.6\linewidth - 0.25mm]{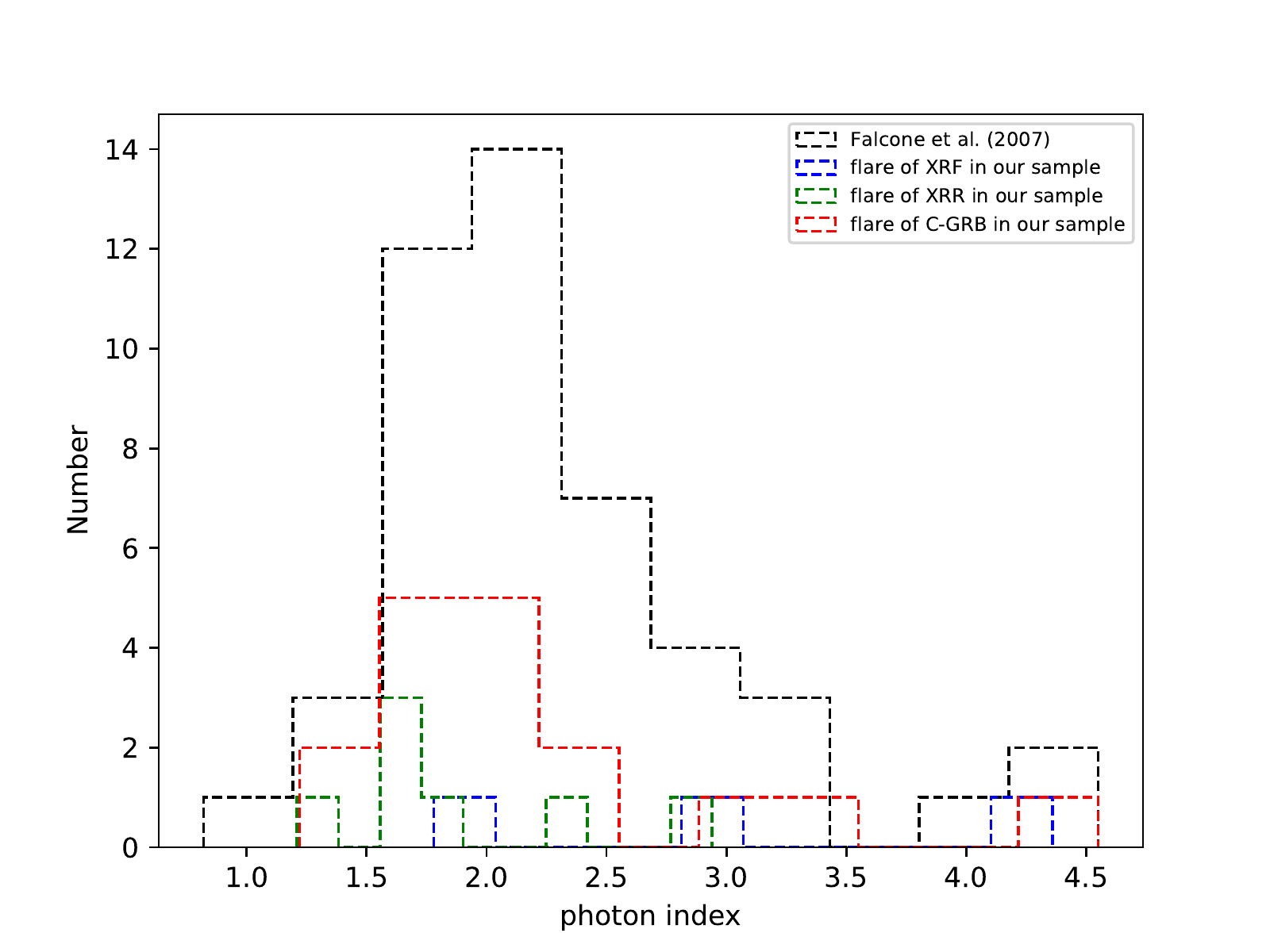}
    \includegraphics[height=.6\linewidth - 0.25mm]{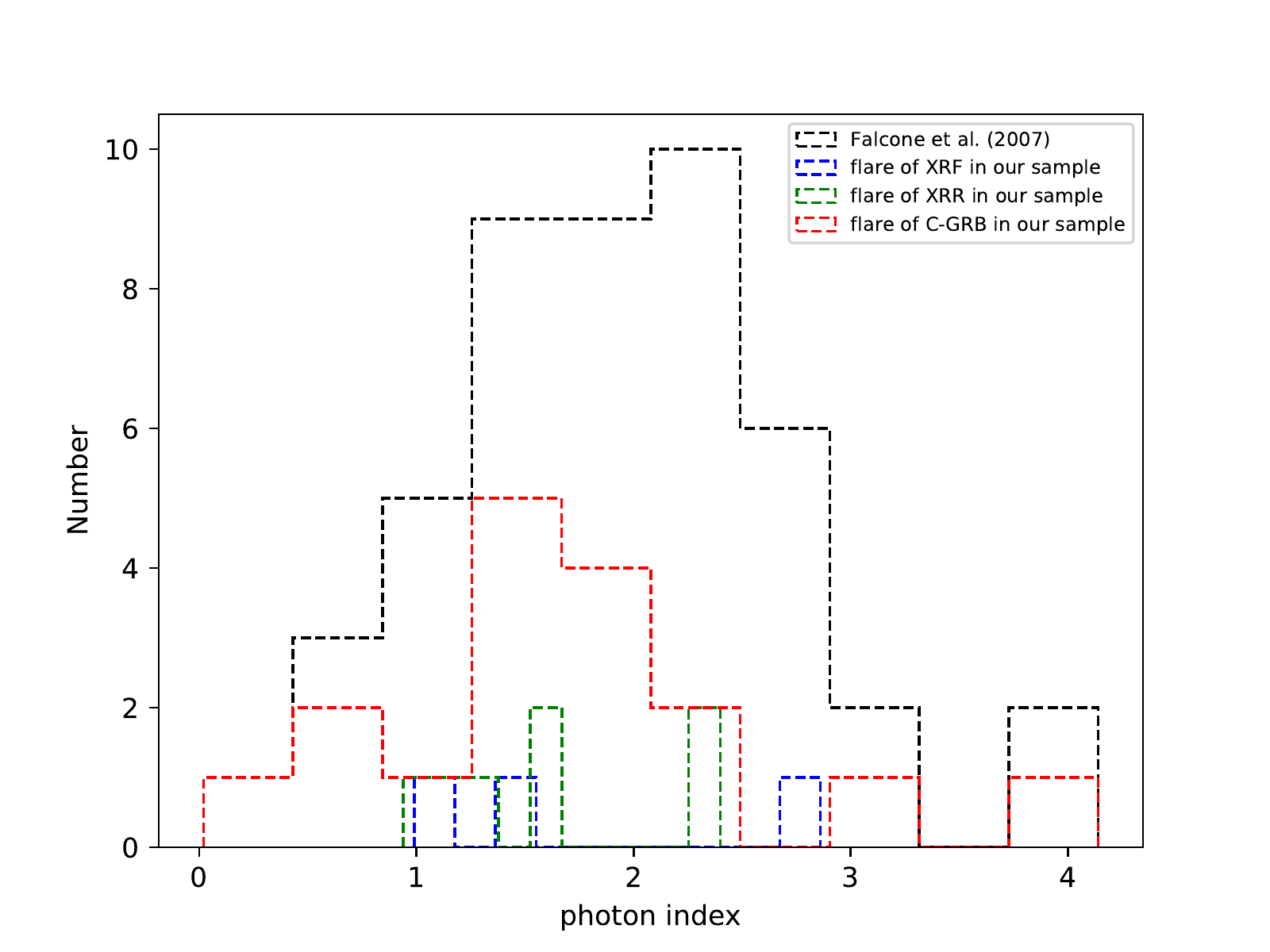}
      \caption{Top panel: the distributions of the X-ray flare power-law spectral fitting photon index for XRFs, XRRs, and C-GRBs. Bottom panel: the distributions of the X-ray flare cutoff power-law spectral fitting photon index for XRFs, XRRs, and C-GRBs.
  }
    \label{Fig:distribution_spec}
\end{figure}

Falcone et al. (2007) presented one detailed spectral analysis on GRB X-ray flares. Here, we select the results of the X-ray flare spectra fitting from the power-law, the cutoff power-law, and the Band function models. Thus, the spectral indices and the peak energy values of the X-ray flares are provided. We obtain the distributions of $\alpha_{\rm{pl}}$ and $\alpha_{\rm{cpl}}$ from the power-law and the cutoff power-law model fittings among XRF, XRR, and C-GRB subclasses in Figure \ref{Fig:distribution_spec}. We also obtain the distributions of $\alpha$, $\beta$, and $E_p$ from the Band function fitting among XRF, XRR, and C-GRB subclasses in Figure \ref{Fig:distribution_band} and Figure \ref{Fig:correlation_ep}. Due to the small sample from Falcone et al. (2007), it is hard to distinguish the differences of the X-ray flare spectral properties among XRF, XRR, and C-GRB subclasses.

\begin{figure}[h!]
    \centering
    \includegraphics[height=.6\linewidth - 0.25mm]{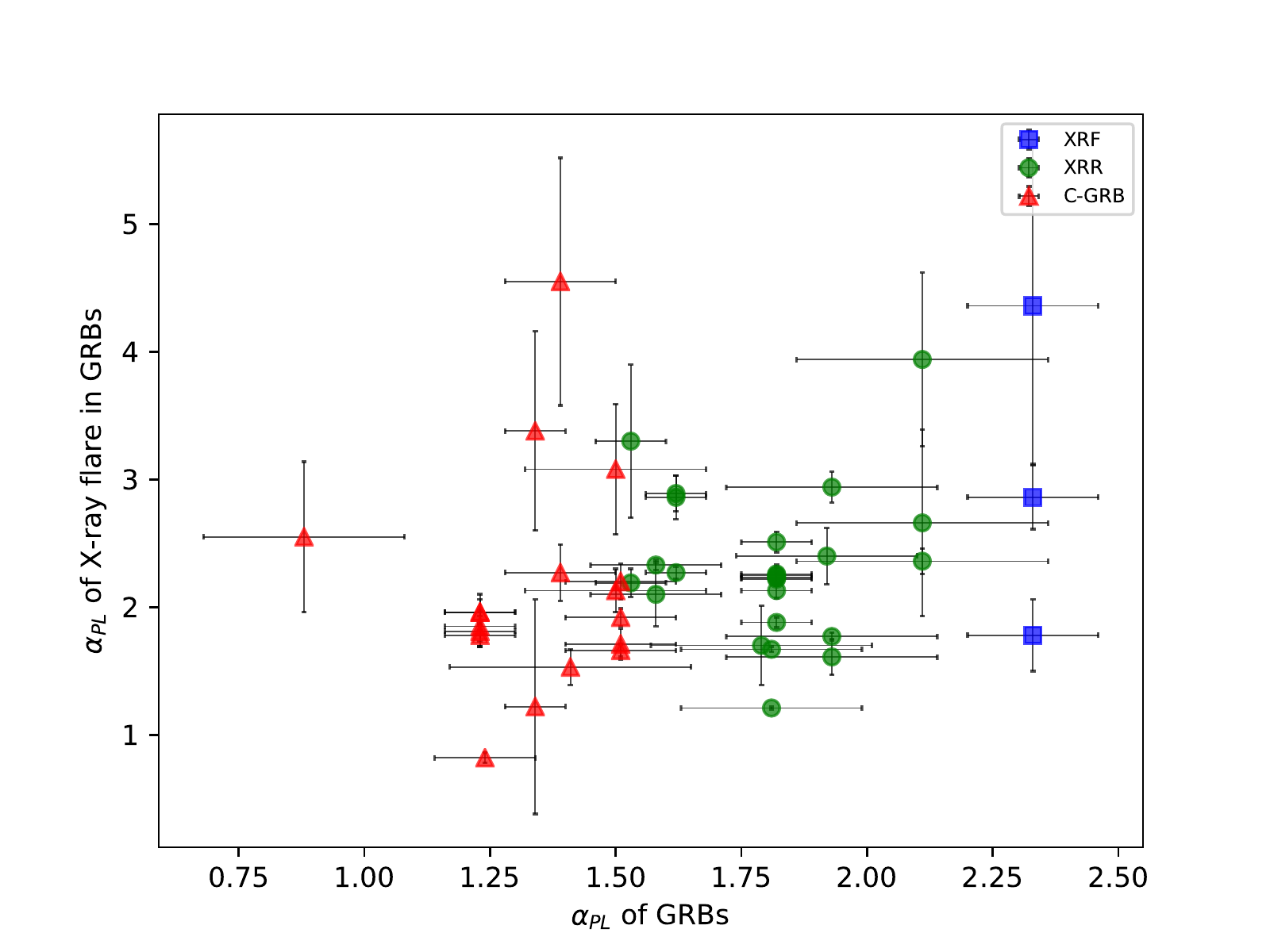}
    \includegraphics[height=.6\linewidth - 0.25mm]{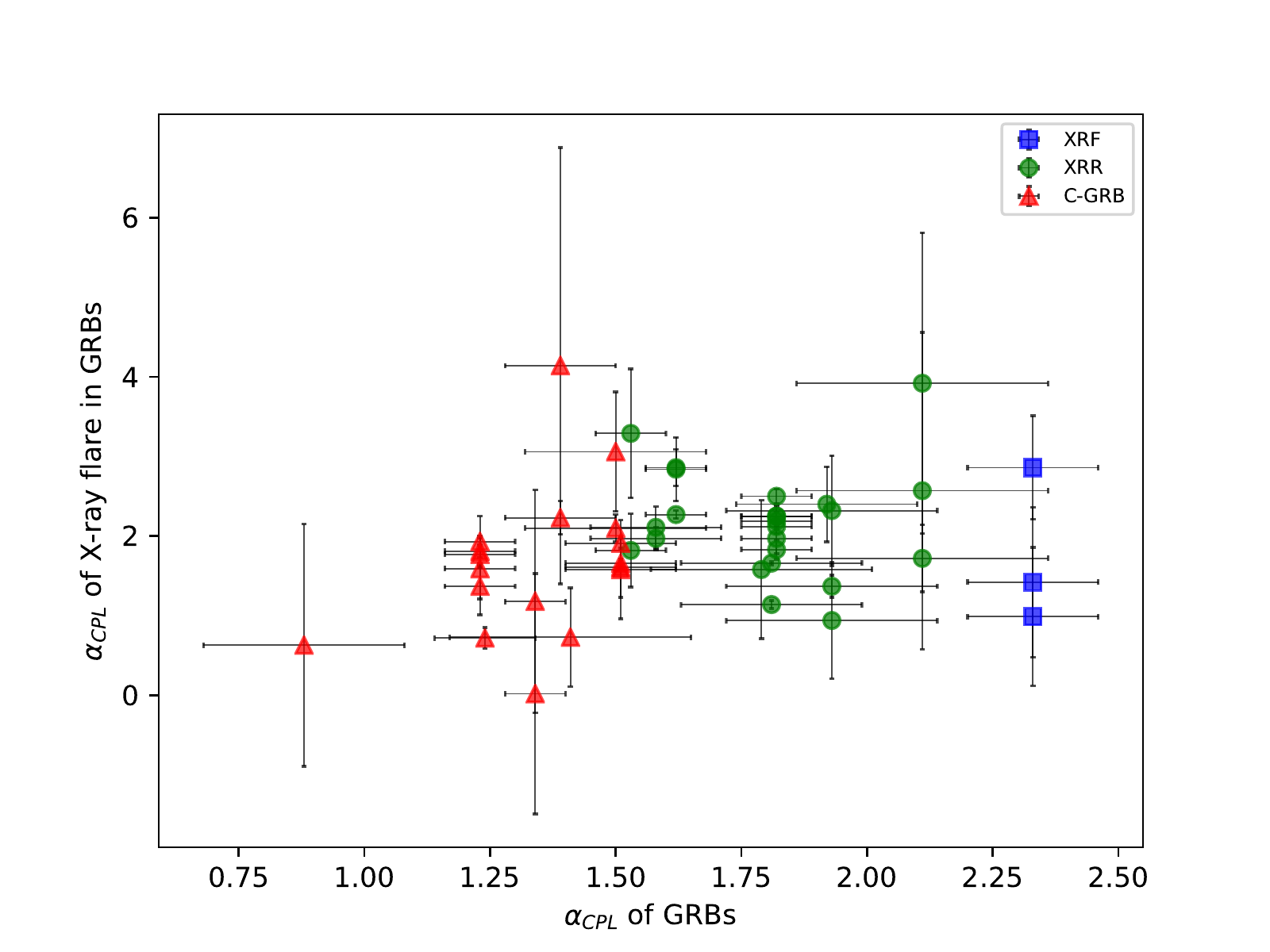}
    \caption{Top panel: the power-law spectral fitting photon index of the X-ray flare vs. the power-law spectral fitting photon index of the prompt emission for XRFs, XRRs, and C-GRBs. Bottom panel: the cutoff power-law spectral fitting photon index of the X-ray flare vs. the cutoff power-law spectral fitting photon index of the prompt emission for XRFs, XRRs, and C-GRBs.}
    \label{Fig:correlation_spec}
\end{figure}

\begin{figure}[h!]
    \centering
    \includegraphics[height=.6\linewidth - 0.25mm]{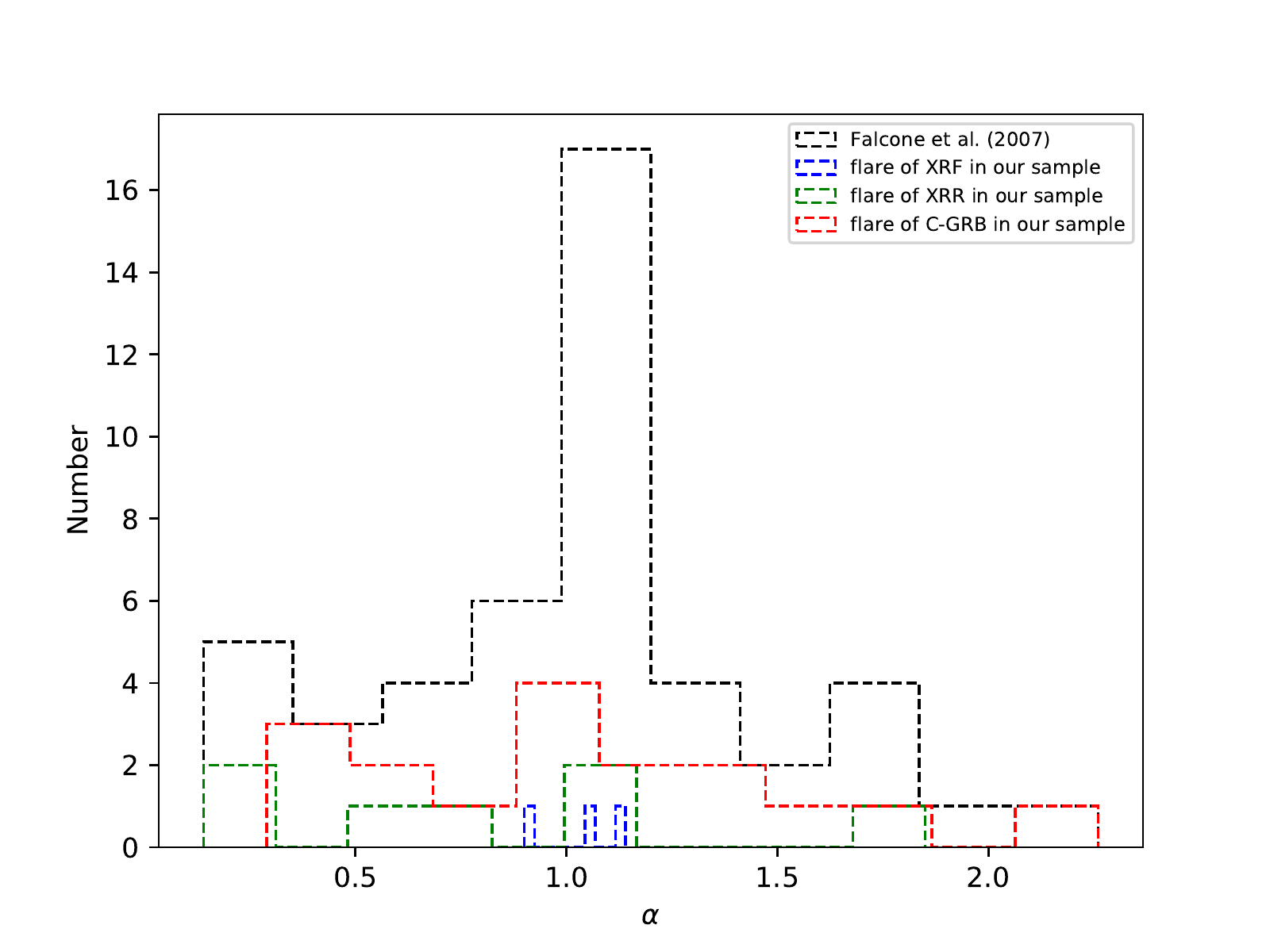}
    \includegraphics[height=.6\linewidth - 0.25mm]{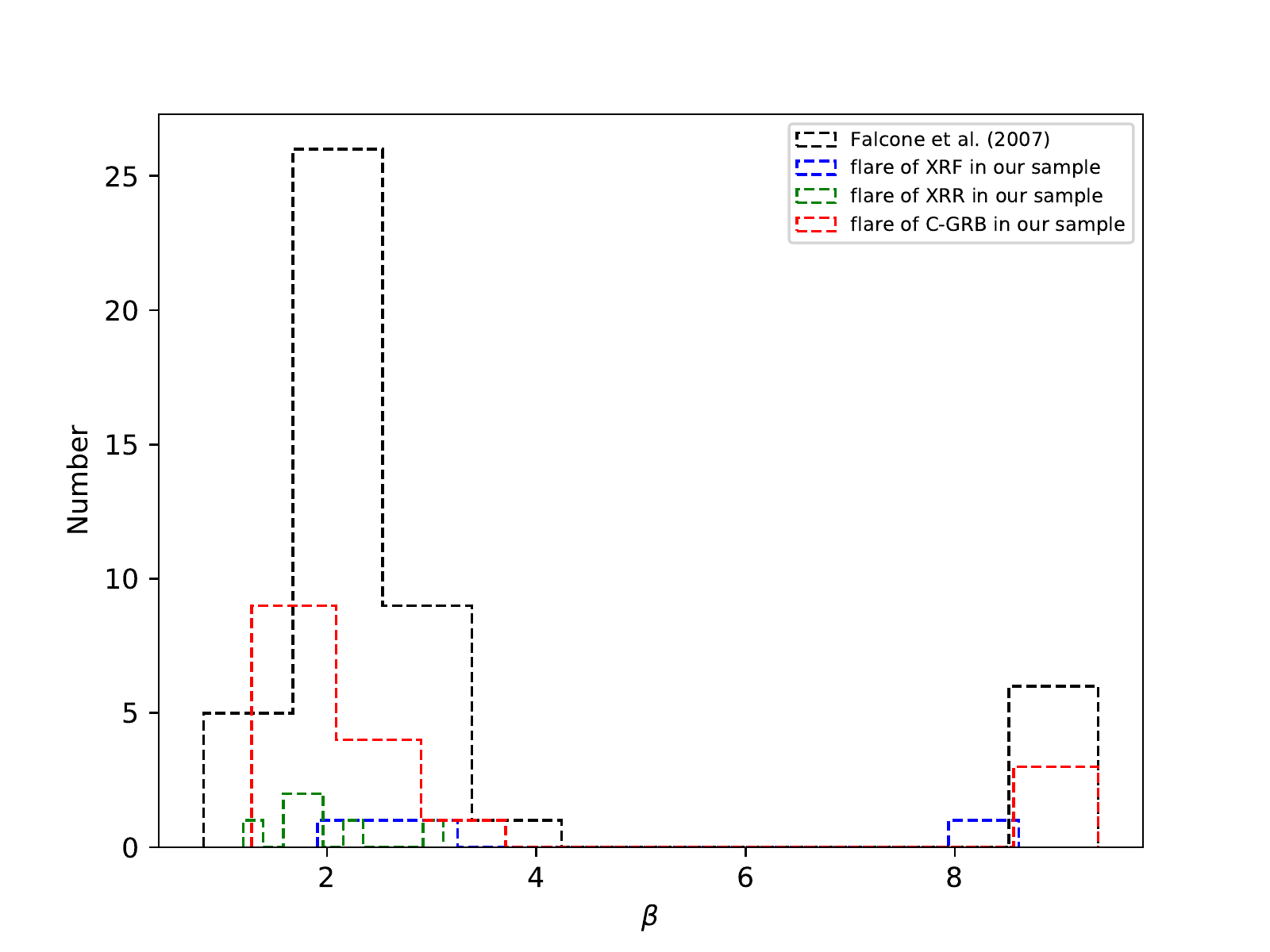}
      \caption{Top panel: the distributions of the X-ray flare low-energy spectral index $\alpha$ of Band function for XRFs, XRRs, and C-GRBs. Bottom panel: the distributions of the X-ray flare high-energy spectral index $\beta$ of Band function for XRFs, XRRs, and C-GRBs.
  }
    \label{Fig:distribution_band}
\end{figure}

\begin{figure}[h!]
    \centering
    \includegraphics[height=.6\linewidth - 0.25mm]{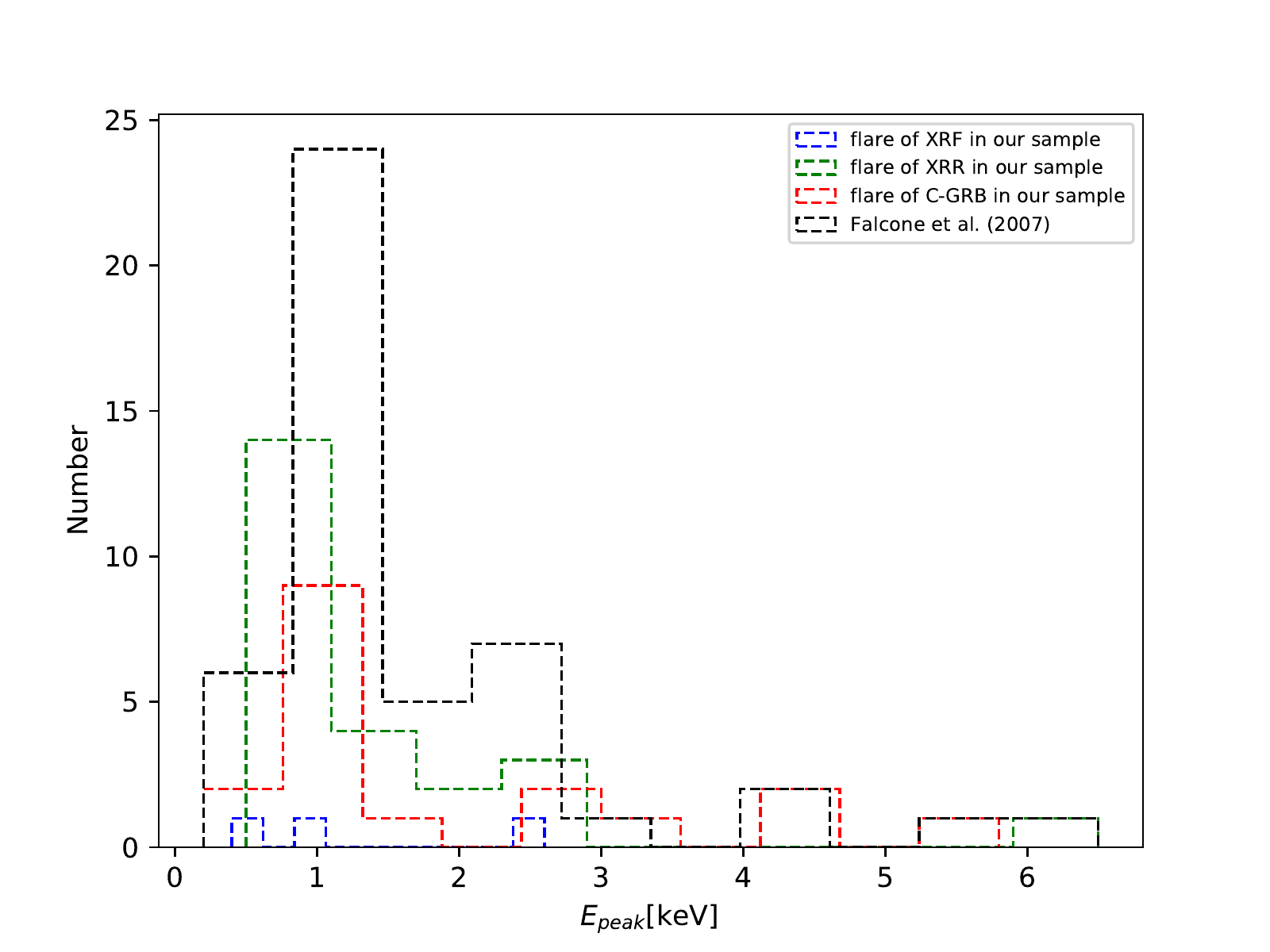}
    \includegraphics[height=.6\linewidth - 0.25mm]{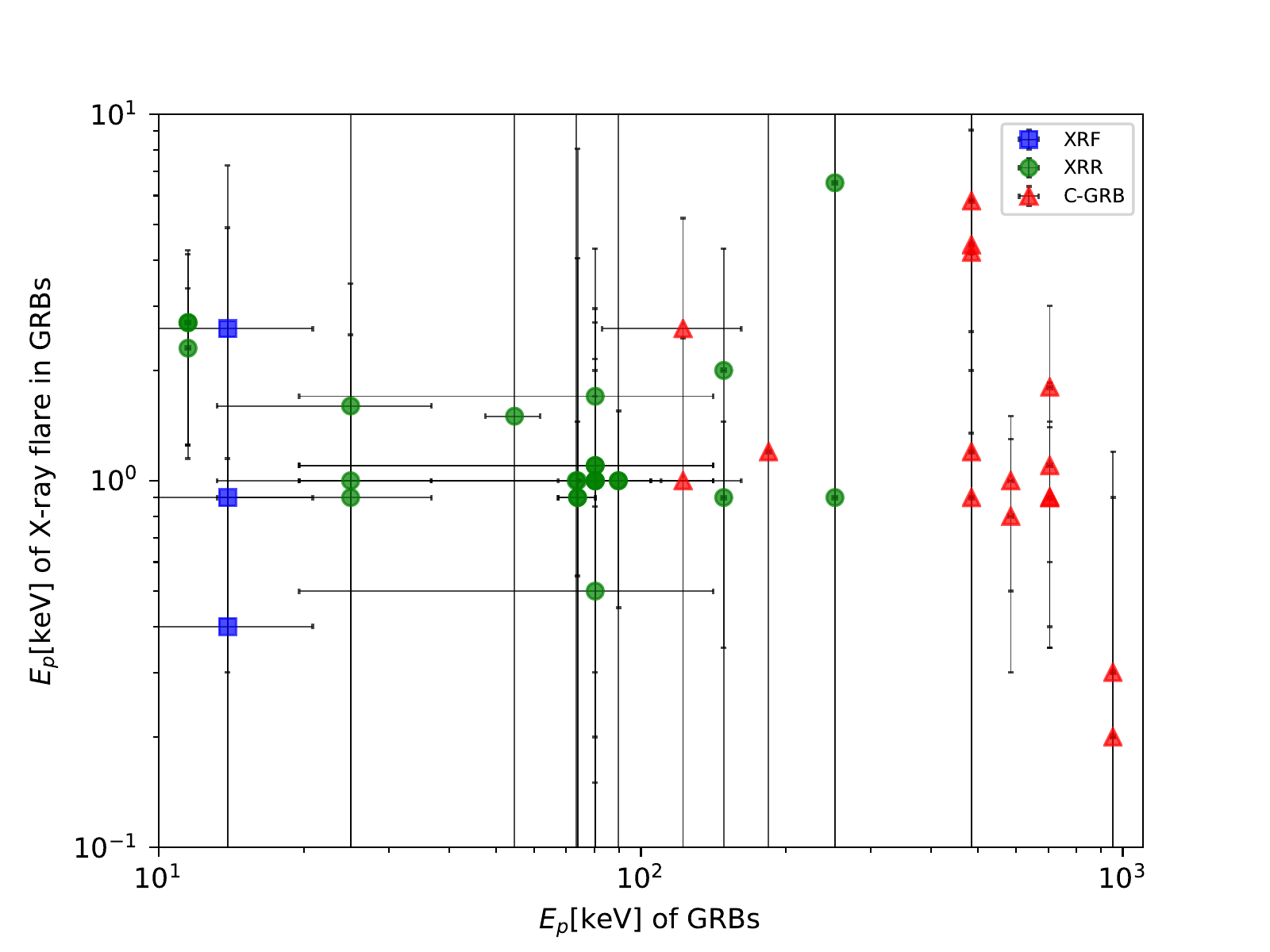}
    \caption{Top panel: the distributions of the X-ray flare peak energy $E_{\rm{peak}}$ of Band function for XRFs, XRRs, and C-GRBs. Bottom panel: the X-ray flare peak energy vs. the prompt emission peak energy for XRFs, XRRs, and C-GRBs.}
    \label{Fig:correlation_ep}
\end{figure}

In order to examine the spectral correlation between the prompt emission and the X-ray flare, we show the spectral index of the X-ray flare vs. the spectral index of the prompt emission in Figure \ref{Fig:correlation_spec}. The spectral indices are obtained from the power-law model fitting and the cutoff power-law modeling fitting. We do not find any possible correlation between the spectral index of the X-ray flare and that of the prompt emission. We also show the X-ray flare peak energy vs. the prompt emission peak energy in Figure \ref{Fig:correlation_ep}. We do not find any possible correlation between the X-ray flare peak energy and the prompt emission peak energy. A large data sample is required for further investigations.

\section{Discussion}
When we mention that GRBs are associated with supernovae, we find only three GRBs that have X-ray flares in this paper. They are XRR 060729, XRR 060904, and C-GRB 111209A. Thus, we do not find any XRFs with X-ray flares to be associated with supernovae. However, a few XRFs without X-ray flares have associations with supernovae. For example, XRF 060218 is associated with SN 2006aj, XRF 081007 is associated with SN 2008hw, and XRF 100316D is associated with 2010bh \citep{Bi2018ApJ}. We expect a large sample in the future to further investigate this issue.

The X-ray flares that are contemporaneous with the early X-ray afterglow are likely due to a late time manifestation of the same emission mechanism as the prompt emission \citep{Zhang2006ApJ}.
\cite{Godet2007A&A} found that the prompt emission of XRF 050822 was followed by three X-ray flares, and the X-ray flares are shown in the deep decay phase of the X-ray light curve. Kazanas et al. (2015) comprehensively investigated the BAT-to-XRT flux ratio of the {\it Swift}-detected GRBs. From the X-ray flare sample of Chincarini et al. (2010), it was suggested that the GRB X-ray flares can be
generated by the central engine, and this suggestion seems to be also valid for the X-ray flares in some short GRBs \citep{Mu2018MN}. Thus, we indicate a possible energy contribution from the bright X-ray flare to the total GRB emission, and these GRBs might be identified as XRFs. In this paper, we identify some observational properties of the X-ray flares among XRF, XRR, and C-GRB subclasses.
We find that XRFs and XRRs have more bright X-ray flares than C-GRBs. When we calculate the fluence ratio between the X-ray flare and the prompt emission for each GRB, we identify that XRFs are different from C-GRBs. The $w-t_p$ correlations are also different between XRF and C-GRB subclasses.
However, the X-ray flares observed by {\it Swift}-XRT are in the energy range of $0.2-10$ keV, while the XRFs identified by {\it Swift}-BAT are in the energy range of 15 to 350 keV. Therefore, when we indicate that there is an inclination to identify the GRBs with bright X-ray flares as XRFs, we should consider that X-ray flare and XRF are detected in different energy bands. Falcone et al. (2006) examined the spectral properties of the X-ray flare in GRB 090502B from the {\it Swift}-XRT data,  and one hardening signature was clearly seen during the giant flare duration. Although we perform a detailed analysis of the X-ray flare spectral properties among XRF, XRR, and C-GRB subclasses in this paper, we expect large samples of the X-ray flare spectral analysis in the future.

The intrinsic X-ray flare properties, such as fluence, width, and peak time, should be corrected by both bulk Lorentz factor and redshift. When we compare the X-ray flare properties among XRF, XRR, and C-GRB subclasses, we do not consider the effect from the bulk Lorentz factor of the X-ray flares.
Thus, we note the bias related to the comparison of fluence, width, and peak time of X-ray flares among XRF, XRR, and C-GRB subclasses, even these temporal parameters are redshift-corrected.
However, when we consider $r_i$, $r_t$, and $w-t_p$ correlation for each GRB, the results are not affected by the bulk Lorentz factor and the redshift.

The selection criteria of the X-ray flare can be different among different X-ray flare samples. The flares with S/Ns larger than 5 were selected in Chincarini et al. (2007) and the flares with S/Ns larger than 3 were selected in Falcone et al. (2007), such that the low-signal X-ray flares were excluded. However, some fluctuations in GRB X-ray light curves were identified as X-ray flares in Chincarini et al. (2010). We list the different sources between Chincarini et al. (2010) and Yi et al. (2016) in Section 2. Moreover, in the sample of Chincarini et al. (2010), some blended flares are included. The blended X-ray flares of one certain GRB in Chincarini et al. (2010) were identified in a different way by Yi et al. (2016). Thus, judging the blended structure of the X-ray flare is a challenge. Compared to Chincarini et al. (2010), Yi et al. (2016) considered the X-ray flares with the redshift measurements, and small fluctuations shown in light curves were also identified as X-ray flares. Chincarini et al. (2010) selected only the early-time (occurring less than $10^3$ s after the trigger) X-ray flares, while Yi et al. (2016) considered both the early-time and the late-time X-ray flares. Therefore, systematic bias can be induced in our results. In this paper, we consider the temporal samples of Chincarini et al. (2010) and Yi et al. (2016), because the two samples have more X-ray flare data than Chincarini et al. (2007).

Different fitting functions have been applied to the temporal structure of the X-ray flare. Chincarini et al. (2007) simply used a Gaussian function to fit the X-ray flare temporal structure. The profile proposed by Norris et al. (2005) was applied in Chincarini et al. (2010). This profile consists of two combined exponentials to be flexible for the fitting of pulse shapes. Yi et al. (2016) performed the fitting with a smooth broken power-law function. Comparison of the different fitting functions is out of the scope of this paper, but different fitting functions may lead to a systematic bias on the fitting results. We list the temporal results of the X-ray flares given by Chincarini et al. (2007), Chincarini et al. (2010), and Yi et al. (2016) in Table 3. We see some X-ray flares with different fitting results and different blended X-ray flare identification in one GRB. However, we note that the log($\Delta F/F$)-log($w/t_p$) plane shown in Chincarini et al. (2007) and that shown in this paper (see \ref{Fig:origin}) have no significant differences. This restates the indication that GRB X-ray flares have a central engine origin.

Although GRBs are usually classified as long/short GRBs, some evidence has shown that GRBs may have an intermediate class. \citet{ho10} performed one analysis on the {\it Swift}-detected GRBs, in which the GRB hardness ratio was considered. de Ugarte Postigo et al. (2011) further examined some GRBs that are belong to the intermediate class. \citet{zi15} suggested the existence of the intermediate group due to the asymmetric distribution of the GRB duration from the {\it Swift}-BAT sample. However, \citet{koen12} performed a careful analysis by the autocorrelation function with the {\it Swift}-BAT data, and the results seem to incline the two-class (long/short) GRB classification. \citet{tarnopolski19} examined the GRB data from BATSE and {\it Fermi} observations, and the results infer that the intermediate class is not physical.

It is important to note that the long/short GRB classification takes effect on the classification of XRFs, XRRs, and C-GRBs, although the investigation of the GRB hardness-duration is not a major topic in this paper. For example, \citet{veres10} found that intermediate GRBs have soft spectra. This indicates that intermediate GRBs are related to XRFs. From the {\it Swift}-BAT3 catalog, Bi et al. (2018) found that a few XRFs are short GRBs. In this paper, we do not find any XRFs having X-ray flares that are short GRBs, but a large data sample is required to further confirm this issue.

The XRF, XRR, and C-GRB samples of Bi et al. (2018) adopted in this paper follow the selection criteria by Sakamoto et al. (2008). Some XRRs are likely to be extensions of C-GRBs: the selection criteria may induce the overlaps of XRFs/XRRs and CRRs/C-GRBs that are shown in the hardness-duration distribution (Figure \ref{Fig:hardness_T90}). The overlaps are also shown in the $E_p$ distributions among XRF, XRR, and C-GRB subclasses (Figure \ref{Fig:GRBEp}). When we further examine the spectral properties of the X-ray flares among XRF, XRR, and C-GRB subclasses, the $E_p$ overlaps of XRFs/XRRs and XRRs/C-GRBs are clearly shown in Figure \ref{Fig:correlation_ep}. Meanwhile, we note that the $E_p$ values of both X-ray flare and prompt emission are derived in narrow spectral ranges. This creates large uncertainty on the $E_p$ values. However, the overlaps of XRFs/XRRs and XRRs/C-GRBs are not shown in the spectral index distributions (Figure \ref{Fig:correlation_spec}). Further investigations are useful when we have more data from the spectral analysis in the future.

Different detectors and instruments have bias on the GRB classification. In the literature, \citet{tar15} summarized long/short classification by several GRB detectors. \citet{tarnopolski19} further noted that the {\it Swift}-BAT detection is more sensitive to soft bands. Thus, some {\it Swift}-detected GRBs might be identified as XRFs by the biased detection, and even the intermediate class is elusive. Spectral cross-calibration from different detectors to simultaneously observe the same GRB should be performed to give a reliable conclusion after enough data are accumulated \citep{saka11}.

\section{Summary}
GRB X-ray flares may have internal origin. In the paper, we investigate the different temporal and spectral properties of the X-ray flares among XRF, XRR, and C-GRB subclasses. We find that: (1) XRFs and XRRs have more bright X-ray flares than GRBs; (2) the ratio between the X-ray flare fluence and the prompt emission fluence has different distributions between XRF and C-GRB subclasses; (3) and the linear correlations between the duration and the peak time of the X-ray flares are different between XRF and C-GRB subclasses. We infer that there is an inclination to identify the GRBs with bright X-ray flares as XRFs. We caution that the classifications of XRFs, XRRs, and C-GRBs may affect our results. Our results might also be affected by the selection effects and the instrument bias.
We expect more observational data to further investigate our results in the future, and some theoretical explanations are also required.

\section*{Acknowledgements}
We utilize the {\it Swift}-BAT/XRT databases and the related webpages for the further data analysis in this paper.
J.M. is supported by the National Natural Science Foundation of China (11673062), the Hundred Talent Program of the Chinese Academy of Sciences, and the Oversea Talent Program of Yunnan Province.

\setlength{\tabcolsep}{1pt}
\begin{deluxetable}{lccccccccccccccccccl}
    \rotate
    \centering
    \tablecolumns{21}
    \tablecaption{Temporal Properties of X-Ray flares among XRFs, XRRs, and C-GRBs \label{Tab:infor}}
    \tablewidth{0pt}
    \tabletypesize{\fontsize{5pt}{7pt}\selectfont}
    \tablehead{
        \colhead{GRB}  &    \colhead{GRB}   &    \colhead{GRB}   &    \colhead{$z$}    &    \colhead{$T_{90}$}   &    \colhead{$E_p^{a}$}    &    \colhead{$\alpha_{\rm{PL}}$}    &    \colhead{$\alpha_{\rm{CPL}}$}    &    \colhead{$S_p$(25-50keV)}    &    \colhead{$S_p$(50-100keV)}    &    \colhead{$S_p^{f}$(15-350keV)}    &    \colhead{$t_p^{b}$}    &    \colhead{$w^{c}$}  &    \colhead{$S_i^{d}$(0.2-10keV)}   &    \colhead{$S_t^{e}$(0.2-10keV)}     &    \colhead{r$_i^{g}$}  &    \colhead{r$_t^{h}$}    &    \colhead{r$_i$}   &    \colhead{r$_t$ }    &    \colhead{Ref.$^k$}\\
        \colhead{Name}  &    \colhead{Type}   &    \colhead{Type}    &    \colhead{ }    &    \colhead{s}   &    \colhead{$10^{2}$keV}    &    \colhead{ }    &    \colhead{ }    &    \colhead{$10^{-8}$erg cm$^{-2}$}    &    \colhead{$10^{-8}$erg cm$^{-2}$}    &    \colhead{$10^{-8}$erg cm$^{-2}$}    &    \colhead{s}    &    \colhead{s}  &    \colhead{$10^{-8}$erg cm$^{-2}$}   &    \colhead{$10^{-8}$erg cm$^{-2}$}     &    \colhead{ }  &    \colhead{ }    &    \colhead{$\ge 0.2$$^i$ }   &    \colhead{ $\ge 0.2$$^j$}    &    \colhead{ }\\
    }
    \startdata
    \multirow{2}{*}{GRB 050406}    &    \multirow{2}{*}{XRF}    &    \multirow{2}{*}{\nodata}    &    \multirow{2}{*}{2.7}    &    \multirow{2}{*}{5.8$\pm$1.4}    &    \multirow{2}{*}{0.3$\pm$6.1e-2}    &    \multirow{2}{*}{2.4$\pm$0.3}    &    \multirow{2}{*}{-4.2e-1$\pm$2.4}    &    \multirow{2}{*}{2.5$\pm$0.5}    &    \multirow{2}{*}{1.9$\pm$0.7}    &    \multirow{2}{*}{142.0$\pm$26.4}    &    218.1$\pm$7.3    &    401.8$\pm$28.9    &    1.3$\pm$0.1    &    \multirow{1}{*}{1.3$\pm$0.1}    &    9.2e-3    &    \multirow{1}{*}{9.2e-3}    &        &    \multirow{1}{*}{}    &    \multirow{1}{*}{Yi16}    \\
\cdashline{12-20}
    &    &    &    &    &    &    &    &    &    &    &    209.6$\pm$4.0    &    70.1$\pm$7.1    &    0.9$\pm$0.6    &    \multirow{1}{*}{0.9$\pm$0.6}    &    6.3e-3    &    \multirow{1}{*}{6.3e-3}    &        &    \multirow{1}{*}{}    &    \multirow{1}{*}{Ch10}    \\
\cline{1-20}
    \multirow{2}{*}{GRB 050714B}    &    \multirow{2}{*}{XRF}    &    \multirow{2}{*}{\nodata}    &    \multirow{2}{*}{2.438}    &    \multirow{2}{*}{49.4$\pm$11.3}    &    \multirow{2}{*}{0.3$\pm$8.3e-2}    &    \multirow{2}{*}{2.4$\pm$0.3}    &    \multirow{2}{*}{9.2e-3$\pm$1.8}    &    \multirow{2}{*}{26.2$\pm$4.7}    &    \multirow{2}{*}{10.8$\pm$5.8}    &    \multirow{2}{*}{75.1$\pm$17.0}    &    382.9$\pm$7.7    &    484.3$\pm$42.5    &    2.2$\pm$0.3    &    \multirow{1}{*}{2.2$\pm$0.3}    &    3.0e-2    &    \multirow{1}{*}{3.0e-2}    &        &    \multirow{1}{*}{}    &    \multirow{1}{*}{Yi16}    \\
\cdashline{12-20}
    &    &    &    &    &    &    &    &    &    &    &    \nodata    &    \nodata    &    \nodata    &    \multirow{1}{*}{\nodata}    &    \nodata    &    \multirow{1}{*}{\nodata}    &        &    \multirow{1}{*}{}    &    \multirow{1}{*}{Ch10}    \\
\enddata
    \tablecomments{\footnotesize ``Int." in the second column of GRB type means that the GRB belongs to the intermediate class. $E_p^{a}$ indicates the GRB peak energy. $t_p^{b}$ is the peak time of the X-ray flare. $w^{c}$ is the duration of the X-ray flare. $S^{d}$(0.2-10keV) is the fluence of the single X-ray flare. $S_t^{e}$(0.2-10keV) is the total fluence of all X-ray flares corresponding to one GRB. $S_p^{f}$(15-350keV) is the fluence of the GRB prompt emission. r$_i^{g}$ is equal to $S$(0.2-10keV)/$S_p$(15-350keV). r$_t^{h}$ is equal to $S_t$(0.2-10keV)/$S_p$(15-350keV).  r$_i\ge 0.2$$^i$ represents that the ratio of the individual flare is greater than 0.2. If it is greater than the threshold, this column is marked as $\surd$. r$_t\ge 0.2$$^j$ represents whether the ratio of total flare is greater than 0.2. If it is greater than the threshold, this column is marked as $\surd$. Ref.$^k$ lists two samples from which the parameters of the flare are obtained. The sample of Yi16 ranges from 2005 April to 2015 March for Yi et al. (2016). The sample of Ch10 ranges from 2005 April to 2008 March for Chincarini et al. (2010).}
\end{deluxetable}

\clearpage

\renewcommand{\multirowsetup}{\centering}
\setlength{\tabcolsep}{4pt}
\begin{deluxetable}{cccccc}
    \centering
    \tablecaption{K-S test for the distributions of $T_{90}$, $E_{grb,p}$, $r_i$, $r_t$, $S_{flare_i}$, $S_{flare_t}$, $E_{\mathrm{flare}_\mathrm{i},\mathrm{iso}}$, $E_{\mathrm{flare}_\mathrm{t},\mathrm{iso}}$, $w$, $w_z$, $t_p$, and $t_{p,z}$ among XRFs, XRRs, and C-GRBs \label{Tab:K-S}}
    \tablewidth{0pt}
    \tabletypesize{\footnotesize}
    \tablehead{
         \colhead{} & \colhead{} & \colhead{C-GRB versus XRF} & \colhead{C-GRB versus XRR} & \colhead{XRF versus XRR} & \colhead{Reference}\\
         }
    \startdata
\multirow{4 }{*}{$T_{90}$} & sample size & C-GRB:15 XRF:7 & C-GRB:15 XRR:34 & XRF:7 XRR:34 & \multirow{2 }{*}{Ch10}\\
    \cdashline{2-5}[0.8pt/2pt]
                        & $p$-value & $0.11$ & $0.84$ & $0.26$ & \\
    \cline{2-6}
      & sample size & C-GRB:64 XRF:16 & C-GRB:64 XRR:118 & XRF:16 XRR:118 & \multirow{2 }{*}{Yi16}\\
    \cdashline{2-5}[0.8pt/2pt]
                        & $p$-value & $0.10$ & $0.82$ & $0.08$ & \\
\hline
    \multirow{2 }{*}{$E_{grb,p}$} & sample size & C-GRB:45 XRF:15 & C-GRB:45 XRR:102 & XRF:15 XRR:102 & \multirow{2 }{*}{Bi18}\\
    \cdashline{2-5}[0.8pt/2pt]
                        & $p$-value & $1.2\times10^{-10}$ & $1.7\times10^{-5}$ & $3.1\times10^{-10}$ & \\
    \hline
    \multirow{4 }{*}{r$_i$} & sample size & C-GRB:31 XRF:12 & C-GRB:31 XRR:69 & XRF:12 XRR:69 & \multirow{2 }{*}{Ch10}\\
    \cdashline{2-5}[0.8pt/2pt]
                        & $p$-value & $2.35\times 10^{-5}$ & $1.27\times 10^{-6}$ & $0.09$ & \\
\enddata
    \tablecomments{\footnotesize
     We list two references of the X-ray flare data in the Reference column. ``Ch10" indicates Chincarini et al. (2010), and ``Yi16" indicates Yi et al. (2016). $E_{grb,p}$ indicates the GRB peak energy.}
\end{deluxetable}

\clearpage

\linespread{1}
\setlength{\tabcolsep}{1mm}
\begin{deluxetable}{lccccccl}
    \centering
    \tablecolumns{8}
    \tablecaption{Comparison of X-Ray Flare Temporal Properties in Different Samples\label{Tab:infor2}}
    \tablewidth{0pt}
    \tabletypesize{\small}
    \tablehead{
        \colhead{GRB Name}    &    \colhead{$\Delta t$}   &    \colhead{$t_p$}   &    \colhead{$\Delta t/t_p$}   &    \colhead{$\Delta F$}    &    \colhead{$F$}    &    \colhead{$\Delta F/F$}    &    \colhead{Ref.}\\
        \colhead{}    &    \colhead{s}   &    \colhead{s}   &    \colhead{}   &    \colhead{count s$^{-1}$}   &    \colhead{count s$^{-1}$}   &    \colhead{}    &    \colhead{}
    }
    \startdata
    \multirow{3}{*}{GRB 050406}    &    184.0    &    208.6    &    0.882    &    4.6    &    0.6        &    7.71    &    \multirow{1}{*}{Ch07}    \\
\cdashline{2-8}
    &    70.1$\pm$7.1    &    209.6$\pm$4.0    &    0.334$\pm$0.034    &    4.6$\pm$0.4    &    0.64$\pm$0.56    &    7.18$\pm$6.34    &    \multirow{1}{*}{Ch10}    \\
\cdashline{2-8}
    &    401.8$\pm$28.9    &    218.1$\pm$7.3    &    1.842$\pm$0.146    &    1.2$\pm$0.1    &    \nodata    &    \nodata    &    \multirow{1}{*}{Yi16}    \\
\cline{1-8}
    \multirow{8}{*}{GRB 050502B}    &    523.6    &    387.3        &    1.352    &    88.0    &    2.3    &    3.55    &    \multirow{3}{*}{Ch07}    \\
    &   \nodata     &    \nodata    &    \nodata    &    0.012    &   0.043     &    0.28    &    \\
    &   \nodata     &    \nodata    &    \nodata    &    0.027    &   0.017     &    1.57    &    \\
\cdashline{2-8}
    &    438.9$\pm$86.6    &    655.7$\pm$101.2    &    0.669$\pm$0.168    &    14.6$\pm$3.7    &    0.12$\pm$0.12    &    119.16$\pm$128.31    &    \multirow{3}{*}{Ch10}    \\
    &    234.9$\pm$31.4    &   683.4$\pm$53.1     &    0.344$\pm$0.053    &    64.2$\pm$4.1    &    0.12$\pm$0.12    &    543.87$\pm$567.90    &    \\
\enddata
    \tablecomments{\footnotesize ``Ch07", ``Ch10", and ``Yi16" in the Ref. column indicate Chincarini et al. (2007), Chincarini et al. (2010), and Yi et al. (2016), respectively.}
\end{deluxetable}

\clearpage

\clearpage

\clearpage

\clearpage

\clearpage

\clearpage

\clearpage

\clearpage

\clearpage

\clearpage

\clearpage

\clearpage

\clearpage

\clearpage

\clearpage

\clearpage

\clearpage

\clearpage

\clearpage

\clearpage

\clearpage

\end{document}